\documentclass[pra,onecolumn,unsortedaddress,superscriptaddress,notitlepage]{revtex4-1}
\usepackage{graphicx}
\usepackage{epsfig,color}
\usepackage{amsmath,amssymb,eufrak}
\usepackage{amsthm}
\usepackage{enumerate}
\usepackage[normalem]{ulem}
\usepackage{hyperref}
\usepackage{amsmath,amssymb,txfonts}

\newcommand{\be}{\begin{equation}}
\newcommand{\ee}{\end{equation}}

\usepackage{mathtools}
\DeclarePairedDelimiter\bra{\langle}{\rvert}
\DeclarePairedDelimiter\ket{\lvert}{\rangle}
\DeclarePairedDelimiterX\braket[2]{\langle}{\rangle}{#1 \delimsize\vert #2}

\newcommand{\mr}{\mathrm}

\newcommand{\mc}{\mathcal}

\newcommand{\Up}{1}
\newcommand{\Dn}{2}

\begin{document}

\title{Projected Entangled Pair States with non-Abelian gauge symmetries: an SU(2) study}

\date{\today}

\author{Erez Zohar}
\affiliation{Max-Planck-Institut f\"ur Quantenoptik, Hans-Kopfermann-Stra\ss e 1, 85748 Garching, Germany.}

\author{Thorsten B. Wahl }
\affiliation{Rudolf Peierls Centre for Theoretical Physics, Oxford, 1 Keble Road, OX1 3NP, United Kingdom.}

\author{Michele Burrello}
\affiliation{Max-Planck-Institut f\"ur Quantenoptik, Hans-Kopfermann-Stra\ss e 1, 85748 Garching, Germany.}

\author{J. Ignacio Cirac}
\affiliation{Max-Planck-Institut f\"ur Quantenoptik, Hans-Kopfermann-Stra\ss e 1, 85748 Garching, Germany.}

\begin{abstract}
Over the last years, Projected Entangled Pair States have demonstrated great power for the study of many body systems, as they naturally describe ground states of gapped many body Hamiltonians, and suggest a constructive way to encode and classify their symmetries. The PEPS study is not only limited to global symmetries, but has also been extended and applied
for local symmetries, allowing to use them for the description of states in lattice gauge theories. In this paper we discuss PEPS with a local, SU(2) gauge symmetry, and demonstrate the use of PEPS features and techniques for the study of a simple family of many body states with a non-Abelian gauge symmetry. We present, in particular, the construction of fermionic PEPS able to describe both two-color fermionic matter and the degrees of freedom of an SU(2) gauge field with a suitable truncation.
\end{abstract}

\maketitle

\tableofcontents

\section{Introduction}

The complete understanding of the standard model is one of the most fascinating challenges for physics. The standard model constitutes a pillar of our scientific knowledge and its analytical study achieved impressive results in more than 60 years of history. Despite that, an exact solution of quantum chromodynamics (QCD) remains an open problem and the non-perturbative phenomena which characterize it continue to be a target of intense research (see \cite{Brambilla2014} for a review on the subject).

Numerical investigations based on non-perturbative approaches are a necessary tool to support the progress in this field and, in the last decades, a broad set of theoretical and computational tools have been developed to overcome the complexity in the description of the continuous quantum field theories at the basis of the standard model. Some of the most successful techniques have been inspired by the ideas of Wilson \cite{Wilson} and are based on the study of lattice gauge theories (LGT) \cite{KogutSusskind,KogutLattice,Kogut1983}. These are lattice representations of gauge theories which rely on the discretization of space and time dimensions and constitute an ideal framework for Monte Carlo calculations. Thanks to the LGT construction, Monte Carlo analyses of lattice QCD achieved many remarkable results, including, for instance, the calculation with high accuracies of the mass spectra of both the hadrons and light mesons (see, for example, the recent reviews \cite{Brambilla2014,FLAG2013}).

In recent years, however, the new experimental results and the advances in the study of the standard model have pushed forward the frontier of the research in QCD and have posed again old questions with renewed vehemence.

A paradigmatic example is offered by the new realizations of a quark-gluon plasma reached at LHC through proton-nucleus collisions \cite{CMS2013,Alice2013,Atlas2013}. These observations underline once again the necessity of understanding the phase diagram of QCD at high densities and temperatures. The analysis of similar experimental results, however, requires the development of new techniques to describe many-body problems with non-Abelian gauge symmetries for arbitrary chemical potentials. Such techniques must provide tools to examine both the correlations and the dynamics of these extreme phases of matter and, ultimately, to understand the thermodynamic phases of QCD, their transitions (or crossovers) and the mechanisms regulating different energy scales from confinement to asymptotic freedom.

In high-density regimes, the usual Monte Carlo approaches traditionally suffer from limitations due to the well-known sign problem for fermions \cite{Troyer2005}, therefore new tactics seem necessary to sustain the traditional strategies in these analyses. A promising possibility in this direction is to borrow the tools developed in the last decades for the study of condensed matter many-body physics. Among them, tensor networks occupy a prominent role and are indeed tailored to describe correlations and real-time dynamics of many-body systems \cite{Verstraete2008,Orus2014}, enabling to examine their properties also in the fermionic case.

In the development of tensor network techniques for the description of LGT, it has been natural to start with the simplest playground, namely one-dimensional systems, where matrix product states, the tensor network ansatz at the basis of the density matrix renormalization group \cite{Schollwock2011}, can be exploited to determine spectra and real-time evolution of these systems. For one-dimensional lattice gauge theories many results have been obtained with matrix product states, concerning both Abelian \cite{Banuls2013,Banuls2013a,Rico2013,Kuhn2014,Buyens2014,Saito2014,Banuls2015,Pichler2015,Buyens2015,Banuls2016,Buyens2016} and non-Abelian \cite{Silvi2014,Kuhn2015,Milsted2016,Silvi2016} gauge theories.
In particular, matrix product states allowed for a fully non-perturbative calculation of the Schwinger model spectrum with accuracies comparable with the best Monte Carlo results \cite{Banuls2013,Banuls2013a}, the analysis of chiral condensates at high temperatures \cite{Saito2014,Banuls2016,Buyens2016}, the study of the phase diagram of related link models \cite{Rico2013}, and the study of dynamical effects such as the Schwinger particle creation mechanism \cite{Buyens2014} and the string breaking in U(1) \cite{Pichler2015,Buyens2015} and SU(2) \cite{Kuhn2015} gauge theories. Furthermore the effects of truncations and finite-size limitations which characterize these approaches have been discussed to verify the efficiency in the simulation of gauge theories with continuous groups \cite{Kuhn2014} (see also the recent review \cite{Dalmonte2016}).

The challenges posed by high-energy physics, however, require to move to the more complicated scenario of higher dimensions, thus generalizing the matrix product states results to projected entangled pair states (PEPS). For 2D systems pioneering works established the possibility of including local gauge symmetries in PEPS \cite{Tagliacozzo2014,Haegeman2014}. It has been shown that it is possible to include in gauge invariant tensor network states both fermionic matter and gauge fields degrees of freedom \cite{Zohar2015b,Zohar2016}. The former require a formulation in terms of fermionic PEPS \cite{Kraus2010} of these tensor network states, the latter must be introduced consistently by exploiting a suitable description of the gauge group elements \cite{Zohar2015}, which allows to address both the Gauss laws enforcing the local gauge symmetry and the magnetic flux terms which, in principle, characterize the Kogut-Susskind Hamiltonian in two dimensions.
Recently the entanglement properties of gauge invariant PEPS have been discussed \cite{Verstraete2015} and an alternative approach based on multi-scale entanglement renormalization ansatz has been proposed as well \cite{Osborne2016}.

Also for two-dimensional systems, the first proof-of-principle numerical investigations have been based on the simplest possible scenarios constituted by Abelian theories.
Discrete Abelian $\mathbb{Z}_2$ models have been analyzed as benchmarks to test the general structure of gauge invariant PEPS describing pure LGT \cite{Tagliacozzo2014} or systems with Higgs matter \cite{Haegeman2014}. Afterwards, the more challenging situation of a continuous U(1) gauge group has been considered \cite{Zohar2015b}.
In this generalization from discrete to continuous groups, however, due to the requirements to implement a numerical simulation, it is necessary to introduce a truncation which allows to approximate the infinite Hilbert space of the gauge field degrees of freedom in terms of a discrete set of states \cite{Zohar2015b}.

Non-Abelian states present an even higher level of complexity, due to the distinction between left and right generators of the gauge symmetry and the consequent rich structure of the gauge field degrees of freedom. Therefore the description of such states in terms of tensor network must introduce additional degrees of freedom in both the matter particles, which must transform non-trivially under the SU(2) group, and in the gauge field degrees of freedom, whose choice must be suitable for a definition of Gass laws related to the generators of the group.

This kind of construction raises the following question: what is the simplest variational ansatz that rigorously integrates the SU(2) gauge symmetry with the other symmetries expected in a well-defined lattice gauge theory (global U(1) symmetry, rotational and translational invariance) and displays the most important non-trivial features expected in a well-defined LGT?

In this work we will address this problem with a constructive approach and we will build fermionic PEPS that will allow us to define different thermodynamic phases, typical of an SU(2) gauge symmetry, and to characterize them in terms of both tensor network properties and observables.

In particular, in the process of defining a simple but non-trivial family of variational states, we must consider two intrinsic kinds of limitations of the tensor network construction. The first has an immediate physical connotation: in order to describe a continuous gauge theory within finite local Hilbert spaces, a suitable and consistent truncation scheme must be adopted \cite{Tagliacozzo2014,Zohar2015,Zohar2016}. The second is intrinsic of the embedding of a many body state in a PEPS: tensor networks are described in terms of  bonds carrying a limited amount of information, quantified by the dimension $D$ of a virtual Hilbert space (called the \emph{bond dimension}). Such dimension is an upper limit to the quantity of entanglement flowing in the network.

In particular, for non-Abelian theories, the physical truncation consists of a proper choice of the irreducible representations of the gauge group which describe not only the physical elements in each site of the physical lattice, but also the virtual states within the links of the tensor network. For a fermionic theory with SU(2) symmetry the simplest non-trivial choice is to consider only the representations $0$ and $1/2$, the first associated to the vacuum state or to fermionic singlets, and the second to physical (and virtual) fermions.

The bond dimension $D$, instead, defines a limit in the correlation of the physical states and on the information connecting the local tensors which constitutes the building blocks of the network; thus it is a quantity related to the number of free parameters in the construction of the variational ans\"atze. By increasing $D$ one can choose either to describe the same number of representations with more parameters or to include more representations in the theory description.

In this paper we will discuss the simplest possible scheme with $D=4$, corresponding to two virtual fermions per link of the network which are indeed sufficient to characterize all the local physical states in terms of the representations $0$ and $1/2$ of the SU(2) group.

In Sec. \ref{sec:global} we review the main formalism for the description of the physical and virtual states, and we describe the most general fermionic Gaussian state with global gauge invariance, rotational and translational symmetry and conservation of the matter particle number. Its mapping into two separate BCS states and its continuous limit is analyzed as well.

In Sec. \ref{sec:gauge} we discuss the local gauge symmetry and the introduction of the gauge field; in Sec. \ref{sec:pure} we examine the phase diagram and features of the pure lattice gauge theory model and we study the properties of its transfer matrices, whereas in Sec. \ref{sec:matter} we consider the full model with fermionic matter.

Throughout this paper, summation of doubly repeated indices is assumed. Representation indices, $j$, are often summed also when they appear once, or more than twice.
In these cases, wherever necessary, summation is explicitly written.

\section{Constructing fermionic PEPS with a global SU(2) symmetry} \label{sec:global}

\subsection{The physical system and its symmetries}

 The aim of this paper is to define a family of tensor network states characterized by all the main symmetries expected in a lattice gauge theory with SU(2)  symmetry. The final goal is therefore to build many-body states that describe both fermionic matter particles and gauge fields fulfilling local SU(2) gauge invariace. To set up a tensor network with these constraints, however, it is useful to start from the simpler realization of a global gauge invariant state modeling the fermionic matter only. This will set the stage for the definition of the main local elements entering the tensor network construction.

Hence, we begin our analysis by considering a fermionic model defined on a two dimensional spatial lattice (a $2+1$ dimensional system; the arguments described in this paper apply for higher dimensions as well, but we will restrict our discussion to $d=2$ for simplicity). On each site, or vertex, of the square lattice $\mathbf{x} \in \mathbb{Z}^2$ we define a two-component \emph{color} spinor $\psi_m\left(\mathbf{x}\right)$ ($m=1,2$). These fermions carry a global SU(2) charge, and are staggered \cite{Susskind1977}, i.e., the ones on even sites correspond to particles and the ones on odd sites - to anti-particles; to obtain a continuum limit, pairs of sites have to be merged together into a Lorentz spinor.

The state we wish to construct, $\left|\psi\right\rangle$, has to satisfy (by construction) the following symmetries:

\begin{enumerate}
  \item \emph{Translational Invariance}.
    Denote the lattice basis vectors by $\left\{\mathbf{\hat{e}}_i\right\}_{i=1}^2$. We can then define the quantum translation operators,
    \begin{equation}
    \mathcal{U}_T\left(\mathbf{\hat{e}}_i\right)\psi^{\dagger}_m\left(\mathbf{x}\right)\mathcal{U}^{\dagger}_T\left(\mathbf{\hat{e}}_i\right) = \psi^{\dagger}_m\left(\mathbf{x+\mathbf{\hat{e}}_i}\right)
    \label{ftrans}
    \end{equation}
    A translationally invariant, zero momentum state $\left|\psi\right\rangle$, in the case of staggered fermions, will satisfy
    \begin{equation}
    \mathcal{U}^2_T\left(\mathbf{\hat{e}}_i\right)\left|\psi\right\rangle = \left|\psi\right\rangle, \quad i=1,2
    \end{equation}
    (the double translation results from the staggering).

    \item\emph{Charge conjugation symmetry.} Due to the staggering, our state has to be invariant under the joint action of a single-site translation and a particle-hole transformation, whose composition corresponds to charge conjugation symmetry.

  \item \emph{Rotational Invariance}.
    On the square lattice, rotations are not defined by continuous transformations, but rather their remnant, corresponding to the discrete $C_4$ group. We denote a counter-clockwise $\pi/2$ rotation by $\Lambda$:
    \begin{equation}
    \Lambda \mathbf{x} = \Lambda \left(x_1,x_2\right) = \left(-x_2,x_1\right)
    \end{equation}
    We define the quantum operators reponsible for such a rotation of the physical operators, $\mathcal{U}_p$, as
    \begin{equation}
    \mathcal{U}_p \psi^{\dagger}_m\left(\mathbf{x}\right) \mathcal{U}^{\dagger}_p = \bar{\eta}_p \psi^{\dagger}_m\left(\Lambda\mathbf{ x}\right)
    \end{equation}
    where the phase $\eta_p^4 = -1$ arises from the staggered fermions prescription \cite{Zohar2015b,Susskind1977}. We will choose, for simplicity, $\eta_p=e^{i \pi /4}$.

    A rotational invariant state $\left|\psi\right\rangle$ satisfies:
    \begin{equation}
    \mathcal{U}_p \left|\psi\right\rangle = \left|\psi\right\rangle.
    \end{equation}

  \item \emph{Global SU(2) Invariance}. This is not a spatial symmetry, but rather one belonging to the internal color degree of freedom of the spinors, which mixes the color indices $m$. Before defining the global transformation generators, let us first define the transformation generators of each spinor locally.
      The spinors $\psi_m\left(\mathbf{x}\right)$ belong to the fundamental representation of SU(2): i.e., under an SU(2) transformation, it transforms with respect to the $j=1/2$ transformation rule.

        To be more explicit, out of any two-component fermionic creation and annihilation spinors $\alpha^{\dagger}_m,\alpha_m$ one may construct two sets of generators,
        \begin{equation}
        R^a\left(\alpha\right) = \frac{1}{2}\alpha^{\dagger}_m \sigma^a_{mn} \alpha_n
        \label{Rgen}
        \end{equation}
        and
        \begin{equation}
        L^a\left(\alpha\right) = \frac{1}{2}\alpha^{\dagger}_m \sigma^a_{nm} \alpha_n
        \label{Lgen}
        \end{equation}
        called the right and left generators respectively, where $\left\{\sigma^a\right\}_{a=1}^3$ are the Pauli matrices. These generators satisfy the right and left SU(2) Lie algebras:
        \begin{equation}
        \left[R^a,R^b\right]=i \epsilon^{abc}R^c
        \end{equation}
        and
        \begin{equation}
        \left[L^a,L^b\right]=-i \epsilon^{abc}L^c
        \end{equation}

        We denote by $\Theta_g$ and $\widetilde{\Theta}_g$ the quantum unitary operators corresponding to right and left transformations, respectively, corresponding to a group element
        $g \in$ SU(2). If the group element is parameterized by a set of three group parameters $\phi_a = \phi_a\left(g\right)$, we can write
        \begin{equation}
        \Theta_g = e^{i \phi_a R^a}
        \label{Rtrans}
        \end{equation}
        and
        \begin{equation}
        \widetilde{\Theta}_g = e^{i \phi_a L^a}
        \label{Ltrans}
        \end{equation}
        Then, the fermionic transformation rules are
        \begin{equation}
        \Theta_g \alpha^{\dagger}_m \Theta_g^{\dagger} = \alpha^{\dagger}_n D_{nm}\left(g\right)
        \end{equation}
        and
        \begin{equation}
        \widetilde{\Theta}_g \alpha^{\dagger}_m \widetilde{\Theta}_g^{\dagger} = D_{mn}\left(g\right) \alpha^{\dagger}_n \,,
        \end{equation}
        introducing the Wigner matrices associated to the representation $j=1/2$ of SU(2):
        \begin{equation}
        D_{mn} \left(g\right)= \left\langle jm \right|\Theta_g\left|jn\right\rangle = {\bar D}_{nm}\left(g^{-1}\right)\,.
        \label{WignerD}
        \end{equation}
        In particular, the physical spinors $\psi_m\left(\mathbf{x}\right)$ satisfy these transformation rules.

        For further discussion of the fermionic transformation rules, see appendix \ref{app1}.

        Now we can finally define the generators of the global SU(2) transformations,
        \begin{equation}
        \mathcal{G}_0^a = \frac{1}{2}\underset{\mathbf{x}}{\sum}\psi^{\dagger}_m\left(\mathbf{x}\right)\tilde{\sigma}^a\left(\mathbf{x}\right)\psi_n\left(\mathbf{x}\right)
        \end{equation}
        where
        \begin{equation}
    \tilde{\sigma}^a\left(\mathbf{x}\right) = \left\{
      \begin{array}{ll}
        \sigma^a, & \hbox{$\mathbf{x}$ is even;} \\
        -\sigma^{a\intercal}, & \hbox{$\mathbf{x}$ is odd.}
      \end{array}
    \right.\,,
        \end{equation}
    where the superscript $\intercal$ labels the transposed matrix.

    As a result of this definition, the spinors on even sites undergo a right transformation, while those on the odd sites undergo an (inverse) left transformation. The different behavior on the two sublattices is motivated by the combination of the staggered representation of the fermionic matter and the construction of fermionic PEPS which is most conveniently realized for paired superconducting states:
    in a staggered SU(2) models, the charges are defined exactly in the same way for both particles and anti-particles \cite{Zohar2015}. However, a fermionic PEPS construction is most conveniently based on Gaussian states \cite{Kraus2010}, which are decomposable into BCS paired states;
    to this purpose it is convenient to use the terminology of superconductivity rather than that of high energy physics, i.e. anti-particles are represented by fermions with a ``negative'' charge as in \cite{Zohar2015}. Thus, the state is constructed such that a particle-hole transformation $\psi^{\dagger}_m \rightarrow \psi_m$ on the odd sites will transform it to its high-energy counterpart (corresponding, for example, to the formulation in \cite{Susskind1977}) . This implies that the charges defined on odd vertices must assume the form used on even vertices after such transformation, which is achieved when the charges are constructed as above.

    A state $\left|\psi\right\rangle$ is an SU(2) scalar (or singlet) if it is invariant under generic gauge transformations:
    \begin{equation}
    e^{i \mathcal{G}_0^a \phi_a}\left|\psi\right\rangle =  \left|\psi\right\rangle\,.
    \label{SU2invglob}
    \end{equation}

    In the next section, we shall gauge this symmetry, i.e. make it local, to have a state which is locally gauge invariant, in a procedure involving the introduction of new physical degrees of freedom - the gauge field.

Note that if a state is a singlet of $\mathcal{G}_0^a$, it will also be a singlet of $\overline{\mathcal{G}}_0^a$, due to the charge conjugation symmetry.

\item \emph{Particle conservation}. We also require, in order to comply with high energy physics models, that the total number of fermions is conserved. As mentioned before, since we are using superconductivity language, the conserved charge will not be the total number of physical fermions in the PEPS, but rather the difference between the population of even and odd sites. The conserved charge is therefore
    \begin{equation}
    \mathcal{N}=\underset{x}{\sum}\left(-1\right)^{x_1+x_2}\psi^{\dagger}_m\left(\mathbf{x}\right)\psi_m\left(\mathbf{x}\right)
    \label{Ndef}
    \end{equation}
    and we will thus build tensor network states satisfying the condition:
    \begin{equation} \label{U1invglob}
    e^{i \phi \mathcal{N}}\left|\psi\right\rangle = \left|\psi\right\rangle \,.
    \end{equation}
        Here, as well, a particle-hole transformation on the odd sites maps such a condition into its standard high-energy description. As a results we observe that, in our case, the ``Dirac sea'' corresponds to the vacuum state, differently from the common picture in which all the odd sites are full and the even empty \cite{Zohar2015}.

    Differently from the previous SU(2) gauge symmetry \eqref{SU2invglob}, the additional U(1) symmetry \eqref{U1invglob}, as we shall see later, will remain a global symmetry and will not be gauged to be made local. In other words, we will build locally gauge invariant states $\ket{\psi}$ with an SU(2) rather than a U(2) symmetry.

\end{enumerate}

After having described the state $\left|\psi\right\rangle$, we can now turn to its construction with fermionic Gaussian PEPS, which will lead to a parametrization of states fulfilling the symmetries introduced above.

\subsection{The fPEPS structure and construction}

\subsubsection{The ingredients}
Our state $\left|\psi\right\rangle$ will be constructed as a PEPS, and thus is composed of local (fiducial) states, lying on each lattice vertex, which involve the physical degrees of freedom introduced above, as well as virtual ones on the legs emerging from the vertex. The two virtual degrees of freedom from the two edges of each links will then be projected out to a maximally entangled state, and we shall remain with a global lattice state of the so-called physical fermions (see below). Choosing the fiducial states and the bond states properly will ensure that this state has the symmetries we are interested in.

\begin{figure}
  \centering
  \includegraphics[width=0.25\columnwidth]{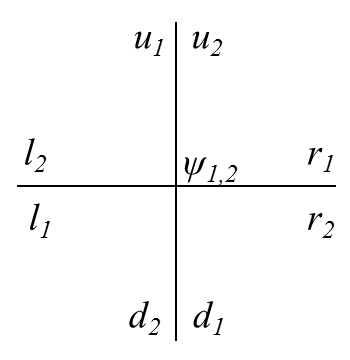}\\
  \caption{The fermionic ingredients of the fiducial state at a each given vertex: two physical modes $\psi^{\dagger}_m\left(\mathbf{x}\right)$, and eight virtual modes $r^{\dagger}_m\left(\mathbf{x}\right),u^{\dagger}_m\left(\mathbf{x}\right),l^{\dagger}_m\left(\mathbf{x}\right)$ and $d^{\dagger}_m\left(\mathbf{x}\right)$.}\label{fiducialfig}
\end{figure}

For each site $\mathbf{x}$, the local fiducial state describes the two physical modes $\psi^{\dagger}_m\left(\mathbf{x}\right)$ and eight virtual fermionic modes, residing on the closest edges of the four surrounding links. On each such edge we will define two virtual fermionic modes, forming similar SU(2) fundamental spinors. We shall name the corresponding creation operators
$r^{\dagger}_m\left(\mathbf{x}\right),u^{\dagger}_m\left(\mathbf{x}\right),l^{\dagger}_m\left(\mathbf{x}\right)$ and $d^{\dagger}_m\left(\mathbf{x}\right)$, for the edges lying on the right, up, left, down sides of the vertex respectively (see Fig. \ref{fiducialfig}).

We will define a local vacuum state $\ket{\Omega\left(\mathbf{x}\right)}=\left|\Omega_p\left(\mathbf{x}\right)\right\rangle\left|\Omega_v\left(\mathbf{x}\right)\right\rangle$
as the state without fermions in the site $\mathbf{x}$; for the sake of simplicity, we decompose such state into the physical and virtual vacua which fulfill the relations:
\begin{align}
&\psi_m\left(\mathbf{x}\right)\ket{\Omega_p\left(\mathbf{x}\right)} = 0 \,, \\
&l_m\left(\mathbf{x}\right)\ket{\Omega_v\left(\mathbf{x}\right)} =
r_m\left(\mathbf{x}\right)\ket{\Omega_v\left(\mathbf{x}\right)} =
u_m\left(\mathbf{x}\right)\ket{\Omega_v\left(\mathbf{x}\right)} =
d_m\left(\mathbf{x}\right)\ket{\Omega_v\left(\mathbf{x}\right)}= 0.
\end{align}
Note that, in principle, the fermionic Hilbert/Fock states should be defined in a more rigorous way accounting for fermionic statistics, and the tensor product structure is, in general, not defined. Based on the fact that we shall only combine operators with an even fermionic parity, however, hereafter we will adopt this simpler notation.

\begin{figure}
  \centering
  \includegraphics[width=0.45\columnwidth]{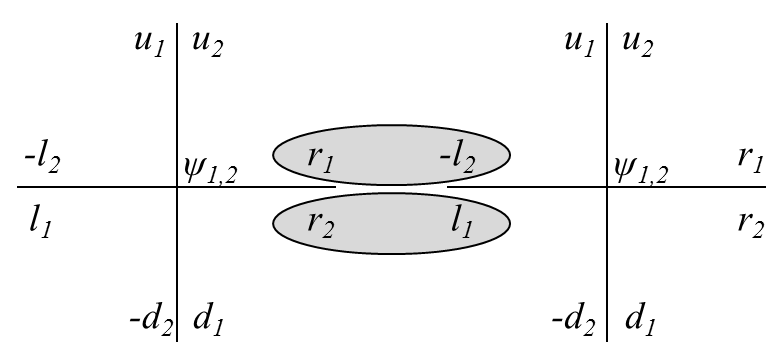}\\
  \caption{The maximally entangled state $\left|H\right\rangle$ on which the fiducial states are projected in the horizontal direction. Similar projection is done to $\left|V\right\rangle$ in the vertical direction. Note the signs, which correspond to the antisymmetric tensor $\epsilon_{mn}$ in the definition of the states (\ref{Hdef},\ref{Vdef}).}\label{figproj}
\end{figure}

The fiducial state on each link will be generated by an operator $A\left(\mathbf{x}\right)$ consisting of the ten fermionic creation operators
$\left\{\alpha^{\dagger}_m\left(\mathbf{x}\right)\right\} =
\left\{\psi^{\dagger}_m\left(\mathbf{x}\right),l^{\dagger}_m\left(\mathbf{x}\right),
r^{\dagger}_m\left(\mathbf{x}\right),u^{\dagger}_m\left(\mathbf{x}\right),d^{\dagger}_m\left(\mathbf{x}\right)\right\}$,
parameterized by a matrix $T$,
\begin{equation}
A\left( T,\mathbf{x}\right)=\text{exp}\left(T_{ij}\alpha^{\dagger}_i \left(\mathbf{x}\right) \alpha^{\dagger}_j \left(\mathbf{x}\right) \right).
\end{equation}
In general, the $T$ matrices may be different for even and odd sites, but, as we shall show, they are the same.
On each link we will project the virtual fermions onto maximally entangled bond states. To this purpose, let us define first the operators
\begin{equation}
H_{\mathbf{x}} = \frac{1}{2}\text{exp}\left(\epsilon_{mn}l^{\dagger}_m\left(\mathbf{x+\hat{e}}_1\right)r^{\dagger}_n\left(\mathbf{x}\right)\right) \, ;
\quad
h_{\mathbf{x}}= l_1\left(\mathbf{x+\hat{e}}_1\right)l^{\dagger}_1\left(\mathbf{x+\hat{e}}_1\right)l_2\left(\mathbf{x+\hat{e}}_1\right)l^{\dagger}_2\left(\mathbf{x+\hat{e}}_1\right)
r_1\left(\mathbf{x}\right)r^{\dagger}_1\left(\mathbf{x}\right)r_2\left(\mathbf{x}\right)r^{\dagger}_2\left(\mathbf{x}\right)
\label{Hdef}
\end{equation}
on horizontal links, and
\begin{equation}
V_{\mathbf{x}} = \frac{1}{2}\text{exp}\left(\epsilon_{mn}u^{\dagger}_m\left(\mathbf{x}\right)d^{\dagger}_n\left(\mathbf{x+\hat{e}}_2\right)\right) \, ;
\quad
v_{\mathbf{x}}= d_1\left(\mathbf{x+\hat{e}}_2\right)d^{\dagger}_1\left(\mathbf{x+\hat{e}}_2\right)d_2\left(\mathbf{x+\hat{e}}_2\right)d^{\dagger}_2\left(\mathbf{x+\hat{e}}_2\right)
u_1\left(\mathbf{x}\right)u^{\dagger}_1\left(\mathbf{x}\right)u_2\left(\mathbf{x}\right)u^{\dagger}_2\left(\mathbf{x}\right)
\label{Vdef}
\end{equation}
on the vertical ones,
where $\epsilon_{mn} = i \sigma^y_{mn}$ is the two dimensional anti-symmetric symbol (see Fig. \ref{figproj}). From these operators, we build the bond projectors
\begin{equation} \label{projectors}
\omega\left(\mathbf{x}\right)=H_{\mathbf{x}}h_{\mathbf{x}}H^{\dagger}_{\mathbf{x}}
\, , \quad \eta\left(\mathbf{x}\right)=V_{\mathbf{x}}v_{\mathbf{x}}V^{\dagger}_{\mathbf{x}} \,,
\end{equation}
which will be used for the contraction of local fiducial states needed for the PEPS construction.

Having defined all its ingredients, we can finally write the final form of our state $\left|\psi\right\rangle$ which will be parameterized by the matrix $T$,
\begin{equation}
\left|\psi\left(T\right)\right\rangle = \left\langle \Omega_v\right| \prod_{\mathbf{x}}\omega\left(\mathbf{x}\right)\prod_{\mathbf{x}}\eta\left(\mathbf{x}\right)\prod_{\mathbf{x}}A\left(T,\mathbf{x}\right)\left|\Omega\right\rangle\,.
\end{equation}
Here $\left|\Omega\right\rangle$ is the global vacuum for all the fermionic modes, whereas $\left|\Omega_v\right\rangle$ is the global vacuum for virtual modes only.
$\ket{\psi(T)}$ is, by construction, a Gaussian state for the physical fermionic modes. It is completely parameterized by $T$, and we shall show now how to do that in a way that satisfies the physical symmetries. For a start, note that translational invariance is already satisfied by our construction, since the matrix $T$ does not depend on the lattice position.

\subsubsection{Global  SU(2) invariance}
First we wish to parameterize $T$ is a way to guarantee the global SU(2) invariance (\ref{SU2invglob}).

\emph{Statement 1}. Define the operators
\begin{equation}
\hat\Theta_g = \widetilde{\Theta}_g^r\widetilde{\Theta}^u_g\Theta^{l\dagger}_g\Theta^{d\dagger}_g\Theta^{\dagger p}_g
= e^{i \phi_a G^a}
\label{hattheta}
\end{equation}
and
\begin{equation}
\widetilde{\hat\Theta}_g = \Theta^{r\dagger}_g\Theta^{u\dagger}_g\widetilde{\Theta}^l_g\widetilde{\Theta}^d_g\widetilde{\Theta}^{p}_g
= e^{-i \phi_a \widetilde{G}^a}
\label{hatthetatilde}
\end{equation}
with
\begin{equation}
G^a = L^a\left(r\right)+ L^a\left(U\right)- R^a\left(l\right)-R^a\left(d\right)-R^a\left(\psi\right)
\label{virtG}
\end{equation}
and
\begin{equation}
\tilde{G}^a = R^a\left(r\right)+ R^a\left(U\right)- L^a\left(l\right)-L^a\left(d\right)-L^a\left(\psi\right)\,.
\label{virtGl}
\end{equation}
Then, if $A$ is invariant under $\hat\Theta_g$, it is also invariant under $\widetilde{\hat{\Theta}}_g$, i.e.
\begin{equation} \label{gausslaw}
\hat\Theta_g A \hat\Theta_g^{\dagger} = A \Leftrightarrow \widetilde{\hat\Theta}_g A \widetilde{\hat\Theta}_g^{\dagger} = A
\end{equation}

\emph{Proof}.
The $G^a$ ($\widetilde{G}^a$) operators are generators of a right (left) SU(2) algebra, and thus the transformations $\hat\Theta_g$ ($\widetilde{\hat\Theta}_g$) may be seen as rotations.  If the operator $A$ is invariant under these rotations, it is block diagonal (by Schur's lemma). In our case invariance does not involve a phase, since we wish to describe states with no static charges, thus $A$ should be a scalar and its only nonvanishing matrix elements are the ones in the subspace of scalars under rotations (group singlets). However singlets for right rotations and left rotations are exactly the same states, which proves statement 1. We also see that the translation invariance is now stronger than the desired one - i.e., the state will be invariant for single site translations by this construction.

\emph{Statement 2}. The bond projectors defined in Eq. \eqref{projectors} satisfy
\begin{equation} \label{omegatr}
\widetilde{\Theta}_g^r\left(\mathbf{x}\right)\widetilde{\Theta}_g^l\left(\mathbf{x+\hat{e}}_1\right)
\omega\left(\mathbf{x}\right) \widetilde{\Theta}_g^{l\dagger}\left(\mathbf{x+\hat{e}}_1\right) \widetilde{\Theta}_g^{r\dagger}\left(\mathbf{x}\right)
=\omega\left(\mathbf{x}\right)
\, ; \quad
\Theta_g^r\left(\mathbf{x}\right)\Theta_g^l\left(\mathbf{x+\hat{e}}_1\right)
\omega\left(\mathbf{x}\right)
\Theta_g^{l\dagger}\left(\mathbf{x+\hat{e}}_1\right)\Theta_g^{r\dagger}\left(\mathbf{x}\right)
=\omega\left(\mathbf{x}\right)
\end{equation}
and
\begin{equation} \label{etatr}
\widetilde{\Theta}_g^d\left(\mathbf{x+\hat{e}}_2\right)\widetilde{\Theta}_g^u\left(\mathbf{x}\right)
\eta\left(\mathbf{x}\right)
\widetilde{\Theta}_g^{u\dagger}\left(\mathbf{x}\right)\widetilde{\Theta}_g^{d\dagger}\left(\mathbf{x+\hat{e}}_2\right)
=\eta\left(\mathbf{x}\right)
\, ; \quad
\Theta_g^d\left(\mathbf{x+\hat{e}}_2\right)\Theta_g^u\left(\mathbf{x}\right)
\eta\left(\mathbf{x}\right)
\Theta_g^{u\dagger}\left(\mathbf{x}\right)\Theta_g^{d\dagger}\left(\mathbf{x+\hat{e}}_2\right)
=\eta\left(\mathbf{x}\right)\,.
\end{equation}

\emph{Proof.} Using the transformation properties in Appendix \ref{app1} we obtain that
\begin{equation}
\widetilde{\Theta}_g^r\left(\mathbf{x}\right)\widetilde{\Theta}_g^l\left(\mathbf{x+\hat{e}}_1\right)
H_{\mathbf{x}} \widetilde{\Theta}_g^{l\dagger}\left(\mathbf{x+\hat{e}}_1\right) \widetilde{\Theta}_g^{r\dagger}\left(\mathbf{x}\right)
=\frac{1}{2}\text{exp}\left(\epsilon_{mn}D_{mk}\left(g\right)l_k^{\dagger}\left(\mathbf{x+\hat{e}}_1\right)D_{nq}\left(g\right)r^{\dagger}_q\left(\mathbf{x}\right)\right) = H_{\mathbf{x}}
\end{equation}
while $h_{\mathbf{x}}$ is trivially invariant (projecting to a singlet state). Then the invariance of $\omega\left(\mathbf{x}\right)$ under left transformations follows immediately. The invariance under right transformations of the horizontal projector results immediately from the same arguments used to prove statement 1. A similar proof applies for the vertical case.

\emph{Statement 3}. If $\hat\Theta_g A \hat\Theta_g^{\dagger} = A$, the state $\left|\psi\left(T\right)\right\rangle$ is invariant under the global SU(2) transformations defined in
(\ref{SU2invglob}).

\emph{Proof}.
\begin{equation}
\begin{aligned}
e^{i \phi \mathcal{N}}\left|\psi\left(T\right)\right\rangle &= \prod_{\mathbf{x}\text{ even}}\Theta^{p}_g\left(\mathbf{x}\right)
\prod_{\mathbf{x}\text{ odd}}\widetilde{\Theta}^{p}_g\left(\mathbf{x}\right)\left|\psi\left(T\right)\right\rangle\\
&=
\left\langle \Omega_v\right| \prod_{\mathbf{x}}\omega\left(\mathbf{x}\right)\prod_{\mathbf{x}}\eta\left(\mathbf{x}\right)
\prod_{\mathbf{x}\text{ even}}\Theta^{p}_g\left(\mathbf{x}\right)A\left(T,\mathbf{x}\right)\Theta^{p\dagger}_g\left(\mathbf{x}\right)
\prod_{\mathbf{x}\text{ odd}}\widetilde{\Theta}^{p}_g\left(\mathbf{x}\right)A\left(T,\mathbf{x}\right)\widetilde{\Theta}^{p\dagger}_g\left(\mathbf{x}\right)\left|\Omega\right\rangle
\end{aligned}
\end{equation}
Using statement 1, we get that $\hat\Theta_g A \hat\Theta_g^{\dagger} = A$ implies that we can transfer the transformation to the virtual degrees of freedom of the bond states. But these are invariant according
to statement 2, which proves statement 3.

As a conclusion, we see that if we find a $T$ for which $\hat\Theta_g A\left(T\right) \hat\Theta_g^{\dagger} = A$, we are done. There are two ways to do that. We will proceed here with a more physical approach, and another, more mathematical approach (using covariance matrices) is presented in appendix \ref{app2}. In the following, whenever we consider local properties of a single site, we will drop the coordinate label $\mathbf{x}$.

Without loss of generality, thanks to statement 1, we will consider only the transformation of an even vertex. There, we will sort the modes according to the type of transformation they undergo with
$\hat \Theta_g$. The right modes $\left\{\alpha_m^{Ri\dagger}\right\}=\left\{\psi_m^{\dagger},l_m^{\dagger},d_m^{\dagger}\right\}$ undergo a right transformation,
\begin{equation}
\hat \Theta_g \alpha_m^{Ri\dagger} \hat \Theta_g^{\dagger} = \Theta_g^{\alpha\dagger} \alpha_m^{Ri\dagger} \Theta_g^{\alpha} = \alpha_n^{Ri\dagger}D_{nm}\left(g^{-1}\right),
\end{equation}
while the left modes $\left\{\alpha_m^{Li\dagger}\right\}=\left\{r_m^{\dagger},u_m^{\dagger}\right\}$ undergo a left one,
\begin{equation}
\hat \Theta_g \alpha_m^{Li\dagger} \hat \Theta_g^{\dagger} = \widetilde{\Theta}^{\alpha}_g \alpha_m^{Ri\dagger} \widetilde{\Theta}_g^{\alpha\dagger} = D_{mn}\left(g\right)\alpha_n^{Li\dagger}.
\end{equation}
We can expand  $A$ as
\begin{equation}
A= \text{exp}\left(\frac{1}{2}\left(\hat{T}^{RR}\right)^{ij}_{mn}\alpha_m^{Ri\dagger}\alpha_n^{Rj\dagger} +
\frac{1}{2}\left(\hat{T}^{LL}\right)^{ij}_{mn}\alpha_m^{Li\dagger}\alpha_n^{Lj\dagger} +
\left(\hat{T}^{RL}\right)^{ij}_{mn}\alpha_m^{Ri\dagger}\alpha_n^{Lj\dagger}\right).
\end{equation}
Performing the transformation, we obtain
\begin{equation}
\begin{aligned}
\hat{\Theta}_g A \hat{\Theta}^{\dagger}_g &=
 \text{exp}\big{(}\frac{1}{2}\left(\hat{T}^{RR}\right)^{ij}_{mn}\alpha_{m'}^{Ri\dagger}D_{m'm}\left(g^{-1}\right)\alpha_{n'}^{Rj\dagger} D_{n'n}\left(g^{-1}\right)+
\frac{1}{2}\left(\hat{T}^{LL}\right)^{ij}_{mn}D_{mm'}\left(g\right)\alpha_{m'}^{Li\dagger}D_{nn'}\left(g\right)\alpha_{n'}^{Lj\dagger} +\\
& \left(\hat{T}^{RL}\right)^{ij}_{mn}\alpha_{m'}^{Ri\dagger}D_{m'm}\left(g^{-1}\right)D_{nn'}\left(g\right)\alpha_{n'}^{Lj\dagger}\big{)}.
\end{aligned}
\end{equation}

Demanding invariance, we get tensor equations for the components of $\hat T$. For example,
\begin{equation}
\left(\hat{T}^{RR}\right)^{ij}_{mn}=D_{mm'}\left(g\right)\left(\hat{T}^{RR}\right)^{ij}_{m'n'}D^\intercal_{n'n}\left(g\right)
\end{equation}
Since $D\epsilon D^\intercal = \epsilon$ (see  appendix \ref{app1}), we obtain that
\begin{equation}
\left(\hat{T}^{RR}\right)^{ij}_{mn} = \lambda^{Rij}\epsilon_{mn}
\end{equation}
Furthermore, $\epsilon_{mn}\alpha_m^{Ri\dagger}\alpha_n^{Rj\dagger}$ is symmetric under exchanging $i$ and $j$, and thus we get that
\begin{equation}
\lambda^{Rji}=\lambda^{Rij}
\end{equation}
Similar arguments lead us to
\begin{equation}
\left(\hat{T}^{LL}\right)^{ij}_{mn} = \lambda^{Lij}\epsilon_{mn}
\end{equation}
with
\begin{equation}
\lambda^{Lji}=\lambda^{Lij}\,.
\end{equation}
As for the mixed block,
\begin{equation}
\left(\hat{T}^{RL}\right)^{ij}_{mn}=D_{mm'}\left(g^{-1}\right)\left(\hat{T}^{RL}\right)^{ij}_{m'n'}D_{n'n}\left(g\right)
\end{equation}
which leads us to
\begin{equation}
\left(\hat{T}^{LL}\right)^{ij}_{mn} = \rho^{ij}\delta_{mn}\,.
\end{equation}

Eventually, we find that $A$ is defined by
\begin{equation} \label{Ageneric}
A=\text{exp}\left(T_{ij}a_{i}^{\dagger}b_{j}^{\dagger}\right)
\end{equation}
with the negative modes $\left\{a_i\right\}_{i=0}^4=\left\{\psi^{\dagger}_1,l^{\dagger}_1,d^{\dagger}_1,r^{\dagger}_2,u^{\dagger}_2\right\}$ are the ones which lower $G_z$
(i.e., $\left[G_z,a_i\right]=-a_i$)
and the positive modes $\left\{b_i\right\}_{i=0}^4=\left\{\psi^{\dagger}_2,l^{\dagger}_2,d^{\dagger}_2,r^{\dagger}_1,u^{\dagger}_1\right\}$ are the ones which raise it
(i.e., $\left[G_z,b_i\right]=b_i$),
and the matrix $T$ has the form
\begin{equation}
T=\left(
    \begin{array}{ccccc}
      \lambda^{R}_{11} & \lambda^{R}_{12} & \lambda^{R}_{13} & \rho_{11} & \rho_{12} \\
      \lambda^{R}_{12} & \lambda^{R}_{22} & \lambda^{R}_{23} & \rho_{21} & \rho_{22} \\
      \lambda^{R}_{13} & \lambda^{R}_{23} & \lambda^{R}_{33} & \rho_{31} & \rho_{32} \\
      -\rho_{11} & -\rho_{21} & -\rho_{31} & \lambda^{L}_{11} & \lambda^{L}_{12} \\
      -\rho_{12} & -\rho_{22} & -\rho_{32} & \lambda^{L}_{12} & \lambda^{L}_{22}\\
    \end{array}
  \right)
  \label{TSU2}
\end{equation}

\subsubsection{Rotational invariance}

\begin{figure}
  \centering
  \includegraphics[width=0.45\columnwidth]{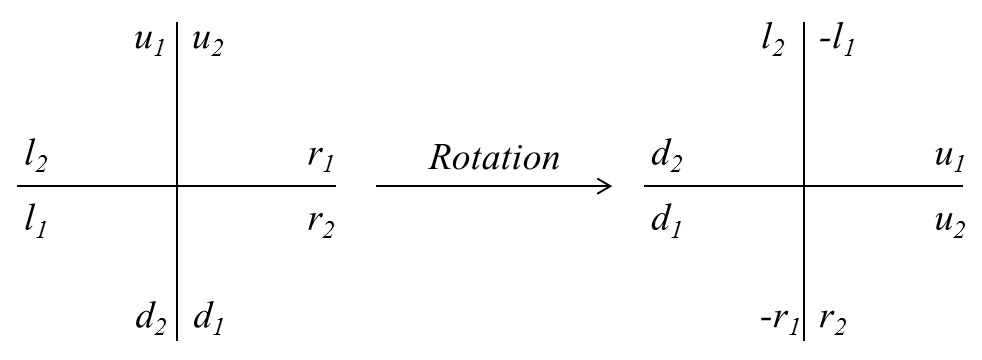}\\
      \caption{Rotation of the virtual modes of a fiducial state, following (\ref{virtrot}). The signs correspond to the $\epsilon_{mn}$ in the transformation.}\label{figrot}
\end{figure}

The next symmetry we wish to impose on our state $\left|\psi\right\rangle$ is rotational invariance.
For that, we have to introduce a rotation transformation for the virtual fermions, $\mathcal{U}_R$. We shall rotate them along with the physical ones, as follows (see Fig. \ref{figrot}):
\begin{equation}
\begin{aligned}
l^{\dagger}_m &\rightarrow \mathcal{U}_R l^{\dagger}_m \mathcal{U}^{\dagger}_R = \eta_u^{-1}d^{\dagger}_m \\
r^{\dagger}_m &\rightarrow \mathcal{U}_R r^{\dagger}_m \mathcal{U}^{\dagger}_R = \eta_u u^{\dagger}_m \\
u^{\dagger}_m &\rightarrow \mathcal{U}_R u^{\dagger}_m \mathcal{U}^{\dagger}_R =\eta_r^{-1}\epsilon_{mn}l^{\dagger}_n \\
d^{\dagger}_m &\rightarrow \mathcal{U}_R d^{\dagger}_m \mathcal{U}^{\dagger}_R =\eta_r \epsilon_{mn} r^{\dagger}_n
\label{virtrot}
\end{aligned}
\end{equation}
where $\eta_{u,r}$ are phases, as this transformation has to be unitary. This transformation leaves the projected states on the bonds invariant, and the $\epsilon_{mn}$ symbols which appear in two of the transformations account for the change of orientation of the rotated link (see appendix \ref{app1} for details).

Rotation invariance is achieved if $\mathcal{U}_p A \mathcal{U}^{\dagger}_p = \mathcal{U}^{\dagger}_R A \mathcal{U}_R$: if this is satisfied,
\begin{equation}
\begin{aligned}
\mathcal{U}_p \left|\psi\left(T\right)\right\rangle & =
\left\langle \Omega_v\right| \prod_{\mathbf{x}}\omega\left(\mathbf{x}\right)\prod_{\mathbf{x}}\eta\left(\mathbf{x}\right) \prod_{\mathbf{x}}\mathcal{U}_p A\left(T,\mathbf{x}\right) \mathcal{U}^{\dagger}_p \left|\Omega\right\rangle \\
& = \left\langle \Omega_v\right| \prod_{\mathbf{x}}\omega\left(\mathbf{x}\right)\prod_{\mathbf{x}}\eta\left(\mathbf{x}\right) \prod_{\mathbf{x}}\mathcal{U}^{\dagger}_R A\left(T,\mathbf{x}\right) \mathcal{U}_R \left|\Omega\right\rangle
= \left|\psi\left(T\right)\right\rangle
\end{aligned}
\end{equation}

One may define matrices $R_A$, $R_B$ such that
\begin{equation}
\begin{aligned}
\mathcal{U}_R a^{\dagger}_i \mathcal{U}^{\dagger}_R& = \left(R_A\right)_{ij} a_j^{\dagger} \\
\mathcal{U}_R b^{\dagger}_i \mathcal{U}^{\dagger}_R& = \left(R_B\right)_{ij} b_j^{\dagger}
\end{aligned}
\end{equation}
and then,
\begin{equation}
\mathcal{U}_p \mathcal{U}_R A\left(T\right) \mathcal{U}^{\dagger}_p \mathcal{U}^{\dagger}_R =  A \left(R_A^\intercal T R_B\right).
\end{equation}
Thus, the state will be rotationally invariant if
\begin{equation}
R_A^\intercal T R_B = T
\end{equation}

This results in
\begin{equation} \label{Tmatrix0}
T=\left(\begin{array}{ccccc}
0 & t & \eta_{u}^{-1}\eta_{p}^{-1}t & -\eta_{p}^{-2}\eta_{r}\eta_{u}^{-1}t & -\eta_{p}^{-3}\eta_{r}t\\
t & x & z/\sqrt{2} & 0 & \eta_{r}\eta_{u}z/\sqrt{2}\\
\eta_{u}^{-1}\eta_{p}^{-1}t & z/\sqrt{2} & \eta_{u}^{-2}x & -\eta_{u}^{-1}\eta_{r}z/\sqrt{2} & 0\\
\eta_{p}^{-2}\eta_{r}\eta_{u}^{-1}t & 0 & \eta_{r}\eta_{u}^{-1}z/\sqrt{2} & -\eta_{r}^{2}\eta_{u}^{-2}x & -\eta_{r}^{2}z/\sqrt{2}\\
\eta_{p}^{-3}\eta_{r}t & -\eta_{r}\eta_{u}z/\sqrt{2} & 0 & -\eta_{r}^{2}z/\sqrt{2} & -\eta_{r}^{2}x
\end{array}\right)
\end{equation}

The phases $\eta_{u},\eta_{r}$ are redundant and may be removed by redefining the virtual degrees of freedom in a suitable way (See Appendix \ref{app:phases}). In this way we can rewrite the most general form of the matrix $T$ as:
\begin{equation}
T=\left(\begin{array}{ccccc}
0 & t & \eta_{p}^{-1}t & -\eta_{p}^{-2}t & -\eta_{p}^{-3}t\\
t & x & z/\sqrt{2} & 0 & z/\sqrt{2}\\
\eta_{p}^{-1}t & z/\sqrt{2} & x & -z/\sqrt{2} & 0\\
\eta_{p}^{-2}t & 0 & z/\sqrt{2} & -x & -z/\sqrt{2}\\
\eta_{p}^{-3}t & -z/\sqrt{2} & 0 & -z/\sqrt{2} & -x
\end{array}\right)
\label{Tgen}
\end{equation}
with $t>0,z\geq0,x\in\mathbb{C}$ and $\eta_p=e^{i\frac{\pi}{4}}$.

\subsubsection{Particle number conservation}
The last symmetry we wish to impose on our fermionic state is the conservation of fermionic number \emph{after} a particle hole transformation, which means, in our terms, invariance under the transformations generated by $\mathcal{N}$ and defined in (\ref{Ndef},\ref{U1invglob}). Such staggered U(1) transformations promote the global gauge symmetry from SU(2) to SU(2)$\times$U(1)$=$U(2), although they correspond to multiplying the physical fermions $\psi^{\dagger}_m$ by a phase $e^{i \phi}$ on even sites, and by the opposite phase $e^{-i \phi}$ on odd sites, thus leading to a traditional U(2) global gauge symmetry only after the particle-hole transformation of the odd sites. We observe that the additional conservation of $\mathcal{N}$ rules out the possibility of describing a ``color superfluid'' or a ``diquark condensate'' \cite{Kogut1999,Kogut2001} in our model, in spite of the fact that such states do not violate the SU(2) gauge symmetry. The fermionic number conservation could be relaxed to describe these more exotic states, but in the following we will limit our analysis to the simplest scenario corresponding to this additional U(1) symmetry.

This symmetry is fulfilled only if we set $x=z=0$: then it is equivalent to shift the phase transformation (\ref{U1invglob}) from the physical to the virtual creation operators in both sublattices,
which leaves the projectors invariant and thus implements the additional symmetry. Hence, we conclude that the state $\left|\psi\left(T\right)\right\rangle$ which respects the particle number symmetry
is parameterized by
\begin{equation}
T=\left(\begin{array}{ccccc}
0 & t & \eta_{p}^{-1}t & -\eta_{p}^{-2}t & -\eta_{p}^{-3}t\\
t & 0 & 0 & 0 & 0\\
\eta_{p}^{-1}t & 0 & 0 & 0 & 0\\
\eta_{p}^{-2}t & 0 & 0 & 0 & 0\\
\eta_{p}^{-3}t & 0 & 0 & 0 & 0
\end{array}\right) \label{T0matrix}
\end{equation}
- a single, real, non-negative parameter $t$. Furthermore, we see that as the virtual fermions are decoupled, we get two decoupled fermionic PEPS, for each color - one for $\psi^{\dagger}_1$ and its four associated virtual fermions, and another for $\psi^{\dagger}_2$ and its four associated ones. The global SU(2) transformations will mix these two states, but will not make them coupled in any way. Only gauging the global SU(2) symmetry, as we shall see in the next section, will make the two colors coupled.

To see this explicitly, note that the fiducial state $\left|A\left(t\right)\right\rangle$ can be written now as
\begin{equation}
\left|A\left(t\right)\right\rangle = \left|A_1\left(t\right)\right\rangle \left|A_2\left(t\right)\right\rangle
\end{equation}
with
\begin{align}
\left|A_{1}\right\rangle &=\left(1+t\psi_{1}^{\dagger}\left(l_{2}^{\dagger}+\eta_{p}^{-1}d_{2}^{\dagger}+ir_{1}^{\dagger}+\eta_{p}u_{1}^{\dagger}\right)\right)\left|\Omega_{1}\right\rangle \label{FST1}\\
\left|A_{2}\right\rangle &=\left(1+t\psi_{2}^{\dagger}\left(-l_{1}^{\dagger}-\eta_{p}^{-1}d_{1}^{\dagger}+ir_{2}^{\dagger}+\eta_{p}u_{2}^{\dagger}\right)\right)\left|\Omega_{2}\right\rangle \label{FST2}
\end{align}
where $\left|\Omega_{1,2}\right\rangle$ are the respective (physical and virtual) local vacua.
Each of these fiducial states belong to a separate PEPS, connected by projectors created with
\begin{equation}
\begin{aligned}
&H_{1,\mathbf{x}} = \frac{1}{\sqrt{2}}\text{exp}\left(-l^{\dagger}_2\left(\mathbf{x+\hat{e}}_1\right)r^{\dagger}_1\left(\mathbf{x}\right)\right)
\quad ; \quad
H_{2,\mathbf{x}}  = \frac{1}{\sqrt{2}}\text{exp}\left(l^{\dagger}_1\left(\mathbf{x+\hat{e}}_1\right)r^{\dagger}_2\left(\mathbf{x}\right)\right)
\\
& V_{1,\mathbf{x}}  = \frac{1}{\sqrt{2}}\text{exp}\left(u^{\dagger}_1\left(\mathbf{x}\right)d^{\dagger}_2\left(\mathbf{x+\hat{e}}_2\right)\right)
\quad ; \quad
V_{2,\mathbf{x}}  = \frac{1}{\sqrt{2}}\text{exp}\left(-u^{\dagger}_2\left(\mathbf{x}_1\right)d^{\dagger}_1\left(\mathbf{x+\hat{e}}_2\right)\right)
\end{aligned}
\end{equation}

These two separate states are, in fact, identical; perform the canonical transformation
\begin{equation}
l^{\dagger}_1  \rightarrow -l^{\dagger}_1 \,,\quad
d^{\dagger}_1  \rightarrow -d^{\dagger}_1
\end{equation}
and obtain that the two PEPS are two copies of a single PEPS created by
\begin{equation}
\left|A\right\rangle =\left(1+t\psi^{\dagger}\left(l^{\dagger}+\eta_{p}^{-1}d^{\dagger}+ir^{\dagger}+\eta_{p}u^{\dagger}\right)\right)\left|\Omega\right\rangle
\end{equation}
and
\begin{equation}
\begin{aligned}
& H_{\mathbf{x}} = \frac{1}{\sqrt{2}}\text{exp}\left(l^{\dagger}\left(\mathbf{x+\hat{e}}_1\right)r^{\dagger}\left(\mathbf{x}\right)\right)\,,
\\
& V_{\mathbf{x}} = \frac{1}{\sqrt{2}}\text{exp}\left(u^{\dagger}\left(\mathbf{x}\right)d^{\dagger}\left(\mathbf{x+\hat{e}}_2\right)\right)\,.
\end{aligned}
\end{equation}

\subsubsection{The BCS state and its parent Hamiltonian} \label{BCSsection}
As the state is Gaussian, the final form of the physical state may be found through a Gaussian mapping (see Appendix \ref{app3}), and one obtains two copies of the BCS
p-wave state
\begin{equation}
\left|\psi\right\rangle =\underset{\mathbf{k}}{\otimes}\left(1+2t^{2}\left(\sin\left(k_{1}\right)-i\sin\left(k_{2}\right)\right)\psi^{\dagger}\left(\mathbf{k}\right)\psi^{\dagger}\left(-\mathbf{k}\right)\right)\left|\Omega\left(\mathbf{k}\right)\right\rangle\,,
\end{equation}
where the product is over half of the Brillouin zone, to avoid double counting.
For each copy we see that the pairing function $g(\mathbf{k})= 2t^{2}\left(\sin\left(k_{1}\right)-i\sin\left(k_{2}\right)\right) $ and corresponds, in real space, to maximally localized Cooper pairs since $\hat{g}(\mathbf{r_1} - \mathbf{r_2}) \neq 0$ only when $\mathbf{r_1}$ and $\mathbf{r_2}$ are nearest neighbors. This behavior is consistent with the constraint $x=z=0$ which hinders the formation of pairs with larger distances.

The correlation functions of the system are given by:
\begin{align}
\left\langle \psi_{\alpha}^{\dagger}\left(\mathbf{k}\right)\psi_{\beta}\left(\mathbf{q}\right)\right\rangle &=\frac{1}{2}\delta_{\alpha\beta}\delta_{\mathbf{k},\mathbf{q}}\left(1-R\left(\mathbf{k}\right)\right) \label{corrfun} \\
\left\langle \psi_{\alpha}\left(\mathbf{k}\right)\psi_{\beta}\left(\mathbf{q}\right)\right\rangle &=-\frac{1}{2}\delta_{\alpha\beta}\delta_{\mathbf{k},\mathbf{-q}}\Delta\left(\mathbf{k}\right)
\end{align}
with
\begin{align}
R\left(\mathbf{k}\right)&=\frac{1-4t^{4}\left(\sin^{2}\left(k_{1}\right)+\sin^{2}\left(k_{2}\right)\right)}{1+4t^{4}\left(\sin^{2}\left(k_{1}\right)+\sin^{2}\left(k_{2}\right)\right)} \,,\\
\Delta\left(\mathbf{k}\right)&\equiv P\left(\mathbf{k}\right)-iI\left(\mathbf{k}\right)=\frac{4t^{2}\left(\sin\left(k_{1}\right)-i\sin\left(k_{2}\right)\right)}{1+4t^{4}\left(\sin^{2}\left(k_{1}\right)+\sin^{2}\left(k_{2}\right)\right)}\,.
\end{align}
It is convenient to adopt a Nambu spinor notation, such that $\Psi\left(\mathbf{k}\right)=\left(\psi\left(\mathbf{k}\right),\psi^{\dagger}\left(-\mathbf{k}\right)\right)^\intercal$ and to define the functions:
\begin{align}
R_{0}\left(\mathbf{k}\right)&=1-4t^{4}\left(\sin^{2}k_{1}+\sin^{2}k_{2}\right)\,,\\
\varDelta_{0}\left(\mathbf{k}\right)&=P_{0}\left(\mathbf{k}\right)-iI_{0}\left(\mathbf{k}\right)=4t^{2}\left(\sin k_{1}-i\sin k_{2}\right)\,.
\end{align}
Then, the parent Hamiltonian, whose ground state is $\left|\psi\left(t\right)\right\rangle$, is given by
\begin{equation}
H=\frac{m}{2}\underset{\mathbf{k}}{\sum}\Psi^{\dagger}\left(\mathbf{k}\right)\mathcal{H}\left(\mathbf{k}\right)\Psi\left(\mathbf{k}\right)
\label{parham}
\end{equation}
for each color separately, with
\begin{equation}
\mathcal{H}\left(\mathbf{k}\right)=R_{0}\left(\mathbf{k}\right)\sigma_{z}+I_{0}\left(\mathbf{k}\right)\sigma_{y}+P_{0}\left(\mathbf{k}\right)\sigma_{x}
\end{equation}
giving rise to the dispersion relation
\begin{equation}
E\left(\mathbf{k}\right)=1+4t^{4}\left(\sin^{2}k_{1}+\sin^{2}k_{2}\right)\,.
\label{disprel}
\end{equation}
which is gapped for every value or $t$, thus entailing the absence of phase transitions.

 The separation of these two states and the strongly localized nature of their Cooper pairs are severe constraints imposed by the condition $x=z=0$, which, in turn, depends on the coexistence of rotational symmetry and conservation of the particle number for the matter fermions. We observe though, that this feature is a characteristic which emerges only due to the low bond dimension we are exploiting in the construction of our PEPS. By increasing the number of virtual modes it would be possible to obtain additional configurations in which the two BCS states would acquire a less localized nature, similarly to what showed in \cite{Zohar2015b} where a single species BCS state fulfilling an Abelian U(1) symmetry was implemented with two virtual fermionic modes per link (thus corresponding to four modes per link in order to obtain a full U(2) symmetry).

\subsubsection{The continuum limit of the parent Hamiltonian}
Let us derive the continuum field theory which is achieved as a limit of the parent Hamiltonian (\ref{parham}). To obtain a continuum limit, we first perform a Fourier transform
to real space, and then  a particle-hole transformation on the odd
sublattice
\begin{equation}
\psi^{\dagger}\left(\mathbf{x}\right)\longrightarrow\frac{1}{2}\left(\left(1+\left(-1\right)^{x_{1}+x_{2}}\right)\psi^{\dagger}\left(\mathbf{x}\right)+\left(1-\left(-1\right)^{x_{1}+x_{2}}\right)\psi\left(\mathbf{x}\right)\right)
\end{equation}
The Hamiltonian we obtain, then, takes the form
\begin{equation}
H=H_{R}+H_{\Delta}.
\end{equation}
The first part is
\begin{equation}
H_{R}=m\underset{\mathbf{x}}{\sum}\left(-1\right)^{x_{1}+x_{2}}\psi^{\dagger}\left(\mathbf{x}\right)\psi\left(\mathbf{x}\right)+mt^{4}\underset{\mathbf{x},i=1,2}{\sum}\left(-1\right)^{x_{1}+x_{2}}\left[\left(\psi^{\dagger}\left(\mathbf{x}+2\mathbf{\hat{e}}_{i}\right)-2\psi^{\dagger}\left(\mathbf{x}\right)+\psi^{\dagger}\left(\mathbf{x}-2\mathbf{\hat{e}}_{i}\right)\right)\psi\left(\mathbf{x}\right)+H.c.\right]
\end{equation}
where we identify a staggered mass term, as well as a staggered second
derivative term. If we block every two neighboring sites as a continuum
Lorentz spinor $\Psi$, with the first component belonging to the
even site and the second one - to the odd, we obtain the Hamiltonian
\begin{equation}
H_{R}=m\underset{\mathbf{x}}{\sum}\Psi^{\dagger}\left(\mathbf{x}\right)\beta\Psi\left(\mathbf{x}\right)-4mt^{4}a^{2}\underset{\mathbf{x}}{\sum}\Psi^{\dagger}\left(\mathbf{x}\right)\hat{\mathbf{k}}^{2}\beta\Psi\left(\mathbf{x}\right)
\end{equation}
with $\beta=\sigma_{z}$, where me made the replacement
\begin{equation}
\psi\left(\mathbf{x}+2\mathbf{\hat{e}}_{i}\right)-2\psi\left(\mathbf{x}\right)+\psi\left(\mathbf{x}-2\mathbf{\hat{e}}_{i}\right)=\Delta_{i}^{2}\psi\left(\mathbf{x}\right)\longrightarrow4a^{2}\underset{i}{\sum}\partial_{i}^{2}\psi\left(\mathbf{x}\right)=-4a^{2}\hat{\mathbf{k}}^{2}\psi\left(\mathbf{x}\right);
\end{equation}
here and in the following $\hat{\mathbf{k}}=-i\boldsymbol{\partial}$ is considered as an operator.

The second part is
\begin{equation}
H_{\Delta}=4imt^{2}\underset{\mathbf{x}}{\sum}\left(\psi^{\dagger}\left(\mathbf{x}\right)\psi\left(\mathbf{x}+\mathbf{\hat{e}}_{1}\right)-H.c.\right)+4mt^{2}\underset{\mathbf{x}}{\sum}\left(-1\right)^{x_{1}+x_{2}}\left(\psi^{\dagger}\left(\mathbf{x}\right)\psi\left(\mathbf{x}+\mathbf{\hat{e}}_{2}\right)+H.c.\right)
\end{equation}
unifying to a 2-component spinor as well, we obtain
\begin{equation}
H_{\Delta}=4imt^{2}\underset{\mathbf{x}}{\sum}\Psi^{\dagger}\left(\mathbf{x}\right)\left(\begin{array}{cc}
0 & \Delta_{1}-i\Delta_{2}\\
-\Delta_{1}+i\Delta_{2} & 0
\end{array}\right)\Psi\left(\mathbf{x}\right)
\end{equation}
however,
\begin{equation}
\Delta_{1}-i\Delta_{2}\longrightarrow a\left(\partial_{1}-i\partial_{2}\right)=-ia\left(\hat{k}_{1}-i\hat{k}_{2}\right)
\end{equation}
and altogether we obtain
\begin{equation}
H_{\Delta}=4imt^{2}a\underset{\mathbf{x}}{\sum}\Psi^{\dagger}\left(\mathbf{x}\right)\boldsymbol{\alpha}\cdot\hat{\mathbf{k}}\Psi\left(\mathbf{x}\right)
\end{equation}
where $\boldsymbol{\alpha}=\left(\sigma_{x},\sigma_{y}\right)$.

Finally, we rescale the fields,
\begin{equation}
\Psi\left(\mathbf{x}\right)\longrightarrow\frac{1}{a}\Psi\left(\mathbf{x}\right)
\end{equation}
to get
\begin{equation}
\left\{ \Psi_{\alpha}\left(\mathbf{x}\right),\Psi_{\beta}^{\dagger}\left(\mathbf{x}\right)\right\} =a^{-2}\delta_{\alpha\beta}\delta_{\mathbf{x},\mathbf{y}}\longrightarrow\delta_{\alpha\beta}\delta^{\left(2\right)}\left(\mathbf{x-y}\right)
\end{equation}
as well as
\begin{equation}
\underset{\mathbf{x}}{\sum}a^{2}\longrightarrow\int d^{2}x
\end{equation}
and obtain:
\begin{equation}
H=m\int d^{2}x\Psi^{\dagger}\left(\mathbf{x}\right)\left[\left(1-4t^{4}a^{2}\hat{\mathbf{k}}^{2}\right)\beta+4t^{2}a\boldsymbol{\alpha}\cdot\hat{\mathbf{k}}\right]\Psi\left(\mathbf{x}\right).
\end{equation}
This Hamiltonian corresponds to the first-quantized Hamiltonian
\begin{equation}
H_{1}=m\left[\left(1-4t^{4}a^{2}\hat{\mathbf{k}}^{2}\right)\beta+4t^{2}a\boldsymbol{\alpha}\cdot\hat{\mathbf{k}}\right];
\end{equation}
if we square it, using the usual ``Dirac trick'', we get
\begin{equation}
H_{1}^{2}=m^{2}\left[\left(1-4t^{4}a^{2}\hat{\mathbf{k}}^{2}\right)^{2}+16t^{4}a^{2}\hat{\mathbf{k}}^{2}\right]=m^{2}\left(1+4t^{4}a^{2}\hat{\mathbf{k}}^{2}\right)^{2}
\end{equation}
and so the energy spectrum is
\begin{equation}
E\left(\mathbf{k}\right)=m\left(1+4t^{4}a^{2}\mathbf{k}^{2}\right)
\end{equation}
unsurprisingly, in accordance with Eq. (\ref{disprel}). We
would like to give this dispersion relation a physical meaning. The
first term is the rest energy, and the second looks like a nonrelativistic
kinetic energy. Thus we can demand
\begin{equation}
4t^{4}a^{2}m=\frac{1}{2m}
\end{equation}
and obtain
\begin{equation} \label{tma}
t=\frac{1}{\sqrt{2^{3/2}ma}}
\end{equation}
this is a dimensionless parameter, setting the relation between the
mass and the lattice spacing - the only physical scales in the problem,
which can be expressed, after rescaling, by one. Thus, if one sets
$a$ to be a constant, increasing $t$ means decreasing the fermion
mass, and vice versa: $t=0$ corresponds to the completely static
fermions case (infinite mass) which makes a lot of physical sense,
as in this limit the physical fermions are completely decoupled from
the virtual ones and the state is simply the vacuum (or the Dirac sea after the particle-hole transformation).
On the
other hand, considering the mass as constant, an infinite $t$ will
correspond to the continuum limit.

After plugging this $t$ also to the other part of the Hamiltonian,
we finally get
\begin{equation}
H=\int d^{2}x\Psi^{\dagger}\left(\mathbf{x}\right)\left[\left(m-\frac{\hat{\mathbf{k}}^{2}}{2m}\right)\beta+\sqrt{2}\boldsymbol{\alpha}\cdot\hat{\mathbf{k}}\right]\Psi\left(\mathbf{x}\right)
\end{equation}
with the spectrum
\begin{equation}
E\left(\mathbf{k}\right)=\pm\left(m+\frac{\mathbf{k}^{2}}{2m}\right)
\end{equation}
This looks like the first approximation of the energy of a nonrelativstic
particle, with $k\ll m$. Thus, for $k\ll m$, we can also write
\begin{equation}
H'=\int d^{2}x\Psi^{\dagger}\left(\mathbf{x}\right)\left[\sqrt{m^{2}-\hat{\mathbf{k}}^{2}}\beta+\sqrt{2}\boldsymbol{\alpha}\cdot\hat{\mathbf{k}}\right]\Psi\left(\mathbf{x}\right)
\end{equation}
this Hamiltonian has exactly the same spectrum of the Dirac Hamiltonian.
Thus it is related to it by a similarity transformation. For $k<m$,
which is definitely our case since for us $k\ll m$, this Hamiltonian
is Hermitian, and thus may be unitarily diagonalized, and so related
to the Dirac Hamiltonian
\begin{equation}
H_{d}=\int d^{2}x\Psi^{\dagger}\left(\mathbf{x}\right)\left[m\beta+\boldsymbol{\alpha}\cdot \hat{\mathbf{k}}\right]\Psi\left(\mathbf{x}\right)
\end{equation}
by a unitary transformation. In fact, both of them belong to the family
of Hamiltonians
\begin{equation}
H\left(\theta\right)=\int d^{2}x\Psi^{\dagger}\left(\mathbf{x}\right)\left[\sqrt{m^{2}-\sinh^{2}\theta\,\hat{\mathbf{k}}^{2}}\beta+\cosh\theta\,\boldsymbol{\alpha}\cdot \hat{\mathbf{k}}\right]\Psi\left(\mathbf{x}\right)
\end{equation}
where $H_{d}=H\left(\theta=0\right)$ and $H'=H\left(\theta=\mathrm{arccosh}\left(\sqrt{2}\right)\right)\underset{k\ll m}{\longrightarrow}H$.

The resulting theory is non-relativisitic, i.e. not Lorentz-invariant in the continuum limit. This may be a feature of all PEPS with small (or finite) bond dimension, and thus it may be possible to approximate a Lorentz invariant theory better by increasing it.

\section{Gauging the symmetry - the local SU(2) case} \label{sec:gauge}

\subsection{The physical system}
We start from the state $\left|\psi\left(T\right)\right\rangle$ introduced in the last section. The local fiducial states
$\left|A\left(\mathbf{x}\right)\right\rangle = A\left(\mathbf{x}\right)\left|\Omega\left(\mathbf{x}\right)\right\rangle$
are invariant under the transformations (\ref{hattheta},\ref{hatthetatilde}), generated by (\ref{virtG},\ref{virtGl}),
which can be seen as virtual Gauss law operators, or generators of virtual gauge transformations.

These operators involve mostly virtual degrees of freedom which are contracted and do not participate physically in the final PEPS $\left|\psi\left(T\right)\right\rangle$,
therefore they may only contribute together to a global symmetry and do not give rise to physical local conservation laws. To overcome this limitation, we have to introduce physical degrees of freedom residing on the links of the lattice, in a way that allows to replace the virtual gauge transformations and Gauss laws (\ref{hattheta},\ref{hatthetatilde},\ref{virtG},\ref{virtGl}) by physical ones \cite{Haegeman2014,Zohar2015b}, involving only physical degrees of freedom, which will generate separate local symmetry transformations and thus gauge the symmetry.

For this, we introduce new physical degrees of freedom, belonging to the \emph{gauge field},
residing in  two more physical Hilbert spaces, ``side'' (s)
and ``top'' (t), on the right and up legs of the fiducial states. Each
of them is infinite, and spanned by SU(2) reprsentation states
$\left|jmn\right\rangle$ \cite{Zohar2015}. On each of these links we define
right and left SU(2) algebras, satisfying
\begin{equation}
\begin{aligned}
&\left[R^{a},R^{b}\right]=i\epsilon^{abc}R^{c} \\
&\left[L^{a},L^{b}\right]=-i\epsilon^{abc}L^{c} \\
&\left[R^{a},L^{b}\right]=0 \\
&L^{a}L^{a}=R^{a}R^{a}\equiv\mathbf{E}^{2}
\end{aligned}
\end{equation}

The representation basis states $\left|jmn\right\rangle$ satisfy
\begin{equation}
\begin{aligned}
\mathbf{E}^{2}\left|jmn\right\rangle &=j\left(j+1\right)\left|jmn\right\rangle \\
L_{z}\left|jmn\right\rangle &=m\left|jmn\right\rangle \\
R_{z}\left|jmn\right\rangle &=n\left|jmn\right\rangle
\end{aligned}
\end{equation}
We also define the ``rotation matrices'' or ``group elements''
$U_{mn}^{j}$, satisfying
\begin{equation}
U_{mn}^{j}\left|000\right\rangle =\frac{1}{\sqrt{2j+1}}\left|jmn\right\rangle\,.
\end{equation}
$U_{mn}^{j}$ are matrices of operators, given in this basis by \cite{Zohar2015}:
\begin{equation}
U_{mn}^{j}=\underset{J,K}{\sum}\sqrt{\frac{\mathrm{dim}\left(J\right)}{\mathrm{dim}\left(K\right)}}\left\langle JMjm|KN\right\rangle \left\langle KN'|JM'jm'\right\rangle \ket{KNN'} \bra{JMM'}\,.
\end{equation}
Right and left transformations of these operators are generated by
the SU(2) generators defined above, as
\begin{equation}
\left[R^{a},U_{mn}^{j}\right]=\left(U^{j}\mathcal{T}^{aj}\right)_{mn}\,;\quad\left[L^{a},U_{mn}^{j}\right]=\left(\mathcal{T}^{aj}U^{j}\right)_{mn}
\end{equation}
where $\mathcal{T}^{aj}$ is the $j$ matrix representation of the $a^{\rm th}$ SU(2)
generator (e.g., $\mathcal{T}^{a,j=1/2}=\sigma^a/2$). Then we can identify the generators as ``electric fields''
- right and left ones (the difference along the link is due to the
fact the group is non-Abelian, and thus the electromagnetic field
carries a charge - the rotation of the electric field along the link),
and $U_{mn}^{j}$ on a link is the connection - the Wilson line along
the link. The electric field generates transformations of the connection,
which will combine together to local gauge transformations.

Eventually, we wish our states to be invariant under local gauge transformations,
generated by the local Gauss law operators,
\begin{equation}
G_{a}\left(\mathbf{x}\right)=L_{a}\left(\mathbf{x},1\right)+L_{a}\left(\mathbf{x},2\right)-R_{a}\left(\mathbf{x-\hat{e}_{1}},1\right)-R_{a}\left(\mathbf{x-\hat{e}_{2}},2\right)-Q_{a}\left(\mathbf{x}\right)
\end{equation}
or, in terms of unitaries,
\begin{equation}
\Theta_{g}^{P}\left(\mathbf{x}\right)\equiv
\left\{
  \begin{array}{ll}
    \widetilde{\Theta}_{g}^{s}\left(\mathbf{x}\right)\widetilde{\Theta}_{g}^{t}\left(\mathbf{x}\right)\Theta_{g}^{s\dagger}\left(\mathbf{x}-\mathbf{\hat{e}}_{1}\right)\Theta_{g}^{t\dagger}\left(\mathbf{x}-\mathbf{\hat{e}}_{2}\right)\Theta_{g}^{p\dagger}\left(\mathbf{x}\right), & \hbox{$\mathbf{x}$ even;} \\
    \widetilde{\Theta}_{g}^{s}\left(\mathbf{x}\right)\widetilde{\Theta}_{g}^{t}\left(\mathbf{x}\right)\Theta_{g}^{s\dagger}\left(\mathbf{x}-\mathbf{\hat{e}}_{1}\right)\Theta_{g}^{t\dagger}\left(\mathbf{x}-\mathbf{\hat{e}}_{2}\right)\widetilde{\Theta}_{g}^{p\dagger}\left(\mathbf{x}\right), & \hbox{$\mathbf{x}$ odd.}
  \end{array}
\right.
\end{equation}
See Appendix \ref{app1} and \ref{app:transfer} for more detail on the definition of $\Theta^{s/t}$.

\subsection{Gauging the fPEPS}

\begin{figure}
  \centering
  \includegraphics[width=0.25\columnwidth]{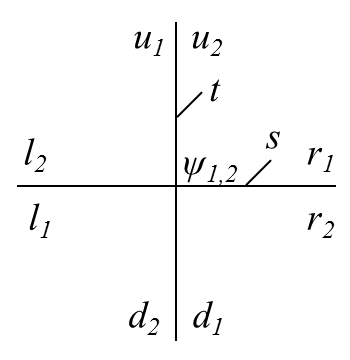}\\
  \caption{The ingredients of the gauged fiducial states: on top of the previously defined fermionic modes, introduce two more physical Hilbert spaces, $s$ and $t$, for the gauge field's degrees of freedom.}\label{gfig}
\end{figure}

Now we are ready to describe the gauging procedure of the state $\left|\psi\left(T\right)\right\rangle$ introduced above. We will denote by ${U}_{mn}^{s}\left(\mathbf{x}\right),{U}_{mn}^{t}\left(\mathbf{x}\right)$, the $j=1/2$ rotation matrices of the side and top physical subspaces of the site $\mathbf{x}$, respectively (see Fig. \ref{gfig}; in general we will omit the representation index $j$ when we refer to the fundamental representation, i.e. $U\equiv U^{j=1/2}$).
We introduce the unitary transformation
\begin{equation} \label{gauging1}
\begin{array}{ccc}
\mathcal{U}^{G}l_{m}^{\dagger}\left(\mathbf{x}\right)\mathcal{U}^{G\dagger} & = & l_{m}^{\dagger}\left(\mathbf{x}\right)\\
\mathcal{U}^{G}d_{m}^{\dagger}\left(\mathbf{x}\right)\mathcal{U}^{G\dagger} & = & d_{m}^{\dagger}\left(\mathbf{x}\right)\\
\mathcal{U}^{G}r_{m}^{\dagger}\left(\mathbf{x}\right)\mathcal{U}^{G\dagger} & = & \begin{cases}
U_{mn}^{s}r_{n}^{\dagger}\left(\mathbf{x}\right) & \hbox{$\mathbf{x}$ even}\\
\overline{U}_{mn}^{s}\left(\mathbf{x}\right)r_{n}^{\dagger}\left(\mathbf{x}\right) & \hbox{$\mathbf{x}$ odd}
\end{cases}\\
\mathcal{U}^{G}u_{m}^{\dagger}\mathcal{U}^{G\dagger} & = & \begin{cases}
U_{mn}^{t}\left(\mathbf{x}\right)u_{n}^{\dagger}\left(\mathbf{x}\right) & \hbox{$\mathbf{x}$ even}\\
\overline{U}_{mn}^{t}\left(\mathbf{x}\right)u_{n}^{\dagger}\left(\mathbf{x}\right) & \hbox{$\mathbf{x}$ odd}
\end{cases}
\end{array}
\end{equation}
and the gauged fiducial state
\begin{equation} \label{GFS}
\left|A^{G}\left(\mathbf{x}\right)\right\rangle =\mathcal{U}^{G}\left|A\left(\mathbf{x}\right)\right\rangle \left|0\right\rangle =\mathcal{U}^{G}A\left(\mathbf{x}\right)\mathcal{U}^{G\dagger}\left|\Omega\right\rangle \left|0\right\rangle \equiv A^{G}\left(\mathbf{x}\right)\left|\Omega\right\rangle \left|0\right\rangle
\end{equation}
where $\left|0\right\rangle $ is the vacuum of the gauge field (both
side and top). $\overline{U}$ is the complex conjugation of the matrix
operator $U$, which may be evaluated easily as $U \in$ SU(2),
by
\begin{equation} \label{ubar}
\overline{U}=U^{\dagger \intercal}=U^{-1\intercal}=\epsilon U\epsilon^\intercal=\mathrm{adj}\left(U^\intercal\right)\,.
\end{equation}

Let us consider the effect of the previous relation. Consider links
emanating from even vertices first. $U_{mn}^{s}r_{n}^{\dagger}$ and
$U_{mn}^{t}u_{n}^{\dagger}$ undergo left transformations of $U$
exactly as $r_{m}^{\dagger},u_{m}^{\dagger}$:
\begin{equation}
\widetilde{\Theta}^s_{g}U_{mn}^{s}\widetilde{\Theta}_{g}^{s\dagger}r_{n}^{\dagger}=D_{mm'}\left(g\right)U_{m'n}^{s}r_{n}^{\dagger}
\end{equation}
and similarly for $u$ (see appendix \ref{app1}). The original parameterization in Eq. \eqref{virtGl} was defining a virtual symmetry transformation as
$\widetilde{\Theta}_{g}^{r}\widetilde{\Theta}_{g}^{u}\Theta_{g}^{l\dagger}\Theta_{g}^{d\dagger}\Theta_{g}^{p\dagger}$; that means that now, instead, we have to rewrite this symmetry of the local (even) fiducial state as:
\begin{equation} \label{mixedgauge}
 \widetilde{\Theta}_{g}^{s}\widetilde{\Theta}_{g}^{t}\Theta_{g}^{l\dagger}\Theta_{g}^{d\dagger}\Theta_{g}^{p\dagger}\ket{A^G(\mathbf{x})}=\ket{A^G(\mathbf{x})}\,,
\end{equation}
where two virtual degrees of freedom have been replaced by physical ones (as
condition (49) in \cite{Zohar2016}). Furthermore, we have that
\begin{equation} \label{Urrel}
\Theta_{g}^{s}U_{mn}^{s}\Theta_{g}^{s\dagger}r_{n}^{\dagger}=U_{mn}^{s}\widetilde{\Theta}_{g}^{r}r_{n}^{\dagger}\widetilde{\Theta}_{g}^{r\dagger}
\end{equation}
(see appendix \ref{app1}) and similarly for $u,t$.

In the odd cases, we have
\begin{equation}
\widetilde{\Theta}_{g}^{s}\overline{U}_{mn}^{s}\widetilde{\Theta}_{g}^{s\dagger}r_{n}^{\dagger}=D_{m'm}\left(g^{-1}\right)\overline{U}_{m'n}^{s}r_{n}^{\dagger}
\end{equation}
similarly to the transformation of $r_{m}^{\dagger}$ before gauging ($\Theta_g^{r\dagger} r_{m}^{\dagger} \Theta_g^r = r_{m'}^{\dagger} D_{m'm}(g^{-1})$), and an analogous relation is fulfilled by $u$ and $t$. Then we can use our parameterization to replace the
symmetry from $\Theta_{g}^{r}\Theta_{g}^{u}\widetilde{\Theta}_{g}^{l\dagger}\widetilde{\Theta}_{g}^{d\dagger}\widetilde{\Theta}_{g}^{p\dagger}$
to $\widetilde{\Theta}_{g}^{s\dagger}\widetilde{\Theta}_{g}^{t\dagger}\widetilde{\Theta}_{g}^{l\dagger}\widetilde{\Theta}_{g}^{d\dagger}\widetilde{\Theta}_{g}^{p\dagger}$
. Besides, also the following relation holds:
\begin{equation} \label{thetasr}
\Theta_{g}^{s\dagger}\overline{U}_{mn}^{s}\Theta_{g}^{s}r_{n}^{\dagger}=\overline{U}_{mn}^{s}\Theta_{g}^{r}r_{n}^{\dagger}\Theta_{g}^{r\dagger}
\end{equation}
Eventually, let us define the gauged physical state
\begin{equation}
\left|\psi^{G}\right\rangle \equiv
\left\langle\Omega_{v}\right|
\underset{\mathbf{x}}{\prod}\omega\left(\mathbf{x}\right)\underset{\mathbf{x}}{\prod}\eta\left(\mathbf{x}\right)\underset{\mathbf{x}}{\prod}A^{G}\left(\mathbf{x}\right)\left|0\right\rangle \left|\Omega_{p}\right\rangle \left|\Omega_{v}\right\rangle
\end{equation}
and consider the action of the physical transformation
on it:
\begin{equation}
\Theta^{P}\left|\psi_{G}\right\rangle \equiv\left(\underset{\mathbf{x}}{\prod}\Theta_{g\left(\mathbf{x}\right)}^{P}\left(\mathbf{x}\right)\right)\left\langle \Omega_{v}\right|\underset{\mathbf{x}}{\prod}\omega\left(\mathbf{x}\right)\underset{\mathbf{x}}{\prod}\eta\left(\mathbf{x}\right)\underset{\mathbf{x}}{\prod}A^{G}\left(\mathbf{x}\right)\left|0\right\rangle \left|\Omega_{p}\right\rangle \left|\Omega_{v}\right\rangle
\end{equation}
This transformation commutes through the projectors to the fiducial
states. From acting on them, we get for even sites the transformation
\begin{equation}
\widetilde{\Theta}_{g\left(\mathbf{x}\right)}^{s}\widetilde{\Theta}_{g\left(\mathbf{x}\right)}^{t}\Theta_{g\left(\mathbf{x}+\mathbf{\hat{e}}_{1}\right)}^{s\dagger}\Theta_{g\left(\mathbf{x}+\mathbf{\hat{e}}_{2}\right)}^{t\dagger}\Theta_{g\left(\mathbf{x}\right)}^{p\dagger}
\end{equation}
which transforms, after acting on the local fiducial sites,  into
\begin{equation}
\Theta_{g\left(\mathbf{x}\right)}^{l}\left(\mathbf{x}\right)\Theta_{g\left(\mathbf{x}\right)}^{d}\left(\mathbf{x}\right)\widetilde{\Theta}_{g\left(\mathbf{x}+\mathbf{\hat{e}}_{1}\right)}^{r\dagger}\left(\mathbf{x}\right)\widetilde{\Theta}_{g\left(\mathbf{x}+\mathbf{\hat{e}}_{2}\right)}^{u\dagger}\left(\mathbf{x}\right) ,
\end{equation}
and for odd  sites
\begin{equation}
\widetilde{\Theta}_{g\left(\mathbf{x}\right)}^{s}\widetilde{\Theta}_{g\left(\mathbf{x}\right)}^{t}\Theta_{g\left(\mathbf{x}+\mathbf{\hat{e}}_{1}\right)}^{s\dagger}\Theta_{g\left(\mathbf{x}+\mathbf{\hat{e}}_{2}\right)}^{t\dagger}\widetilde{\Theta}_{g\left(\mathbf{x}\right)}^{p}
\end{equation}
which transforms into
\begin{equation}
\widetilde{\Theta}_{g\left(\mathbf{x}\right)}^{l\dagger}\widetilde{\Theta}_{g\left(\mathbf{x}\right)}^{d\dagger}\Theta_{g\left(\mathbf{x}+\mathbf{\hat{e}}_{1}\right)}^{r}\Theta_{g\left(\mathbf{x}+\mathbf{\hat{e}}_{2}\right)}^{u} .
\end{equation}
now we are left only with virtual transformations, which, once acted
to the left on the projectors, leave them invariant and the local
symmetry holds.

What about the other symmetries?
If on top of the fermionic translation (\ref{ftrans}) we add
\begin{equation}
\mathcal{U}_T\left(\mathbf{\hat{e}}_i\right)U^{l}_{mn}\left(\mathbf{x}\right)\mathcal{U}^{\dagger}_T\left(\mathbf{\hat{e}}_i\right) = U^{l}_{mn}\left(\mathbf{x+\hat{e}}_i\right) ,\quad i=1,2,l=t,s
\end{equation}
that one has to shift and conjugate the $U$ operators, translation
invariance becomes charge conjugation symmetry.

Rotations have to be considered more carefully; Once rotated, the
side degree of freedom becomes a top one, and the top degree of freedom
becomes the side degree of freedom of the left nearest neighbor -
but this is not enough, as when the top link is rotated to the left,
one has to exchange the left and right components of the field, as the orientation is inverted.

Thus, the rotation of $U^s$ is simpler, and is simply defined by
\begin{equation}
\mathcal{U}_p U^s_{mn}\left(\mathbf{x}\right) \mathcal{U}_p^{\dagger} = U^t_{mn}\left(\Lambda\mathbf{ x}\right)
\end{equation}
we can verify that it is satisfied by the parametrization; let us consider an even site, for example, where we have
\begin{equation}
\mathcal{U}_p \mathcal{U}_R U^s_{mn}\left(\mathbf{x}\right)r^{\dagger}_n\left(\mathbf{x}\right) \mathcal{U}_R^{\dagger}\mathcal{U}_p^{\dagger} = \eta_u U^t_{mn}\left(\Lambda\mathbf{ x}\right)u^{\dagger}_n\left(\Lambda \mathbf{ x}\right)
\end{equation}
- the same phase of (\ref{virtrot}), which tells us that the parametrization, in this case, holds after gauging.

For the top operators, we define
\begin{equation}
\mathcal{U}_p U^t_{mn}\left(\mathbf{x}\right) \mathcal{U}_p^{\dagger} = \bar{U}^{s}_{nm}\left(\Lambda\mathbf{ x - \hat{e}}_1\right)
\end{equation}
which takes the change of orientation (right $\leftrightarrow$ left) into account. Let us see this explicitly. By acting on $A^G$ with both $\mathcal{U}_p,\mathcal{U}_R$, we obtain the transformation (for an even $\mathbf{x}$, without loss of generality)
\begin{equation}
\begin{aligned}
l^{\dagger}_m\left(\mathbf{x}\right) &\rightarrow  \eta_u^{-1}d^{\dagger}_m \left(\Lambda \mathbf{x}\right) \\
U^s_{mn}\left(\mathbf{x}\right)r^{\dagger}_n\left(\mathbf{x}\right) &\rightarrow  \eta_u U^{t}_{mn} \left(\Lambda \mathbf{x}\right) u^{\dagger}_n \left(\Lambda \mathbf{x}\right) \\
U^t_{mn}\left(\mathbf{x}\right)u^{\dagger}_n\left(\mathbf{x}\right) &\rightarrow \eta_r^{-1} \bar{U}^s_{nm}\left(\Lambda\mathbf{ x - \hat{e}}_1\right) \epsilon_{nk}l^{\dagger}_k \left(\Lambda \mathbf{x}\right) \\
d^{\dagger}_m\left(\mathbf{x}\right) &\rightarrow \eta_r \epsilon_{mn} r^{\dagger}_n \left(\Lambda \mathbf{x}\right)
\label{virtrot2}
\end{aligned}
\end{equation}
However, since we are interested in rotating the whole physical state $\left|\psi^G\right\rangle$, in which the fiducial states, or the operators $A$, are acted  from the left
by the $\omega\left(\mathbf{x}\right)$ operators, we can consider the effect of these projections and obtain that the rotation is, effectively (for an even site again),
\begin{equation}
\begin{aligned}
l^{\dagger}_m\left(\mathbf{x}\right) &\rightarrow  \eta_u^{-1}d^{\dagger}_m \left(\Lambda \mathbf{x}\right) \\
U^s_{mn}\left(\mathbf{x}\right)r^{\dagger}_n\left(\mathbf{x}\right) &\rightarrow  \eta_u U^{t}_{mn} \left(\Lambda \mathbf{x}\right) u^{\dagger}_n \left(\Lambda \mathbf{x}\right) \\
U^t_{mn}\left(\mathbf{x}\right)u^{\dagger}_n\left(\mathbf{x}\right) &\rightarrow \eta_r^{-1}  \epsilon_{mn}l^{\dagger}_n \left(\Lambda \mathbf{x}\right) \\
d^{\dagger}_m\left(\mathbf{x}\right) &\rightarrow \eta_r \epsilon_{mn} U^s_{nk}\left(\Lambda\mathbf{ x }\right) r^{\dagger}_k \left(\Lambda \mathbf{x}\right)
\label{virtrot3}
\end{aligned}
\end{equation}
Similar results may be obtained for odd sites. Altogether we deduce that (\ref{virtrot3}) is a generalization of (\ref{virtrot}) which takes into account (and commutes with)
the gauging procedure of $\mathcal{U}^G$, and thus gauging $A$ into $A^G$ does not spoil the rotational invariance and the same parametrization of $T$ can be used also in the gauged case.

\subsection{Truncation of the Hilbert space}
The gauged PEPS introduced so far might seem, at first sight, problematic
for computational purposes: this is due to the simple fact that the
local Hilbert spaces on the links are infinite. This problem is avoided automatically, however, since we only have access to a finite part of each
of this Hilbert spaces, thanks to the virtual fermionic construction.

When we exapand $A^G$, we can get terms which involve products
of at most two elements of the same $U$ matrix. This is since the
gauging procedure makes sure that every appearance of such a matrix
element will be multiplied by a fermionic creation operator, and we
have two fermions on each bond. Further powers do not contribute thanks
to the fermionic statistics. However, fermionic statistics also implies
anti-symmetrization: thus, acting with a product of two elements of
the same $U$ must result in an anti-symmetric angular momentum state.
Since we start from the singlet, we deduce that the first action will
take us to the $j=1/2$, and the second can only bring us back to
the singlet, as $j=1$ is a symmetric representation.

We can see it explicitly: up to some global sign, such terms
will take the form
\begin{equation}
\underset{nk}{\sum}U_{mn}U_{lk}r_{n}^{\dagger}r_{k}^{\dagger}
\end{equation}
after acting on the fermionic part and considering anti-symmetrization, we get the gauge field contribution
\begin{equation}
\left(U_{m1}U_{l2}-U_{m2}U_{l1}\right)\left|000\right\rangle =\frac{1}{\sqrt{2}}\left(U_{m1}\left|\frac{1}{2},l,-\frac{1}{2}\right\rangle -U_{m2}\left|\frac{1}{2},l,\frac{1}{2}\right\rangle\right)
\end{equation}
each of the summands will involve a singlet contribution, and a $j=1$
contribution. The $j=1$ must vanish thanks to the anti-symmetry inherited
from the fermions. Let us calculate this contribution explicitly.
It is proportional to
\begin{equation}
\underset{NN'}{\sum}\braket*{\frac{1}{2},l,\frac{1}{2},m}{1N}  \left(\braket*{ 1N'}{\frac{1}{2},\frac{1}{2},\frac{1}{2},-\frac{1}{2}} -\braket*{ 1N'}{\frac{1}{2},-\frac{1}{2},\frac{1}{2},\frac{1}{2}} \right)\left|1,N,N'\right\rangle
\end{equation}
however, thanks to the symmetry of the Clebsch-Gordan coefficients \cite{Rose1995},
$\braket*{ 1N'}{\frac{1}{2},\frac{1}{2},\frac{1}{2},-\frac{1}{2}} =\braket*{ 1N'}{\frac{1}{2},-\frac{1}{2},\frac{1}{2},\frac{1}{2}} $,
and the whole $j=1$ contribution vanishes as expected.

Thus, we can effectively replace the infinite Hilbert spaces by finite
ones, with dimension five, spanned by the five states $\left|jmn\right\rangle $,
with $j=0,\frac{1}{2}$, $\left|m\right|,\left|n\right|\leq j$ .
The $U$ operator will be then given by
\begin{equation} \label{Uop}
U_{mn}=\frac{1}{\sqrt{2}}\left(\begin{array}{cc}
\left|++\right\rangle \left\langle 0\right|+\left|0\right\rangle \left\langle --\right| & \left|+-\right\rangle \left\langle 0\right|-\left|0\right\rangle \left\langle -+\right|\\
\left|-+\right\rangle \left\langle 0\right|-\left|0\right\rangle \left\langle +-\right| & \left|0\right\rangle \left\langle ++\right|+\left|--\right\rangle \left\langle 0\right|
\end{array}\right)
\end{equation}
the gauging procedure and symmetries will remain the same, if we define
$\overline{U}=\epsilon U\epsilon^\intercal$ (rather than complex
the conjugate, which holds only in the case $U$ is unitary, but unitarity
is lost in the truncation).

This correspondence also holds in the case of expectation values of operators,
such as the Wilson loops which will be later defined and addressed.
As long as the observable whose expectation values is calculated doesn't
involve more than a single power of $U_{mn}$ on a single link, there is no problem
with the truncation.
$U_{mn}$, acting on a ket which involves $j=0,1/2$, may change the $j$ value on the link to any value between $0-1$, but within an expectation value
one has to consider the bra as well. This will also contain $j=1/2$ at most, and thus
the $j=1$ contribution of the observables's $U_{mn}$ will never contribute.
Thus, indeed, expectation values of observables which do not involve more than a single power of $U_{mn}$ on each link,
when calculated in the truncated, finite Hilbert space approach introduced here, will be exactly identical to these
calculated within the full Hilbert space scheme.

That implies, eventually, that we do not work with an approximated,
truncated Hilbert space for each link: the choice of representing
the virtual fields by fermions forbids our state to leave this $j\leq1/2$
sector.

\section{The pure gauge theory} \label{sec:pure}
The first case we would like to study is the pure-gauge theory, i.e. when there is no dynamical matter. Practically, this means that the $\psi^{\dagger}_m$ modes are not included in the state $\left|\psi\right\rangle$ and $t=0$. On the other hand, since in this case the global U(1) symmetry is completely irrelevant, we can choose again nonzero values for the parameters $x,z$ and therefore these set the parametrization of the pure gauge states. The relevant parametrization is thus given by (\ref{Tgen}) with $t=0$. Furthermore, however, since now $\phi_t$ is not a relevant parameter and may be arbitrarily set to zero (or simply ignored), one may revisit the parametrization \eqref{Tmatrix0} and use the virtual symmetries described in Appendix \ref{app:phases} to simplify the definition of the local fiducial state in such a way that the final parametrization for the pure gauge states is given by (\ref{Tgen}), with $t=0;x,z \in\mathcal{R};z,x \geq 0$. That implies, in particular, that the signs of $x,z$ are not important, i.e. the PEPS is invariant under $x \rightarrow -x$ and $z \rightarrow -z$ transformations, which we shall use in the following.

\subsection{The phase diagram from a virtual PEPS symmetry} \label{sec:ph}
We continue with reducing further the set of parameters $x,z$. This is the result of a further symmetry of the virtual space which corresponds to a particle-hole symmetry of the virtual fermionic modes. Such a transformation is in general antiunitary, thus it may be represented as a product of a unitary operator and a complex conjugation  as customary in the description of superconducting states (see, for example, the classification in \cite{Ludwig2008}). However, below we shall only use it on operators which involve no complex numbers, and thus we can effectively represent it as a unitary transformation $\mathcal{U}^{PH}$.
\begin{equation}
\begin{array}{ccc}
\mathcal{U}^{PH}l_{m}^{\dagger}\mathcal{U}^{PH\dagger} & = & -\epsilon_{mn}l_{n}\\
\mathcal{U}^{PH}d_{m}^{\dagger}\mathcal{U}^{PH\dagger} & = & -\epsilon_{mn}d_{n}\\
\mathcal{U}^{PH}r_{m}^{\dagger}\mathcal{U}^{PH\dagger} & = & \epsilon_{mn}r_{n}\\
\mathcal{U}^{PH}u_{m}^{\dagger}\mathcal{U}^{PH\dagger} & = & \epsilon_{mn}u_{n}
\end{array}
\end{equation}
this transformation leaves the projectors invariant. For example,
\begin{equation}
\mathcal{U}^{PH}\left|H\right\rangle =\mathcal{U}^{PH}\exp\left(\epsilon_{mn}l_{m}^{\dagger}r_{n}^{\dagger}\right)\mathcal{U}^{PH\dagger}l_{1}^{\dagger}l_{2}^{\dagger}r_1^{\dagger}r_2^{\dagger}\left|\Omega\right\rangle =\exp\left(-\epsilon_{mn}l_{m}r_{n}\right)l_{1}^{\dagger}l_{2}^{\dagger}r_1^{\dagger}r_2^{\dagger}\left|\Omega\right\rangle =\exp\left(\epsilon_{mn}l_{m}^{\dagger}r_{n}^{\dagger}\right)\left|\Omega\right\rangle =\left|H\right\rangle \,.
\end{equation}

In the pure gauge theory, the physical fermions are absent ($t=0$) and the gauge degrees
of freedom are the only physical ones. The fermionic fiducial state, $\left|A\right\rangle$, is a fermionic Gaussian state of the virtual particles, and thus may be brought to a canonical BCS form. Such a state has a particle-hole symmetry, thus:
\begin{equation}
\mathcal{U}^{PH}\left|A\right\rangle = \left|A\right\rangle.
\end{equation}
However, although the physical state is invariant under the particle-hole transformation, the parameters $(x,z)$ on which it depends transform under $\mathcal{U}^{PH}$ to another pair of parameters, $(x',z')$ which describes the same physical state. We shall derive an explicit form for this mapping in parameter space and find its fixed points. This result has no significance for the pure fermionic state, but as it applies to the gauged state as well, this analysis shall provide a description of the phase diagram of the pure gauge theory.

We start by bringing the fermionic fiducial state $\left|A\right\rangle$ to a BCS form. For that, we perform a singular
value decomposition of the matrix $\tau$, which is the virtual-virtual block of $T$, to obtain
\begin{equation}
\tau=U_{L}\Lambda U_{R}^{\dagger}
\end{equation}
that defines a unitary transformation $\mathcal{U}^{BCS}$ acting on the negative and positive virtual modes:
\begin{equation}
\widetilde{a}_{k}^{\dagger}=\mathcal{U}^{BCS}a_{k}^{\dagger}\mathcal{U}^{BCS\dagger}=a_{i}^{\dagger}\left(U_{L}\right)_{ik}
\end{equation}
\begin{equation}
\widetilde{b}_{k}^{\dagger}=\mathcal{U}^{BCS}b_{k}^{\dagger}\mathcal{U}^{BCS\dagger}=\left(U_{R}^{\dagger}\right)_{kj}b_{j}^{\dagger}
\end{equation}
The fermionic vacuum is invariant under this transformation, and thus
\begin{equation}
\left|A\right\rangle =\exp\left(\underset{k}{\sum}\lambda_{k}\widetilde{a}_{k}^{\dagger}\widetilde{b}_{k}^{\dagger}\right)\left|\Omega\right\rangle
\end{equation}
Next, let us see how the particle-hole transformation works in this
basis. Since
\begin{equation}
\begin{array}{ccc}
\mathcal{U}^{PH}a_{i}^{\dagger}\mathcal{U}^{PH\dagger} & = & -b_{i}\\
\mathcal{U}^{PH}b_{j}^{\dagger}\mathcal{U}^{PH\dagger} & = & a_{i}
\end{array}
\end{equation}
we simply obtain that
\begin{equation}
\begin{array}{ccc}
\mathcal{U}^{PH}\widetilde{a}_{i}^{\dagger}\mathcal{U}^{PH\dagger} & = & -\widetilde{b}_{i}\\
\mathcal{U}^{PH}\widetilde{b}_{j}^{\dagger}\mathcal{U}^{PH\dagger} & = & \widetilde{a}_{i}
\end{array}
\end{equation}
and thus
\begin{equation}
\mathcal{U}^{PH}\left|A\right\rangle =\exp\left(\underset{k}{\sum}\lambda_{k}\widetilde{a}_{k}\widetilde{b}_{k}\right)\underset{k}{\prod}\widetilde{a}_{k}^{\dagger}\widetilde{b}_{k}^{\dagger}\left|\Omega\right\rangle.
\end{equation}
Here we may treat the particle hole transformation as unitary, as $\lambda_k$ are real numbers (singular values), and $U_R,U_L$ are real matrices, thus the transformed creation operators, when expanded in terms of the original ones, include no complex coefficients.

Now we can consider each of the $k$ modes separately:
\begin{equation}
\left(1+\lambda_{k}\widetilde{a}_{k}\widetilde{b}_{k}\right)\widetilde{a}_{k}^{\dagger}\widetilde{b}_{k}^{\dagger}\left|\Omega\right\rangle =\left(\widetilde{a}_{k}^{\dagger}\widetilde{b}_{k}^{\dagger}-\lambda_{k}\right)\left|\Omega\right\rangle =-\lambda_{k}\left(1-\lambda_{k}^{-1}\widetilde{a}_{k}^{\dagger}\widetilde{b}_{k}^{\dagger}\right)\left|\Omega\right\rangle.
\end{equation}
This implies that we now have a state whose singular values are given
by $\Lambda^{-1}$, plus a renormalization of $\det\left(\Lambda\right)$.
\footnote{This renormalization is irrelevant, especially if we had normalized
the original state: then each mode should have had a normalization
factor of $\frac{1}{\sqrt{1+\lambda_{k}^{2}}}$, and
\begin{equation}
\frac{\lambda_{k}}{\sqrt{1+\lambda_{k}^{2}}}=\frac{1}{\sqrt{1+\lambda_{k}^{-2}}}
\end{equation}
thus we obtain the proper normalization factor, consistently with the transformation being unitary.}

Altogether, we obtain that our new fiducial state is described by a different virtual block $\widetilde{\tau}$ of its $T$ matrix given by:
\begin{equation}
\widetilde{\tau}=-U_{L}\Lambda^{-1}U_{R}^{\dagger}.
\end{equation}
This corresponds (after taking into account that the signs of $x$ and $z$ are irrelevant thanks to the virtual phase transformation)
to the transformation
\begin{equation}
x\longrightarrow x'=\frac{x}{x^{2}-z^{2}},\quad z\longrightarrow z'=\frac{z}{x^{2}-z^{2}}
\label{Hyptran}
\end{equation}
which is an inversion over the hyperbolae $x^{2}-z^{2}=\pm1$. Thus,
combining it with the virtual phase symmetry,
we obtain that the region of the real $xz$
plane enclosed by the two hyperbolae in the
first quadrant includes all the physical states defined by our PEPS construction for $t=0$.
The region outside the hyberbolae constitutes an analogous copy of the same states.

\begin{figure}
  \centering
  \includegraphics[width=0.4\columnwidth]{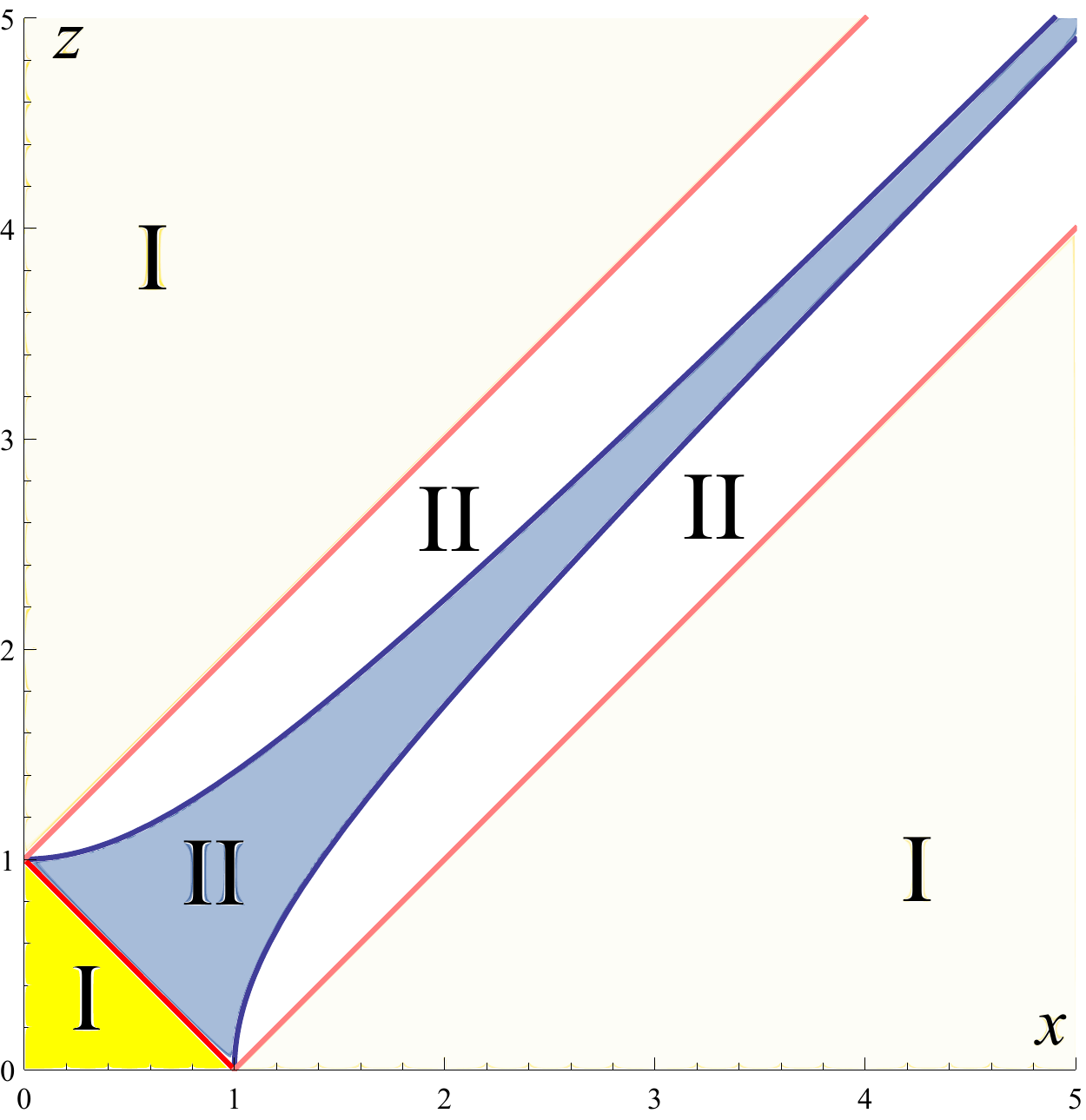}
  \caption{The two phases of the pure gauge theory. All the physical states lie between the blue hyperbolae $x^{2}-z^{2}=\pm1$ and the axes $x=0,z=0$, within the colored regions. The white regions contain copies of states within the colored regions, as points in the white regions are mapped to points in the colored ones by the transformation (\ref{Hyptran}). The red lines $\left|x\pm z\right|=1$ are invariant under (\ref{Hyptran}). Phase I is gapped, while phase II is gapless.}\label{phfig}
\end{figure}

Finally, the lines $\left|x\pm z\right|=1$ are mapped by (\ref{Hyptran})
among themselves, which
means that points from the physical region above and below the line
$z=1-x$ cannot be mapped from one to another. Thus these may correspond
to two different physical phases of the model, which we label I and II (see Fig. \ref{phfig}).

\subsection{The phase diagram as described by the transfer matrix}

Given our PEPS construction, all the states as a function of $x,z$ are ground states of local parent Hamiltonians \cite{PerezGarcia2008}; to evaluate the explicit form of this Hamiltonian is beyond our scope, but the study of the so-called transfer matrix \cite{Verstraete2008,Orus2014,Zauner2015} can help clarifying whether the PEPS at given $x,z$ is characterized by a mass gap or not. In particular, in the presence of a mass gap for the gauge fields, we expect that all the two-point correlation functions will decay exponentially with the distance, whereas if the system is gapless we expect to find algebraically decaying two-point correlations as well.

The two-point correlation functions, however, are not the only observable we can use to identify the properties of the pure gauge states as a function of $x$ and $z$. Other relevant order parameters are given by gauge-invariant string operators. In the following we will define the transfer matrices associated to two-point correlation functions and gauge-invariant Wilson lines and we will characterize the phase diagram of the system starting from their numerical analysis.

\subsubsection{The transfer matrix for two-point correlation functions}

The analytical study of the virtual particle-hole symmetry allows us to define invariant lines in the parameter space $(x,z)$, which are natural candidates to observe phase transitions, as discussed in Sec. \ref{sec:ph}. To corroborate this claim, we study the phase diagram numerically as described by the transfer matrices characterizing the PEPS construction.  Such transfer matrices map the virtual states associated to the modes $d_\Up$ and $d_\Dn$ of the vertical bonds from one row to the following.
The simplest case of transfer matrix is the one associated to two-point correlation functions and we will begin our analysis by commenting the connection between the eigenvalues of this transfer matrix and the decay of the correlations in the system.

For the evaluation of the transfer matrix, we consider a cylindrical geometry with periodic boundary conditions along the horizontal direction. We follow the approach presented in \cite{Zohar2015b} in the context of U(1) gauge symmetries and previously developed in \cite{Yang2015} in the context of fermionic chiral PEPS.

\begin{figure}[ht]
 \includegraphics[width=0.9\textwidth]{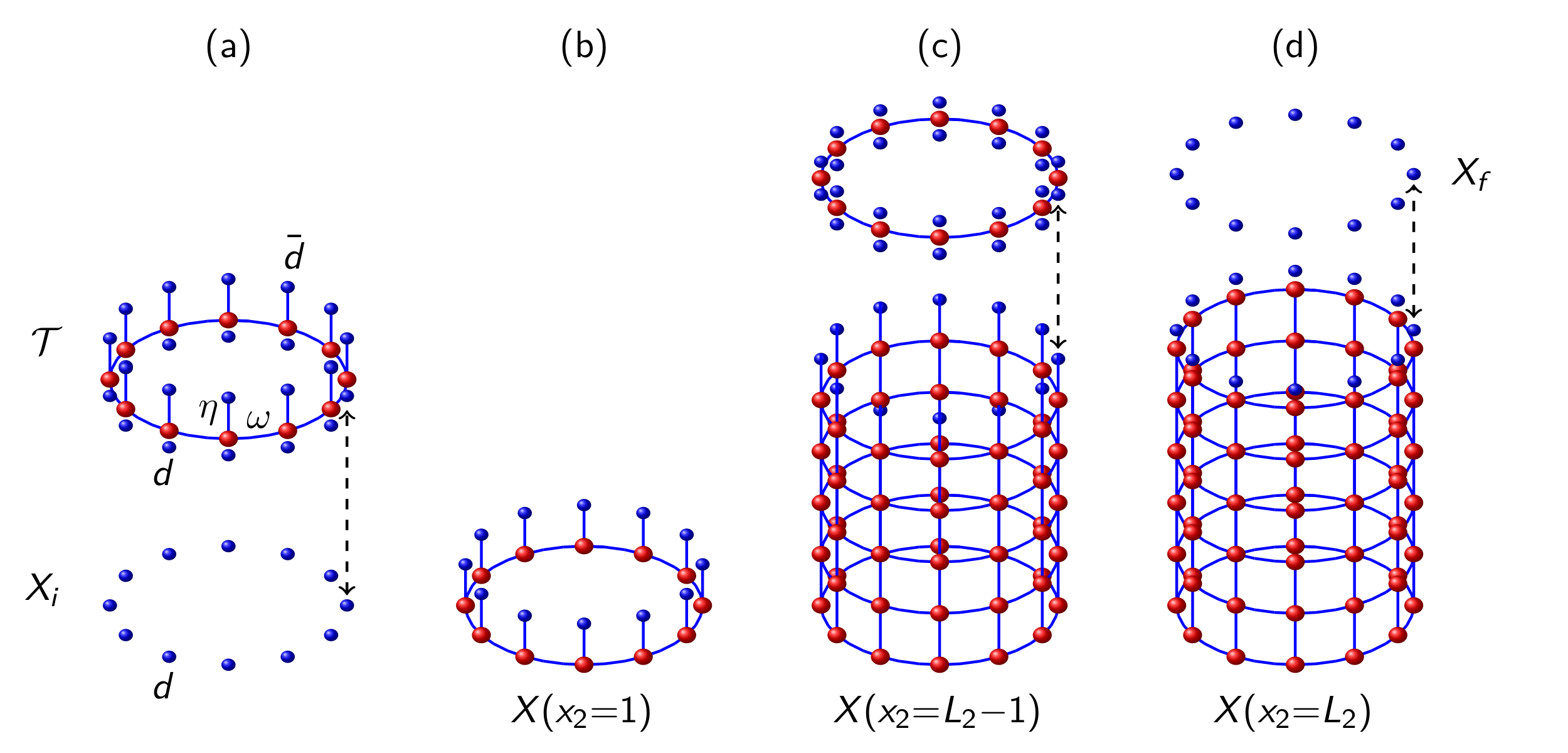}
 \caption{Contraction of the PEPS using the transfer matrix $\mc{T}$. (a) The transfer matrix $\mc{T}$ as defined in Eq. \eqref{transfer} (upper part) gets contracted with $X_i$ (lower part) according to Eq. \eqref{contract-transfer} to obtain $X(x_2 = 1)$ (b). (c) The procedure is repeated until the PEPS has $L_2-1$ rows, whereafter the final transfer matrix is applied, which acts on its down and up virtual particles (as opposed to the down virtual particles of the following row). (d) After that, the up virtual particles of the top row are contracted with those of the chosen final state $X_f$ resulting in the overall contraction of the PEPS.} \label{fig:transfer}
\end{figure}

For each row we define, starting from its local gauged fiducial states \eqref{GFS} and the projectors (\ref{projectors}), the transfer matrix $\left(\mathcal{T}\right) _{\tilde{d},\tilde{d}'}^{d,d'}$, 
\begin{equation} \label{transfer}
 \mathcal{T}(x_2)= {\rm tr}_{\psi,t,s,r,l,u}\left[\left(\prod_{x_1}\eta \omega A^G \right) \ket{\Omega(x_2)}\bra{\Omega(x_2)}\left( \prod_{x_1}A^{G\dag}\omega\eta\right)  \right]\,,
\end{equation}
where the trace is taken over all the physical and virtual modes in the row $x_2$ with the exception of the virtual modes $d_\Up$ and $d_\Dn$ and $\ket{\Omega(x_2)}\bra{\Omega(x_2)}$ is a shorthand notation for the projector over the empty subspace for all the modes of the row $x_2$  (see \cite{Yang2015} for more detail). In this way, $\mathcal{T}$ is an operator acting on the virtual modes $d_\Up$ and $d_\Dn$ in row $x_2$, and $\tilde{d}_\Up$ and $\tilde{d}_\Dn$ in row $x_2+1$, which are introduced through the projectors $\eta$, cf. Fig. \ref{fig:transfer}. In principle, the transfer matrix for even and odd rows could be different, due to the staggering which alternates even and odd fiducial states. However, the resulting transfer matrices are identical for even and odd rows because all the physical degrees of freedom are traced out in Eq. \eqref{transfer}. This can be verified by considering that even and odd fiducial states differ only for the use of the operators $U$ and $\bar{U}=\epsilon U \epsilon^\intercal$ in their definition [see Eqs. (\ref{gauging1},\ref{GFS})] and, when no observable acts on the gauge field degrees of freedom, the antisymmetric matrix $\epsilon$ has no effect when tracing the link states. The transfer matrix of the uppermost row $x_2 = L_2$ is slightly different, as it cannot be contracted with the $\eta$'s anymore, i.e., it is defined on its down and up virtual indices, see Fig. \ref{fig:transfer}.

$\mathcal{T}$ maps the density matrix $X(x_2)_{d,d'}$ of the vertical virtual modes of the row $x_2$ into the density matrix $X(x_2+1)_{\tilde{d},\tilde{d}'}$ associated to the following row (see \cite{Yang2015,Zohar2015b} for more detail):
\begin{equation} \label{contract-transfer}
 X(x_2+1)_{\tilde{d},\tilde{d}'} = \sum_{d,d'} \left(\mathcal{T}\right)_{\tilde{d},\tilde{d}'}^{d,d'} X(x_2)_{d,d'}.
\end{equation}
The transfer matrix naturally appears if one evaluates the two-point correlation function of observables located on single rows and separated vertically by $\delta y$ sites. As we see below, the $(d-1)$-th power of the transfer matrix occurs in this evaluation. This is why the gap between the largest and the second largest eigenvalues of the transfer matrix is connected to the decay of correlations in the thermodynamic limit.

Let us consider a finite system with $L_1$ columns (with periodic boundaries) and $L_2$ rows. For observables $O_{y}$ and $O_{y'}'$ restricted to columns $y$ and $y'$, respectively, one obtains
\begin{equation} \label{corr}
 \left\langle O_{y} O_{y'}' \right\rangle = \frac{{\rm tr}\left[X_i \left( \prod_{x_2=1}^{y-1} \mathcal{T}(x_2)\right)  \tilde{\mathcal{T}}_O(y) \left( \prod_{x_2=y+1}^{y'-1} \mathcal{T}(x_2)\right)  \tilde{\mathcal{T}}_{O'}(y') \left( \prod_{x_2=y'+1}^{L_2} \mathcal{T}(x_2) \right) X_f \right] }{{\rm tr}\left[X_i \left( \prod_{x_2=1}^{L_2} \mathcal{T}(x_2)\right)  X_f\right] }
\end{equation}
where we introduced the projectors $X_{i,f}$ which account for the boundary conditions of the vertical virtual modes on the first and last row respectively. The numerator in this expression describes a system in which observables are introduced only in the rows $y$ and $y'$, whereas the physical degrees of freedom for all the other values of $x_2$ are directly traced out (without prior insertion of physical operators) giving rise to the transfer matrices $\mathcal{T}$. For the special rows $y$ and $y'$ we adopted, instead, a modified transfer matrix which accounts for the measurements $O, O'$, 
\begin{equation} \label{transfer2}
 \tilde{\mathcal{T}}_{O}(y)= {\rm tr}\left[O_y\left(\prod_{x_1}\eta \omega A^G \right) \ket{\Omega(y)}\bra{\Omega(y)}\left( \prod_{x_1}A^{G\dag}\omega\eta\right)  \right].
\end{equation}
Let us analyze better the behavior of $ \left\langle O_{y} O_{y'}' \right\rangle$; we can approximate the transfer operator $\mathcal{T}$ by taking into account its two largest eigenvalues only, which determine the behavior of the correlations for large separations $y' - y$. Thus, under the assumption that $\mathcal{T}$ is  diagonalizable we consider
\begin{equation} \label{Tappr}
 \mathcal{T} \approx \lambda_0 \ket{l_0}\bra{r_0}+ \lambda_1 \ket{l_1}\bra{r_1}.
\end{equation}
Here $\ket{l_a}$ and $\bra{r_b}$ are left and right eigenvectors of $\mathcal{T}$ such that $\braket{r_b}{l_a} = \delta_{a,b}$.
If $\lambda_0>\lambda_1$, the correlation function $\left\langle O_{y} O_{y'}' \right\rangle$ will exponentially approach the value $\left\langle O_{y} \right\rangle \left\langle O_{y'}' \right\rangle$ as $(\lambda_1/\lambda_0)^{(y'-y-1)}$ decreases exponentially as a function of $y'-y$. To see that, we can first  assume that the PEPS is normalised such that $\lambda_0 = 1$ and rewrite Eq. \eqref{corr} as
\begin{equation}
\left\langle O_{y} O_{y'}' \right\rangle = \frac{\mathrm{tr}\left(X_i \mathcal{T}^{y-1} \mathcal{T}_O(y) \mathcal{T}^{y'-y-1} \mathcal{T}_{O'}(y')\mathcal{T}^{L_2-y'} X_f\right)}{\mathrm{tr}\left(X_i \mathcal{T}^{L_2} X_f\right)}.
\end{equation}
The expression for $\left\langle O_y\right\rangle$ or $\left\langle O_{y'}'\right\rangle$ is obtained by replacing $\mathcal{T}_{O'}(y')$ or $\mathcal{T}_{O}(y)$ by $\mathcal{T}$. For large $L_2$ (thermodynamic limit) $\mathcal{T}^{L_2-y'}$ can be replaced by $|L_0\rangle \langle R_0|$ and likewise $\mathcal{T}^y$ if we take $y$ large (of the order of $L_2$). Using \eqref{Tappr}, the two-point correlation reduces to
\begin{equation}
\left\langle O_y O_{y'}' \right\rangle - \left\langle O_y  \right\rangle \left\langle O_{y'}' \right\rangle
\propto
\lambda_1^{y'-y-1} \frac{\langle l_0| \mathcal{T}_O(y) |r_1\rangle \langle l_1| \mathcal{T}^{O'}(y') |r_0\rangle}{\mathrm{tr}\left(X_i \mathcal{T}^{L_2} X_f\right)}
\end{equation}
up to corrections in $\left(\lambda_2/\lambda_1\right)^{y'-y-1}$. The extension to degenerate $\lambda_1 = \lambda_2$ is straightforward. However, this demonstrates only an exponential decay of correlations in the
thin cylinder limit, which corresponds to the exponential decay of correlations in Matrix Product States\cite{Fannes1992}. If $L_1$ increases with $L_2$, the number of eigenvalues also increases exponentially, so even in the presence of a gap of the transfer operator (i.e., all $\lambda_i$ are upper bounded by $c < 1$), the eigenvalues could in principle proliferate such as to give rise to algebraically decaying correlations. However, in numerical calculations it has been observed that a gap in the transfer operator implies exponentially decaying correlations (see, for example \cite{Cirac2011,Schuch2013,Yang2015}), though the converse is not in general true.

Hence, if we are interested in the phase diagram of the local parent Hamiltonian that can be constructed from our PEPS at each point in parameter space ($t$, $x$, $z$), the transfer operator tells us which points correspond to the same phase: If one can connect them without closing the gap of the transfer operator, they lie in the same physical phase, as all PEPS along the path have exponentially decaying correlations and thus a gapped local parent Hamiltonian \cite{Hastings}. Thus, studying the gap of the transfer operator as it approaches the thermodynamic limit (of infinite cylinder circumference) allows us to characterize the phase diagram of the parent Hamiltonian. The main caveat is that phase boundaries of the transfer operator might not be phase boundaries of the physical Hamiltonian, since the PEPS might retain exponentially decaying correlations as the gap of the transfer operator closes.

\begin{figure}[ht]
 \includegraphics[width=0.49\textwidth]{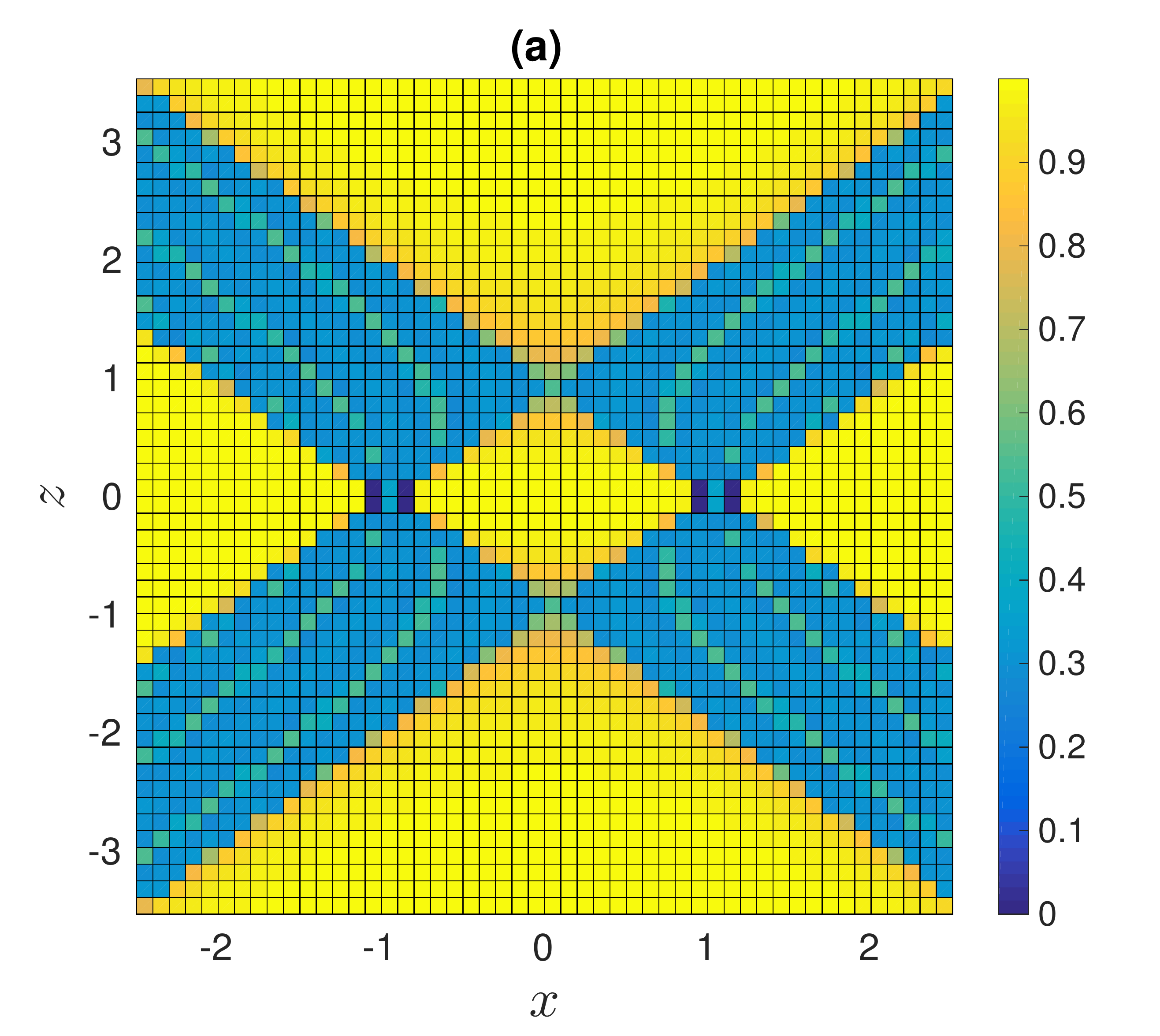}
 \includegraphics[width=0.49\textwidth]{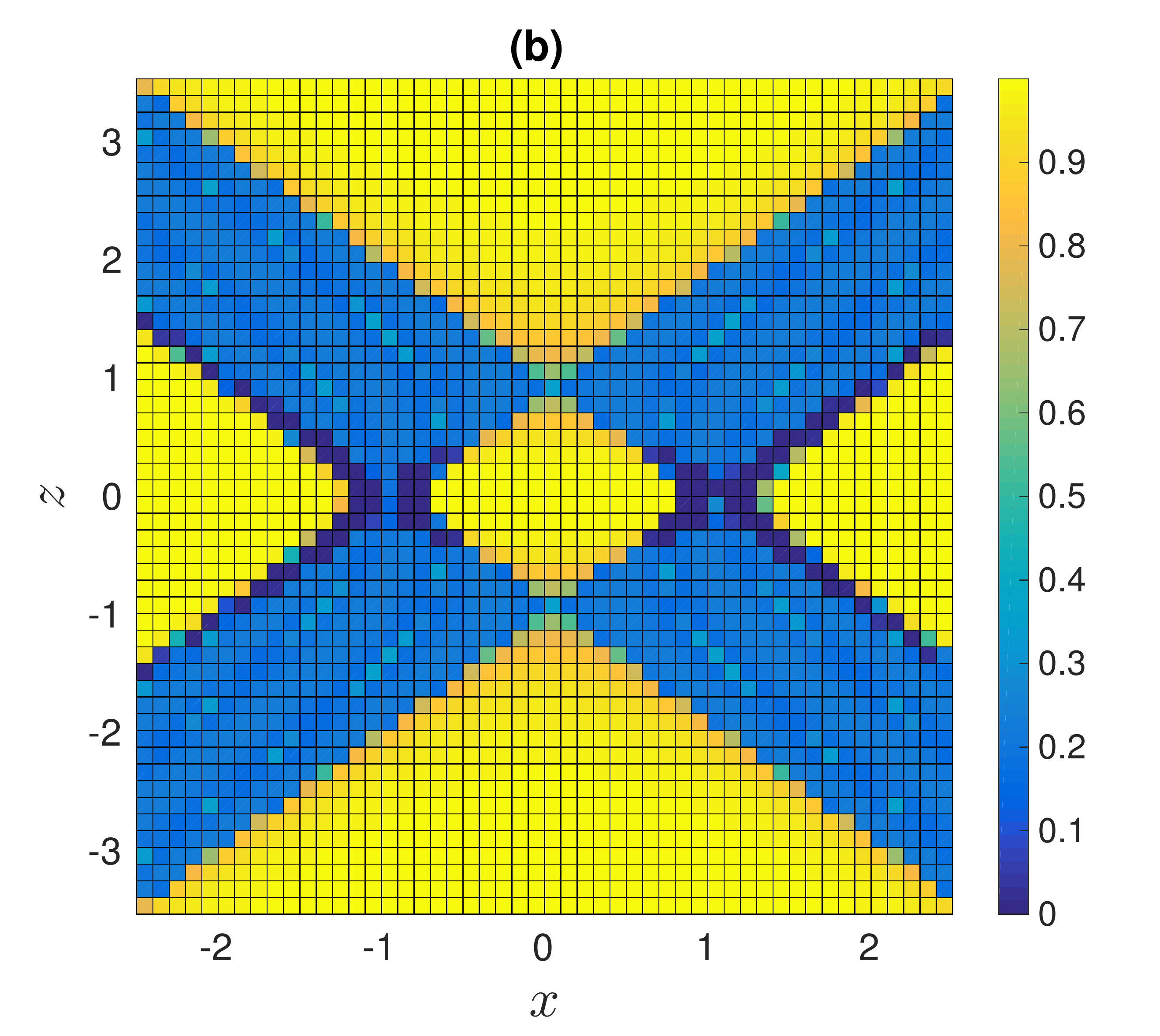}
 \caption{The gap $\varDelta$ between the two largest eigenvalues of the transfer matrix $\mathcal{T}$ is depicted as a function of $x$ and $z$ for the pure gauge theory $t=0$ for $L_1 = 4$ (a) and $L_1 = 8$ (b). The value of the gap decreases with $L_1$ in the blue regions, indicating a gapless $\mathcal{T}$ in the thermodynamic limit. On the other hand, yellow regions correspond to a gapped transfer matrix. The results are perfectly consistent with the separations of the phases I and II in Fig. \ref{phfig}.} \label{fig:diagramt0}
\end{figure}

We verified the previous conjecture about the phase diagram, obtained through the analysis of the virtual particle-hole symmetry, by numerically evaluating the gap $\varDelta \equiv \lambda_0-\lambda_1$ ($\lambda_0 := 1$) of the transfer matrix for cylinders of various widths $L_1\le 8$. The results are shown in Fig. \ref{fig:diagramt0}.
Our numerical analysis clearly confirms the separation of phases I and II previously found and presents a clear distinction between the two phases: phase I displays a clear gap of the transfer matrix, whereas phase II appears to be gapless. In particular, the numerical data at different system sizes show consistently that in phase II the gap $\varDelta$ decreases with the system size. It is this signature which identifies phase II as an extended critical phase.

\begin{figure}[tb]
 \includegraphics[width=0.49\textwidth]{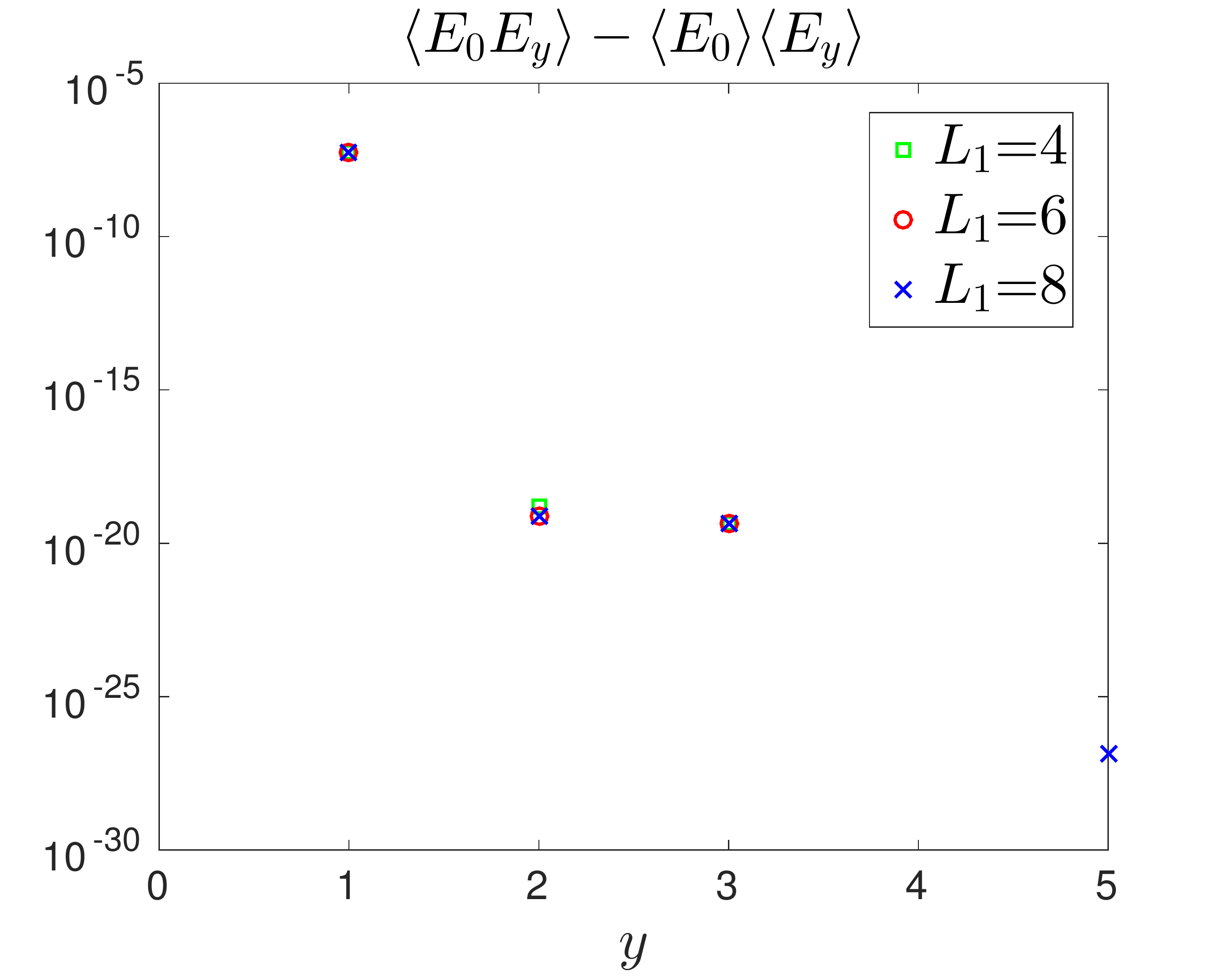}
 \includegraphics[width=0.49\textwidth]{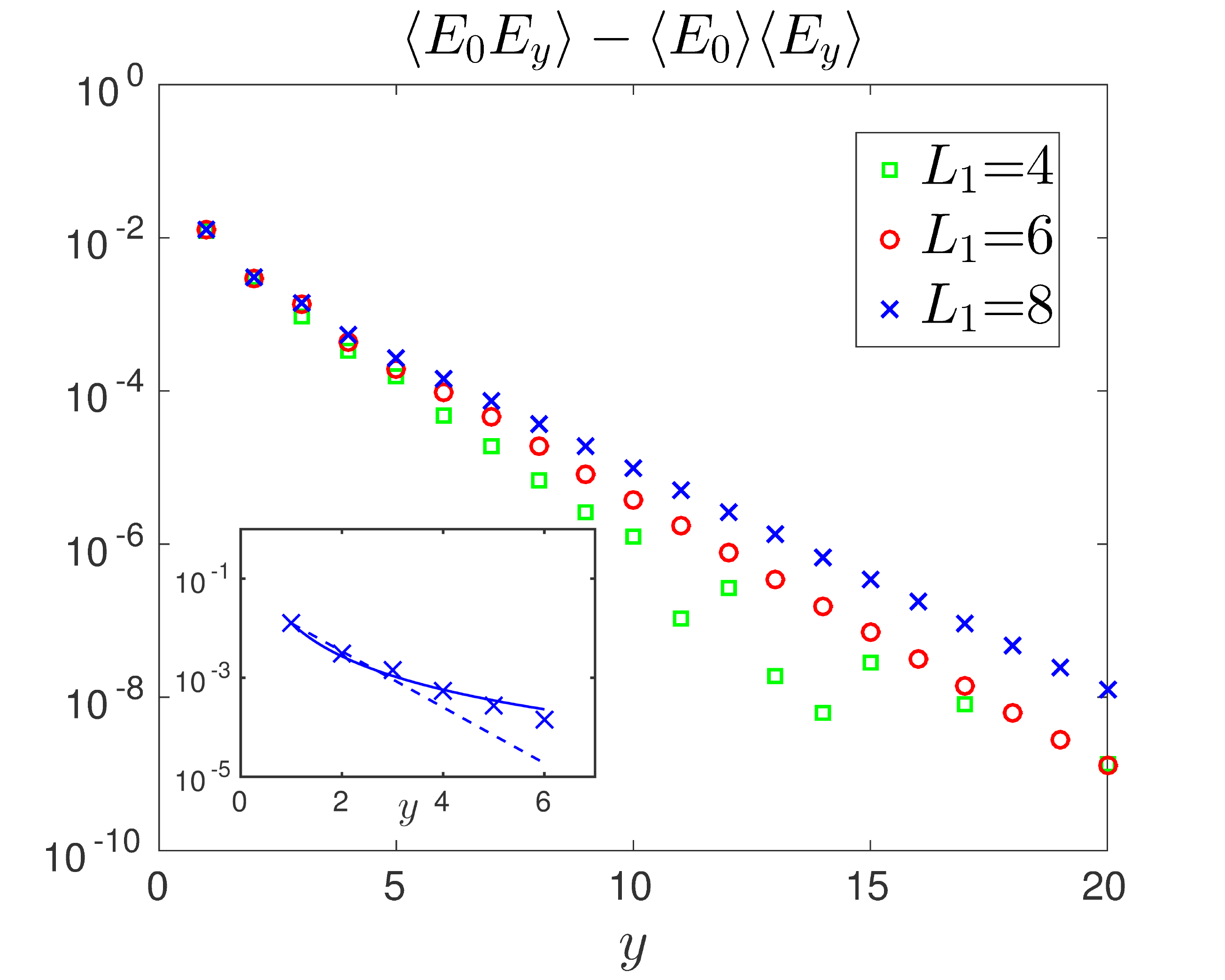}
 \caption{Two-point correlation function between the electric field $E_0$ at $(1,0)$ and $E_y$ at $(1,y)$ for cylinder circumferences $L_1 = 4,6,8$ in the gapped phase at $x = z = 0.2$ (left) and the gapless phase at $x = z = 2.5$ (right). The correlations in the gapped phase decay very abruptly, independently of the system size, and they reach our numerical zero after about 4 rows. On the other hand, the correlations in the gapless phase decay at a rate that decreases with the system size indicating a power-law decay in the thermodynamic limit. Right inset: Exponential (dashed line) and power law (solid line) fit to the first six data points for $L_1 = 8$ (for details, see main text).} \label{fig:correlations}
\end{figure}

However, the gap in phase II decreases relatively slowly as $L_1$ increases. Thus, in order to further corroborate the claim that phase II is gapless in the thermodynamic limit and has in fact algebraically decaying correlations, we evaluated the two-point correlator of the representation index of the gauge field states in two links separated by a vertical distance $y$. We evaluated it for system widths $L_1=4,6,8$ and parameters $x = z = 0.2$ (gapped phase) and $x=z=2.5$ (gapless phase), see Fig.~\ref{fig:correlations}.
In the gapped phase, the correlations decay very abruptly and basically independent of the cylinder circumference $L_1$. In contrast, in the gapless phase, the decay of the correlations depends strongly on $L_1$. For $y \gg L_1$ the decay must be exponential according to the arguments in the beginning of this subsection. From Fig.~\ref{fig:correlations} we gather that the corresponding decay length increases with $L_1$, indicating a polynomial decay in the thermodynamic limit $L_1 \rightarrow \infty$. We corroborated this by fitting the correlations for $y = 1, \ldots, 6$ and $L_1 = 8$ to both an inverse power law,
$a y^{-b}$ and an exponential decay law, $a' e^{- b' y}$ obtaining $a = 0.0128\pm 0.008, \, b = 2.24\pm 0.30$ and $a'  = 0.046\pm 0.014, \, b' = 1.30\pm 0.27$ ($95 \%$ confidence intervals), with the power law coming off as the more accurate description.

Therefore, our numerical estimation of the correlation function, despite the strong finite size limitations, seems to confirm the gapless nature of the region II.

Let us finally mention that the local gauge symmetry yields important implications in the structure of the transfer matrix $\mathcal{T}$: to explicitly evaluate the effect of the local gauge symmetries of the fiducial states on the transfer matrix, let us consider the following formulation, in terms of a state, for one of the blocks of the transfer matrix:
\begin{equation} \label{transferket}
 \mathcal{E}(\mathbf{x}) =  \bra{\Phi_t}\bra{\Phi_s}P_\psi \varpi\zeta A^G(\mathbf{x})\varpi'\zeta'A^{G\prime}(\mathbf{x})\ket{0}\,;
\end{equation}
here we are doubling the Hilbert space of the system with respect to Eq. \eqref{transfer} and all the primed operators refer to the a second PEPS layer playing the role of the state $ \bra{\Omega}A^{G\dag}\omega\eta$ in Eq. \eqref{transfer}. $\ket{0}$ is the global vacuum (of both fermionic modes and the gauge field), and the trace over the physical states $t,s$ and $\psi$ is obtained through the introduction of the maximally entangled states $\bra{\Phi_t}$ and $\bra{\Phi_s}$, which are defined as
\begin{equation} \label{triplet}
 \bra{\Phi_{t/s}} = \frac{1}{\sqrt{5}} \sum_{jmnj'm'n'} \bra{jmn}_{t/s}\bra{j'm'n'}_{t/s} \,\delta_{jj'}\delta_{mm'}\delta_{nn'}\,, 
\end{equation}
and the projector $P_\psi=(1/2)\prod_n\left(\psi_n\psi^\dag_n\psi'_n\psi^{\prime\dag}_n \right)  \prod_m\left(1+\psi_m\psi^{\prime}_m \right)$ (The introduction of $P_\psi$ is not strictly necessary for the analysis of the pure gauge case, but we include it for the sake of generality). The operators $\varpi$ and $\zeta$, instead, are meant to implement the trace over the $r$ and $u$ degrees of freedom in a way consistent with $\omega$ and $\eta$ respectively. Such operators are defined as:
\begin{equation}
 \varpi = \left( \prod_{m}r_mr^\dag_m\right) \frac{1}{2}\exp[\tilde{l}^\dag_m \epsilon_{mn}  r_n] \,,\quad \zeta = \left( \prod_{m}u_mu^\dag_m\right) \frac{1}{2}\exp\left[\tilde{d}^\dag_m \epsilon_{mn}  u_n \right]\,,
\end{equation}
such that the operators of the kind $r^\dag_n$ and $u^\dag_n$ entering into $A^G$ are mapped into $r^\dag_n \to \epsilon_{mn}\tilde{l}^\dag_m$ and $u^\dag_n \to \epsilon_{mn}\tilde{d}^\dag_m$, consistently with the definition of the projectors $\omega$ and $\eta$.
In particular, it is useful to consider the transformation of these objects:
\begin{align}
 \Theta^{\tilde{d}}_g \exp\left[\tilde{d}^\dag_m \epsilon_{mn} u_n \right] \Theta^{\tilde{d}\dag}_g = \exp\left[\tilde{d}^\dag_{m'} D_{m'm}(g) \epsilon_{mn}  u_n \right]=\exp\left[\tilde{d}^\dag_{m}  \epsilon_{mn'} (D^\intercal(g^{-1}))_{n'n}  u_n \right]= \widetilde{\Theta}^{u}_g\exp\left[\tilde{d}^\dag_m \epsilon_{mn} u_n \right] \widetilde{\Theta}^{u\dag}_g\,,
\end{align}
such that:
\begin{equation} \label{transfproj2}
  \Theta^{\tilde{l}}_g \varpi \Theta^{\tilde{l}\dag}_g = \varpi \widetilde{\Theta}^{r\dag}_g  \,,\qquad \widetilde{\Theta}^{\tilde{l}}_g \varpi \widetilde{\Theta}^{\tilde{l}\dag}_g = \varpi \Theta^{r\dag}_g  \,,\qquad \Theta^{\tilde{d}}_g \zeta \Theta^{\tilde{d}\dag}_g = \zeta \widetilde{\Theta}^{u\dag}_g \,,\qquad \widetilde{\Theta}^{\tilde{d}}_g \zeta \widetilde{\Theta}^{\tilde{d}\dag}_g = \zeta \Theta^{u\dag}_g \,.
\end{equation}

\begin{figure}[h!]
 \includegraphics[width=0.8\textwidth]{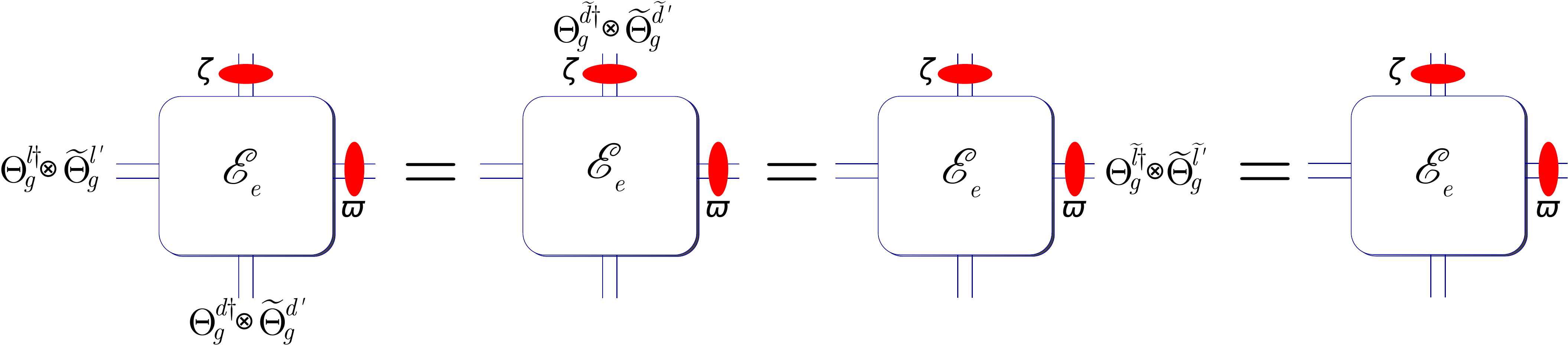}
 \includegraphics[width=0.8\textwidth]{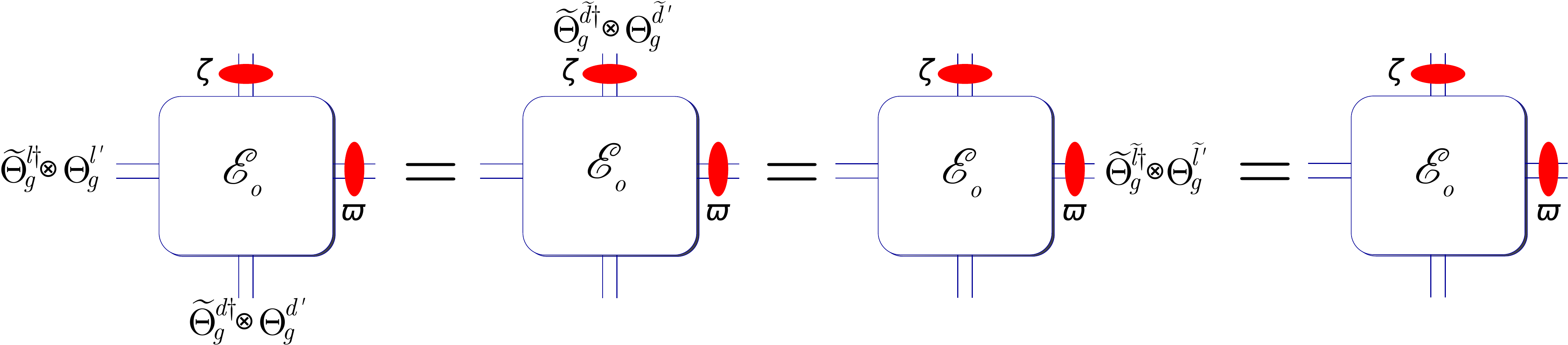}
 \caption{Symmetries of the local transfer matrix $\mathcal{E}$. We represent schematically the effects of the symmetries (\ref{mixedgauge}-\ref{thetasr}) on an even (first row) or odd (second row) block of the transfer matrix $\mathcal{E}$ in Eq. \eqref{transferket}. The double blue lines departing from the central blocks represent pairs of fermionic modes related to the operators $A^G$ and $A^{G\prime}$ in eqref{transferket}. The red ellipses depict the bond operators $\zeta$ and $\varpi$.}
 \label{Fig:symmT}
\end{figure}

Considering the definition \eqref{transferket}, and the previous transformation properties, we can show that the transfer matrix block $\mathcal{E}$ fulfills the following symmetries:
\begin{align} \label{treeven}
 &\Theta_g^{\tilde{d}\dag}(x_1,x_2+1)\otimes \widetilde{\Theta}_g^{\tilde{d}'}(x_1,x_2+1) \mathcal{E}= \Theta_g^{\tilde{l}\dag}(x_1+1,x_2)\otimes \widetilde{\Theta}_g^{\tilde{l}'}(x_1+1,x_2) \mathcal{E}=\mathcal{E}\,, \quad \forall (x_1,x_2) \; {\rm even}\,; \\
 &\widetilde{\Theta}_g^{\tilde{d}\dag}(x_1,x_2+1)\otimes \Theta_g^{\tilde{d}'}(x_1,x_2+1) \mathcal{E}=\widetilde{\Theta}_g^{\tilde{l}\dag}(x_1+1,x_2)\otimes \Theta_g^{\tilde{l}'}(x_1+1,x_2) \mathcal{E}=\mathcal{E}\,, \quad \forall (x_1,x_2) \; {\rm odd}\,, \label{treodd}
\end{align}
and
\begin{align}
 &{\Theta}_g^{l\dag}(\mathbf{x})\otimes \widetilde{\Theta}_g^{l'}(\mathbf{x}) \otimes {\Theta}_g^{d\dag}(\mathbf{x})\otimes \widetilde{\Theta}_g^{d'}(\mathbf{x}) \mathcal{E}= \mathcal{E}\,, \quad \forall (x_1,x_2) \; {\rm even}\,; \label{treeven2} \\
 &\widetilde{\Theta}_g^{l\dag}(\mathbf{x})\otimes {\Theta}_g^{l'}(\mathbf{x}) \otimes \widetilde{\Theta}_g^{d\dag}(\mathbf{x})\otimes {\Theta}_g^{d'}(\mathbf{x}) \mathcal{E}= \mathcal{E}\,, \quad \forall (x_1,x_2) \; {\rm odd}\,. \label{treodd2}
\end{align}
The previous symmetries derive from the relations (\ref{mixedgauge}-\ref{thetasr}) and the definition \eqref{transferket} by taking into account also \eqref{transfproj2}. In particular it is important to notice that the state $\bra{\Phi_{t/s}}$ is invariant under the transformation $D(g) \otimes D^{\prime \intercal}(g^{-1})$ and the transposition appearing on the operator acting on the primed degrees of freedom implies that right and left transformations are exchanged for these modes. We summarize these symmetries in Fig. \ref{Fig:symmT}.

The previous symmetries of the local transfer matrix $\mathcal{E}$ imply very strong constraints on its structure. Each bond of $\mathcal{E}$ is characterized by four fermionic modes (for example, on the vertical links on the bottom of $\mathcal{E}$ we find the modes $d_1,d_2,d_1',d_2'$). Concerning the upper and right bonds, for example, Eqs. (\ref{treeven},\ref{treodd}) imply that only states with an even number of fermions may appear on this links, otherwise $\mathcal{E}$ would transform non-trivially under the virtual gauge transformations $\tilde{\Theta}_g^{\dag}\otimes \Theta_g^{\prime}$. The even fermionic parity of these links allows us to define local Hilbert spaces $\mathcal{S}_{e/o}$ for even and odd sites which, due to Eqs. (\ref{treeven},\ref{treodd}) must be invariant under the group transformations:
\begin{equation}
 \mathcal{S}_e = \left\lbrace \ket{s} \quad {\rm s.t.} \quad{\Theta}_g^{c\dag}\otimes \widetilde{\Theta}_g^{c'} \ket{s} = \ket{s}\right\rbrace  \,, \qquad \mathcal{S}_o = \left\lbrace \ket{s} \quad {\rm s.t.} \quad \widetilde{\Theta}_g^{c\dag}\otimes {\Theta}_g^{c'} \ket{s} = \ket{s}\right\rbrace \,,
\end{equation}
where $c=\tilde{d},\tilde{l}$ for the upper and the right bonds respectively. Considering more carefully the previous definition we find that the spaces $\mathcal{S}_e$ and $\mathcal{S}_o$ coincide. The previous equations state indeed that $\left[ R^a(c) - L^a(c')\right] \ket{s}=0$ for the even case, and for the odd case left and right generators are exchanged, but considering the definitions \eqref{Rgen} and \eqref{Lgen}, the two sets of requirements are equivalent. Therefore, in both cases, there are only five independent fermionic states satisfying these conditions. They can be represented as:
\begin{equation}
 \mathcal{S} = {\rm span}\left\lbrace \frac{\ket{\uparrow,\uparrow}+\ket{\downarrow,\downarrow}}{\sqrt{2}}\,,\, \ket{0_i,0_j}\,; \; {\rm with} \; i=1,2\,,\, j=1,2 \right\rbrace\,; 
\end{equation}
these states are written in a basis distinguishing the $c$ and $c'$ degrees of freedom. The first state is the only fulfilling the previous constraints which belongs to the representation $\frac{1}{2} \otimes \frac{1}{2}$ for the transformations (\ref{treeven},\ref{treodd}); in the previous expression we adopted the shorthand notation in terms of the inner degree of freedom of the fermions for the $c$ and $c'$ modes. The other four states are instead the product of singlet states for both the $c$ and $c'$ modes. In particular we label by $\ket{0_1}$ and $\ket{0_2}$ the empty and fully occupied states respectively. All the states in the space $\mathcal{S}$ have even fermionic parity.

Concerning the left and lower bonds, the situation is more complicated because Eqs. (\ref{treeven2},\ref{treodd2}) do not separate the degrees of freedom of the two links. Therefore we must define  different Hilbert spaces $\mathcal{R}_{e/o}$, hosting eight fermionic modes (which, again, is allowed by the total even fermionic parity of these bonds states) which are invariant under the joint action of the group transformation on the left and lower bonds:
\begin{equation}
 \mathcal{R}_e=\left\lbrace \ket{r} \quad {\rm s.t.} \quad {\Theta}_g^{l\dag}\otimes \tilde{\Theta}_g^{l'} \otimes {\Theta}_g^{d\dag}\otimes \tilde{\Theta}_g^{d'} \ket{r}=\ket{r} \right\rbrace \,, \qquad
 \mathcal{R}_o=\left\lbrace\ket{r} \quad {\rm s.t.} \quad  \tilde{\Theta}_g^{l\dag}\otimes {\Theta}_g^{l'} \otimes \tilde{\Theta}_g^{d\dag}\otimes {\Theta}_g^{d'} \ket{r}= \ket{r}\right\rbrace \,. 
\end{equation}
Analogously to the case of $\mathcal{S}$, also the spaces $\mathcal{R}_e$ and $\mathcal{R}_o$ coincides due to the definitions of left and right generators \eqref{Rgen} and \eqref{Lgen}. In particular the states in $\mathcal{R}$ must satisfy the relations:
\begin{equation}
 \left[ R^a(l)-L^a(l')+R^a(d)-L^a(d')\right] \ket{r}=0\,.
\end{equation}

The invariance relations (\ref{treeven},\ref{treodd2}) state that the local transfer matrices $\mathcal{E}$ are defined in the Hilbert space $\mathcal{S}_{\tilde{d}} \otimes \mathcal{S}_{\tilde{l}} \otimes \mathcal{R}_{l,d}$. In turn, by multiplying together the local transfer matrices along a row,  this implies that the row transfer matrix can be defined as a state in $\mathcal{S}^{\otimes L_1}_{d} \otimes \mathcal{S}^{\otimes L_1}_{\tilde{d}}$. Given the characterization of $\mathcal{S}$, the transfer matrix $\mathcal{T}$ can be written as a matrix of dimension $5^{L_1} \times 5^{L_1}$, which allows us to considerably simplify its numerical evaluation.
Furthermore this decomposition could open the way for an analytical calculation of the fixed points of the transfer matrix.

\subsubsection{The transfer matrix for Wilson lines}

Beside the two-point correlation function of gauge-invariant observables, the other broad family of observables which are customarily adopted to study lattice gauge theories are gauge-invariant string and loop operators. The most common of them are Wilson lines and loops which can be considered as a discretized version of the exponential of the path-ordered integral of the gauge connection $\mathbb{P}\exp\left[\int_{\mathcal{P}} A_\mu dx^\mu \right]$. In order to have a gauge invariant quantity in the pure gauge theory, the path $\mathcal{P}$ must be either a closed loop, or a string connecting two boundaries of the system with suitable boundary conditions. The previous exponential is an operator acting on the gauge field degrees of freedom along the path and it describes the effect of a matter particle tunneling along the path $\mathcal{P}$. If $\mathcal{P}$ is an open path, due to our choice of representing the matter with a spin $1/2$ representation, such connection operators posses two physical indices, $i$ and $f$, corresponding to the initial and final state of the elementary matter before and after its transport.

\begin{figure}
 \includegraphics[width=7cm]{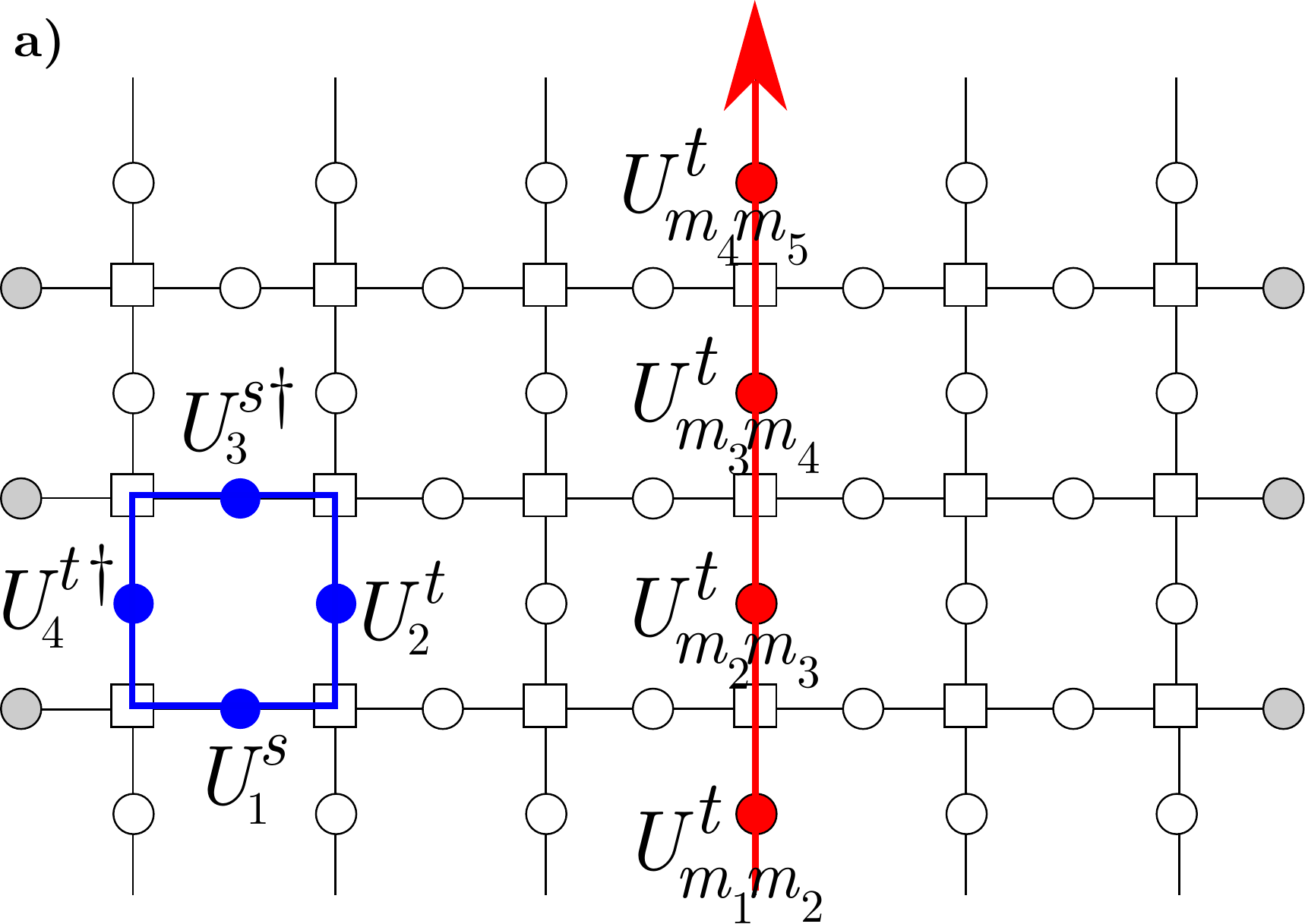} \hspace{0.3cm} \includegraphics[width=7cm]{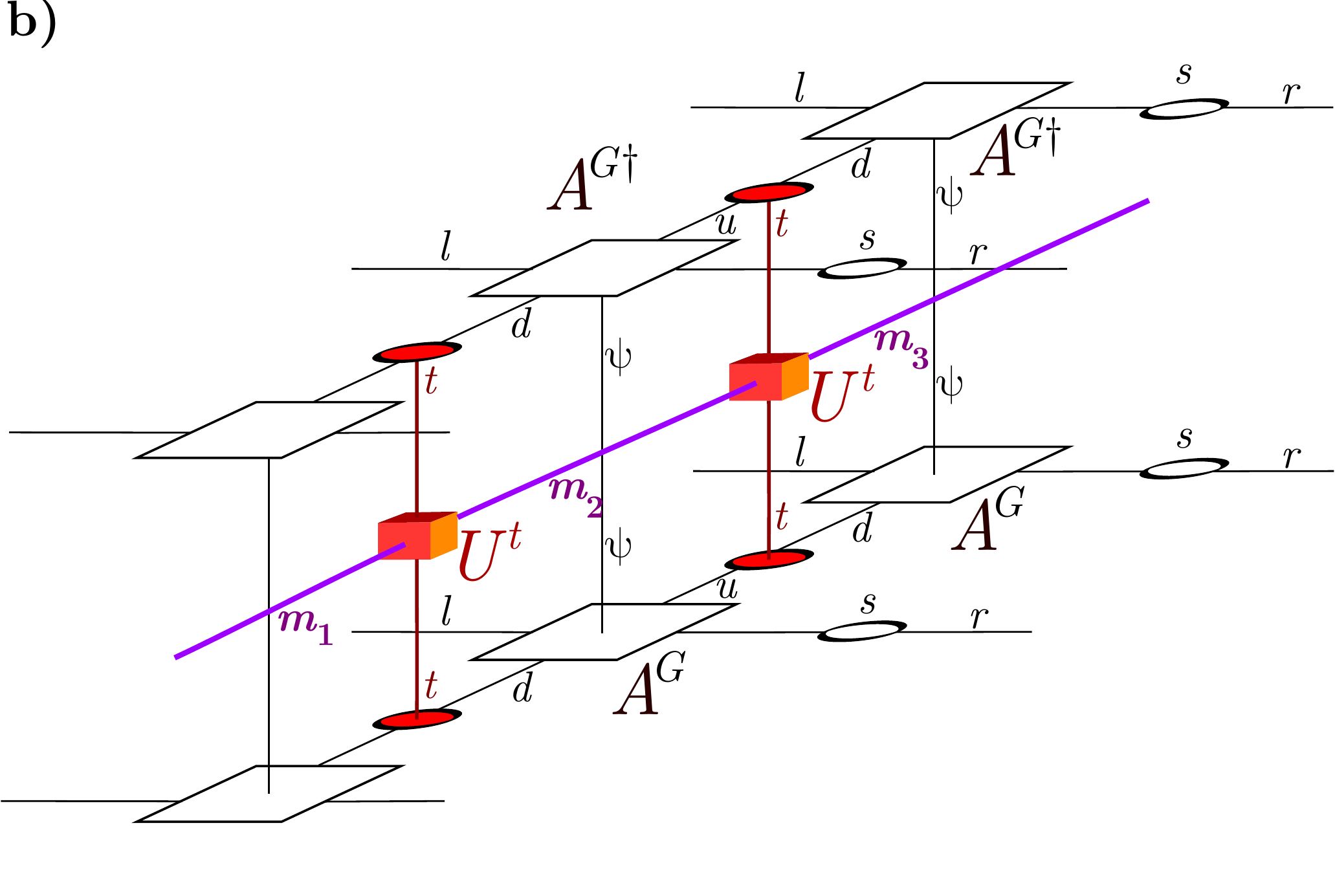}
 \caption{\textbf{a)} Schematic representation of a Wilson line (red) and a $1\times 1$ Wilson loop (blue) acting on the gauge field degrees of freedom of the physical state. The squares represent the matter sites of the lattice gauge theory, whereas the circles represent the gauge fields. The represented system is a cylinder with width $L_1=6$. The red Wilson line acts only on the $t$ degrees of freedom along a vertical line, analogously to the Wilson line introduced in the definition of transfer matrix $\Upsilon_U$ in \eqref{transfer3}. Its orientation specifies that the Wilson line is constituted by $U$ operators (the opposite orientation would correspond to $U^\dag$ operators acting on the same sites). \textbf{b)} Graphical representation of the MPO implementing the Wilson line in the calculation of the transfer matrix $\Upsilon_U$ in Eq. \eqref{transfer3}. The vertical lines correspond to the physical degrees of freedom $\psi,t,s$ which are traced out in Eq. \eqref{transfer3}. The squares represent the operators $A^G$ (and $A^{G\dag}$) defined in Eq. \eqref{GFS} which are necessary to the definition of the local fiducial state; the circles correspond to gauge field degrees of freedom. The Wilson line MPO acts on the the $t$ degrees of freedom along a column and the matrix operators in the MPO are represented by orange boxes connected by purple lines which depict the contraction of the MPO indices $m_i$. } \label{fig:MPO}
\end{figure}

In the lattice case, the role of the integral of the exponential of the connection along a link is played by the operator $U$ in equation \eqref{Uop}:
\begin{equation}
  \mathbb{P}\exp\left[\int_{\mathbf{x}}^{\mathbf{x} + \mathbf{e}_i} A_i(x_i') dx_i \right] \to U_{mn}(\mathbf{x},\mathbf{e}_i)\,.
\end{equation}
In this way, a generic oriented Wilson line over a path $\mathcal{P}$ reads:
\begin{equation} \label{Wline}
 \mathcal{L(\mathcal{P})}_{fi}= [U^{p(\mathbf{l}_N)}(\mathbf{l}_N)]_{f m_{N-1}} \; [U^{p(\mathbf{l}_{N-1})}(\mathbf{l}_{N-1})]_{m_{N-1} m_{N-2}} \, \ldots \, [U^{p(\mathbf{l}_{2})}(\mathbf{l}_{2})]_{m_2 m_{1}}\; [U^{p(\mathbf{l}_{1})}(\mathbf{l}_{1})]_{m_1 i}
\end{equation}
where $\mathbf{l}_k$ labels the $k^{\rm th}$ links along the path $\mathcal{P}$; $U^{p(\mathbf{l}_k)}=U,U^{\dagger}$, depending on the orientation of $\mathcal{P}$ of the link $\mathbf{l}_k$: when the path is oriented leftwards or downwards, the operator $U^\dag$ must be adopted and the adjoint conjugation refers both to the group indices $m_{i}$ and to the operators acting on the bosonic Hilbert space as specified in \eqref{Uop}. Finally the contraction of all the indices $m_k$ of the operators $U$ is assumed with the exception of the initial index $i$ and the final index $f$ (see Fig. \ref{fig:MPO}a for a graphical representation).

The Wilson lines $\mathcal{L}_{fi}$ are central objects in the study of the gauge-invariant states, therefore let us discuss how to evaluate their expectation value for the PEPS.  Given their product form \eqref{Wline} it is natural to consider them as matrix product operators (MPOs) acting on the gauge field degrees of freedom and concatenated through the physical tensor indices $m_k$ (see Fig. \ref{fig:MPO}b). Each string $\left\lbrace m_k \right\rbrace$ represents indeed a different observable associated to the path $\mathcal{P}$ and the Wilson line corresponds to the sum of all the string observables beginning and ending with physical indices $i$ and $f$.

Let us consider, for the sake of simplicity, Wilson lines propagating upwards in the vertical direction only. In this case $p(\mathbf{l})=1$ for each link in the line. Let us assume that we want to evaluate the vertical string-operator $\mathcal{L}(y,y')$ generating from the vertex $(x_w,y)$ and ending in $(x_w,y')$; on the practical side, this operator is gauge-dependent, therefore its expectation is zero, but it is useful to show its explicit calculation because it constitutes one of the building blocks of more complicated gauge-invariant objects. The expectation value of $\mathcal{L}(y,y')$ is given by:
\begin{equation} \label{Wline2}
 \left\langle \mathcal{L}(y,y') \right\rangle = \frac{{\rm tr}\left[X_i \left( \prod_{x_2=1}^{y-1} \mathcal{T}(x_2)\right)  \left( \prod_{x_2=y}^{y'-1} \Upsilon_U (x_2)\right)  \left( \prod_{x_2=y'}^{L_2} \mathcal{T}(x_2) \right) X_f \right] }{{\rm tr}\left[X_i \left( \prod_{x_2=1}^{L_2} \mathcal{T}(x_2)\right)  X_f\right] }
\end{equation}
where we introduced a new transfer matrix $(\Upsilon_U)^{\tilde{m}\tilde{d}\tilde{d'}}_{mdd'}$ that accounts for the $U^t_{\tilde{m}m}$ operator acting on the 'top' physical gauge field state of its vertex $x$ and carries the additional $m$ and $\tilde{m}$ indices. Such MPO indices, once contracted among subsequent rows, enable to account for all the string operators composing $\mathcal{L}(y,y')$. In more detail, $\Upsilon_U$ is given by:
\begin{equation} \label{transfer3}
 \Upsilon_{U}(x_2)_{\tilde{m}m}= {\rm tr}\left[U^t(x_w,x_2)_{m\tilde{m}}\left(\prod_{x_1}\eta \omega A^G \right) \ket{\Omega(x_2)}\bra{\Omega(x_2)}\left( \prod_{x_1}A^{G\dag}\omega\eta\right)  \right].
\end{equation}
We notice that a transposition of the group indices is necessary in the definition of $\Upsilon_{U\tilde{m}m}$ for consistency with the left and right gauge transformations of the operator $U$ \footnote{This transposition is required, since connecting transfer matrices upwards corresponds to a left multiplication, while the $U$ string in the upward direction corresponds to right multiplication. }.
Analogously to the transfer matrix $\mathcal{T}$, the transfer matrix $\Upsilon_{U}$ dictates the behavior of a vertical Wilson line as a function of its length.
There is a main difference, though, between the unperturbed transfer matrix $\mathcal{T}$ and $\Upsilon_{U}$: since $\Upsilon_{U}$ is generated by the introduction of an observable on one of the links, we must distinguish between rows with even and odd $x_2$ and the corresponding transfer matrices $\Upsilon_{U}^e$ and $\Upsilon_{U}^o$ are different.
Therefore, it is convenient to block a pair of rows and define a double transfer matrix $\Upsilon_{U}^e\Upsilon_{U}^o$ associated to the pair.
The analysis of the expectation value of the string operator now follows from Eq. \eqref{Wline2} by generalizing the argument adopted for the two-point correlation functions. When the initial and final rows have different parities, it is sufficient to block in pairs the rows hosting the Wilson line to obtain, in the gapped case, a decaying behavior of the Wilson line of the kind $\left( {\lambda'_1}/{\lambda'_0}\right)^{(y'-y-1)/2}$, where $\lambda_0'$ and $\lambda_1'$ are the two largest eigenvalues of  $\Upsilon_{U}^e\Upsilon_{U}^o$. If the initial and the final rows share the same parity, some additional and inconsequential corrections must be introduced.

In conclusion, when the spectrum of the transfer matrix $\Upsilon_{U}^e\Upsilon_{U}^o$ is gapped, the Wilson lines are expected to decay exponentially, though as in the case of two-point correlations this relationship is not established rigorously. We emphasize, though, that the operator $\Upsilon_{U}^e\Upsilon_{U}^o$ carries the two MPO additional indices. Due to the local gauge symmetry of the PEPS, this implies the presence of additional degeneracies of its full spectrum. The gap, in this case, must be considered as the gap between the first eigenvalue, appearing with this characteristic degeneracy, and the second largest, which is characterized by the same degeneracy.

\begin{figure}[ht]
 \includegraphics[width=0.49\textwidth]{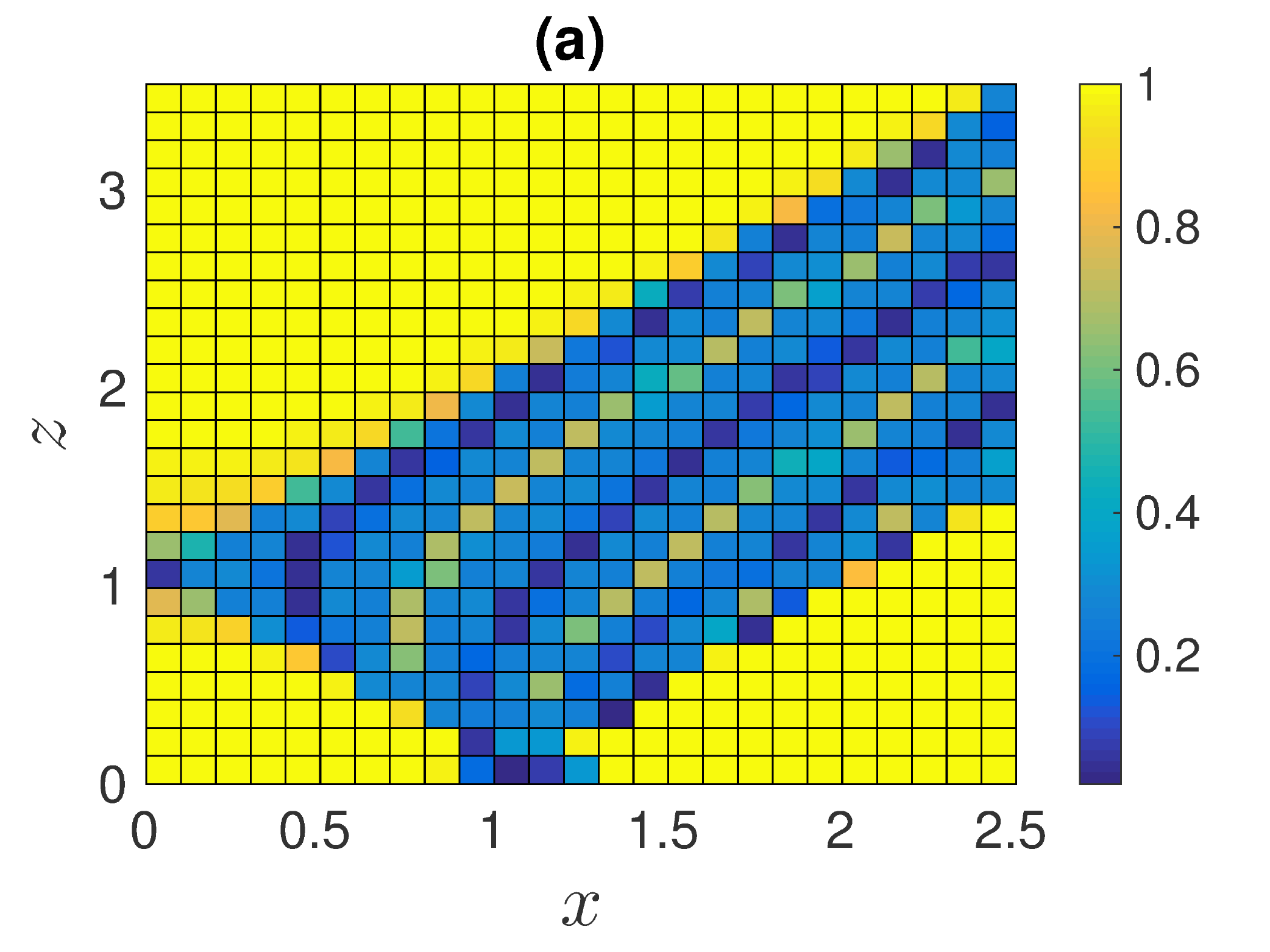}
 \includegraphics[width=0.49\textwidth]{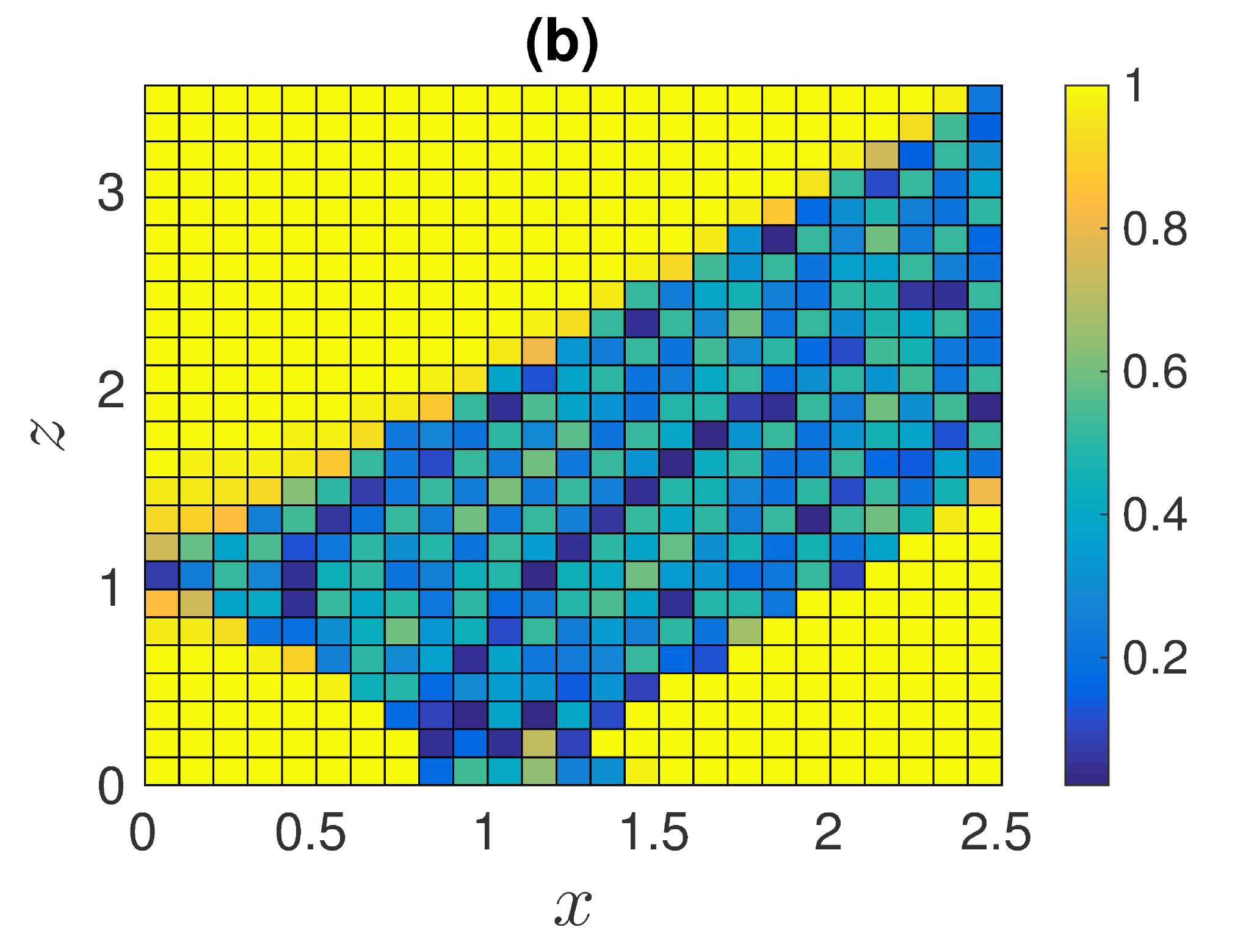}
 \caption{Gap of the transfer matrix $\Upsilon_{U}^e\Upsilon_{U}^o$ for $L_1 = 4$ (a) and for $L_1 = 6$ (b). In the critical phase the gap is smaller for $L_1 = 6$ indicating that this phase is also critical in terms of Wilson loop operators.\label{fig:Ugap}}
\end{figure}

We evaluated the spectrum of $\Upsilon_{U}^e\Upsilon_{U}^o$ as a function of the variational parameters $x$ and $z$ for the cylinder geometry with width $L_1=6$. The qualitative behavior of the gap obtained by $\Upsilon_{U}^e\Upsilon_{U}^o$ is the same as the one of $\mathcal{T}$, (compare the results in Figures \ref{fig:diagramt0} and \ref{fig:Ugap}): also in the presence of the MPO string, we can clearly distinguish a gapped and a gapless phase, which are identical to phase I and phase II of the pure transfer operator. We observed that the spectrum of $\Upsilon_{U}^e\Upsilon_{U}^o$ is real and characterized by an exact four-fold degeneracy in the gapped phase, whereas the gapless phase displays four degenerate pairs of complex conjugate eigenvalues, thus the degeneracy of the absolute values of the eigenvalues becomes 8, cf. Fig. \ref{fig:Ugap}. 

Also in this case, it is useful to consider the behavior of the transfer matrices $\Upsilon_{U}$ under local gauge transformations, which is inherited by the relations in Eqs. (\ref{mixedgauge}-\ref{thetasr}). Differently from the case of the transfer matrix $\mathcal{T}$, $\Upsilon_{U}$ is not invariant under gauge transformations applied on the site in which the $U$ operator is introduced. In all the other sites, instead, the transformation rules coincide with the ones for $\mathcal{E}$ which are depicted in Fig. \ref{Fig:symmT}. Therefore, the row transfer matrix can be obtained by multiplying local transfer matrices $\mathcal{E}$ for all the sites with $x_1 \neq x_w$ with a different local transfer matrix $\mathcal{F}_U(x_w)$:
\begin{equation}\label{transferketUrow}
 {\rm tr}\left[ \mathcal{E}(x_1=1,x_2) \ldots \mathcal{E}(x_1=x_w-1,x_2)\, \mathcal{F}_U(x_w,x_2)\, \mathcal{E}(x_1=x_w+1,x_2) \ldots \mathcal{E}(x_1=L_1,x_2)\right]  \to \Upsilon_U(x_2) \,,
\end{equation}
where the trace indicates the contraction of all the horizontal bonds through suitable projectors. The local transfer matrix $\mathcal{F}_U$ is defined by:
\begin{equation} \label{transferketU}
 \left( \mathcal{F}_{U}\right) _{\tilde{m}m} =  \bra{\Phi_t}\bra{\Phi_s}P_\psi U^t_{m\tilde{m}} \varpi\zeta A^G(\mathbf{x})\varpi'\zeta'A^{G\prime}(\mathbf{x})\ket{0}\,,
\end{equation}
where the operator $U^t$ acts on the $t$ degrees of freedom only. The action of the gauge transformations involving the upper vertical bonds results in:
\begin{align} \label{trfeven}
 &{\Theta}_g^{\tilde{d}\dag}(x_w,x_2+1)\otimes \widetilde{\Theta}_g^{\tilde{d}'}(x_w,x_2+1) \left( \mathcal{F}^e_{U}(x_w,x_2)\right)_{\tilde{m}m}= \left( \mathcal{F}^e_{UD(g^{-1})}(x_w,x_2)\right) _{\tilde{m}m}=D^\intercal_{\tilde{m}\tilde{m}'}(g^{-1})(\mathcal{F}_U^e(x_w,x_2))_{\tilde{m}'m}\,; \\
&\widetilde{\Theta}_g^{\tilde{d}}(x_w,x_2+1)\otimes {\Theta}_g^{\tilde{d}'\dag}(x_w,x_2+1) \left( \mathcal{F}^o_{U}(x_w,x_2)\right)_{\tilde{m}m}= \left( \mathcal{F}^o_{UD(g^{-1})}(x_w,x_2)\right) _{\tilde{m}m}=D^\intercal_{\tilde{m}\tilde{m}'}(g^{-1})(\mathcal{F}_U^o(x_w,x_2))_{\tilde{m}'m}\,, \label{trfodd}
\end{align}
for even and odd sites respectively. Concerning the transformation on left and lower links we have instead:
\begin{align}
 &{\Theta}_g^{l\dag}\otimes \widetilde{\Theta}_g^{l'} \otimes {\Theta}_g^{d\dag}\otimes \widetilde{\Theta}_g^{d'} (\mathcal{F}_U^e)_{\tilde{m}m}= (\mathcal{F}_{D(g)U}^e)_{\tilde{m}m}=(\mathcal{F}_U^e)_{\tilde{m}m'} D^\intercal_{m'm}(g) \,, \label{trfeven2}\\
 &\widetilde{\Theta}_g^{l}\otimes {\Theta}_g^{l'\dag} \otimes \widetilde{\Theta}_g^{d}\otimes {\Theta}_g^{d'\dag} (\mathcal{F}_U^o)_{\tilde{m}m}= (\mathcal{F}_{D(g)U}^o)_{\tilde{m}m}=(\mathcal{F}_U^o)_{\tilde{m}m'}D^\intercal_{m'm}(g) \label{trfodd2}\,.
\end{align}

\begin{figure}[h!]
 \includegraphics[width=0.8\textwidth]{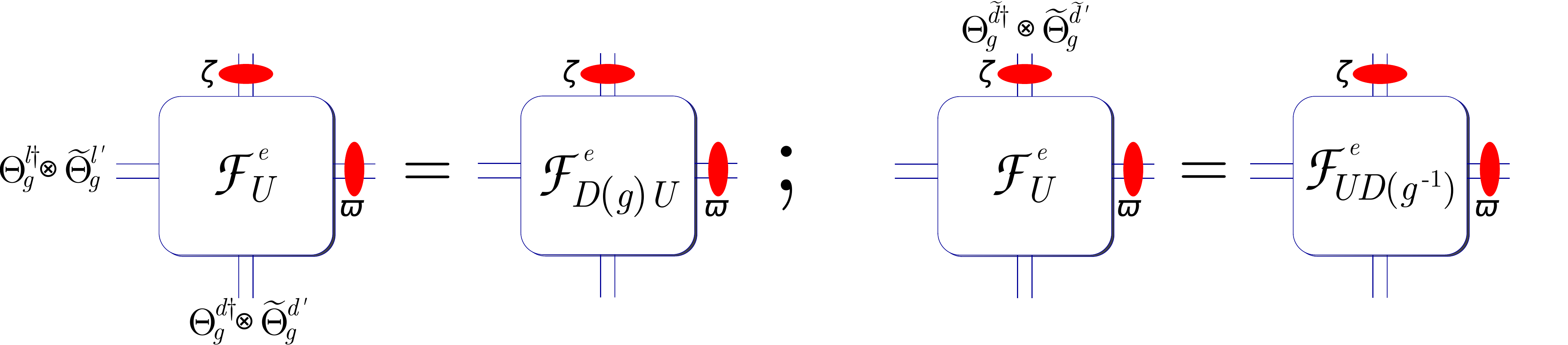}
 \includegraphics[width=0.8\textwidth]{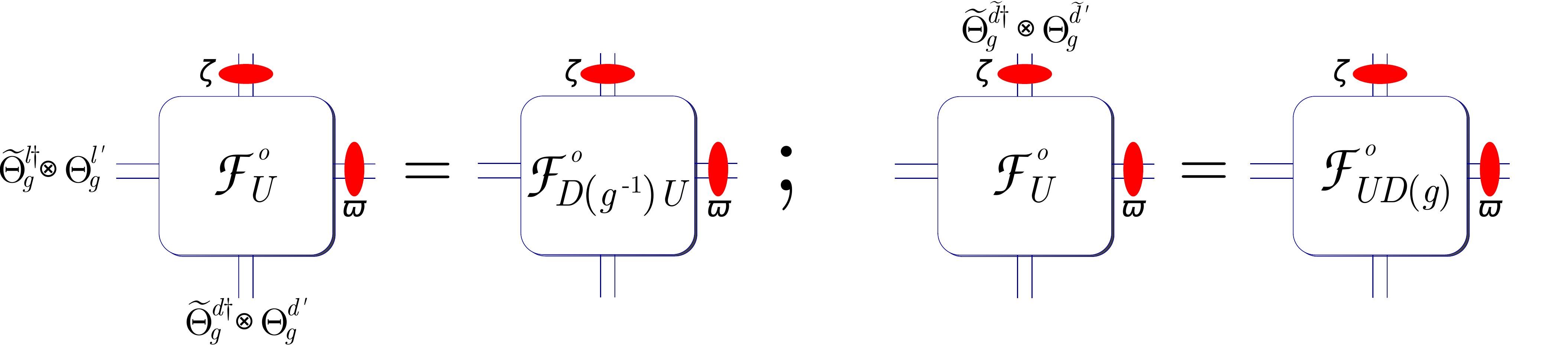}
 \caption{Symmetries of the transfer matrix $\mathcal{F}_{U}$ in Eqs. (\ref{trfeven},\ref{trfodd},\ref{trfeven2},\ref{trfodd2}). We represent schematically the effects of the symmetries (\ref{mixedgauge}-\ref{thetasr}) on the site $x_w$ of the transfer matrix hosting the operator $U$, as defined in Eq. \eqref{transferketU}. The first row depicts the transformation for even sites, the second for odd.}
 \label{fig:symmT2}
\end{figure}

In Fig. \ref{fig:symmT2} we summarize these transformation rules for $\mathcal{F}_U$ on even and odd sites. The action of gauge transformations on the virtual degrees of freedom $(d,d',l,l')$ or $(\tilde{d},\tilde{d}')$ is translated into a left or right gauge transformation of the operator $U^t$ in Eq. \eqref{transfer3}. In particular, this shows that the transfer matrix $\mathcal{F}_U$, and thus $\Upsilon_{U}$ in the site crossed by the Wilson line, transform as objects in the $j=1/2$ representation inherited from the observable $U^t$ (with respect to the MPO-string indices).

Building the row transfer matrix $\Upsilon_U$, we observe that the transfer matrices for even and odd rows are related by:
\begin{equation} \label{UeUe}
(\Upsilon_U^o)^{\tilde{m}\tilde{d}\tilde{d'}}_{mdd'} = \epsilon_{\tilde{m}\tilde{m}'}(\Upsilon_U^e)^{\tilde{m}'\tilde{d}\tilde{d'}}_{m'dd'}\epsilon^\intercal_{m'm}
\end{equation}
which can be easily derived from equations \eqref{gauging1},\eqref{ubar} and \eqref{transfer3}.
Furthermore, based on the transformation relations for $\mathcal{F}$, we can verify that the contraction of two consecutive rows fulfills gauge invariance. The transformation rules in Fig. \ref{fig:symmT2} are indeed consistent with the gauge invariance of the Wilson line. By concatenating two rows
$\Upsilon_{U}^e(x_2+1)\Upsilon_{U}^o(x_2)$, for example, one can derive that:
\begin{equation}
  \Upsilon_{U}^e(x_2+1)\Upsilon_{U}^o(x_2)=\Upsilon_{D(g)U}^e(x_2+1)\Upsilon_{UD(g^{-1})}^o(x_2)\,;
\end{equation}
this corresponds to the construction of an upward Wilson string of the kind $U^t(x_2)U^t(x_2+1)$. To obtain this result, however, it is necessary to consider suitable projectors $P'$ into maximally entangled fermionic states which allow us to transform $\mathcal{F}_U$ into the form required by the operator $\Upsilon_U$, thus enabling a direct multiplication of the transfer matrices. These projectors are similar to $P_\psi$ and are indeed invariant under the joint action of ${\Theta}_g^{\tilde{d}\dag}(x_w,x_2+1)\otimes \widetilde{\Theta}_g^{\tilde{d}'}(x_w,x_2+1)$ and $\widetilde{\Theta}_g^{d}(x_w,x_2+1)\otimes {\Theta}_g^{d'\dag}(x_w,x_2+1)$ taken from the left hand sides of Eqs. \eqref{trfeven} and \eqref{trfodd2} respectively.

\begin{figure}
 \includegraphics[width=0.8\textwidth]{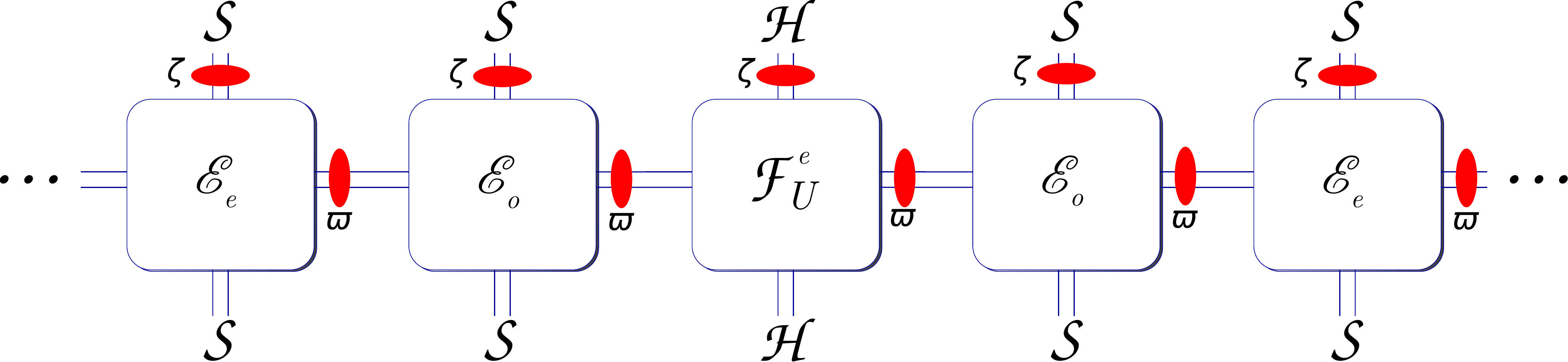}
 \caption{The transfer matrix $\Upsilon_U$ is schematically represented based on Eq. \eqref{transferketUrow}. The labels $\mathcal{S}$ and $\mathcal{H}$ specify the Hilbert space of each vertical bond.} \label{fig:upsilon}
\end{figure}

We can analyze the structure of the row transfer matrix in the left hand side of Eq. \eqref{transferketUrow} analogously to what we did for the transfer matrix $\mathcal{T}$. In the presence of the MPO the row transfer matrix lives in a different vector space which presents a structure of the kind $\mathcal{S}_d^{\otimes L_1-1} \otimes \mathcal{H}(x_w)_{d,m} \otimes \mathcal{S}_{\tilde{d}}^{\otimes L_1-1} \otimes \mathcal{H}(x_w)_{\tilde{d},\tilde{m}}$ (see Fig. \ref{fig:upsilon}). Here the subspaces of the kind $\mathcal{H}(x_w)$ include four fermionic modes and the two dimensional space hosting an MPO index (either $m$ or $\tilde{m}$). To analyze explicitly the role of this additional MPO Hilbert space, it is convenient to express first the matrix $\mathcal{F}_{U}$ in the operatorial form $\left( \mathcal{F}_{U}\right)_{\tilde{m}m} \to \left( \mathcal{F}_{U}\right)_{\tilde{m}m} \ket{m}\bra{\tilde{m}}$ inherited from the matrix $U_{m\tilde{m}}$ in Eq. \eqref{transferketU}; then it is useful to map this operator into a state by defining:
\begin{equation} \label{Fprime}
 \mathcal{F'}_{U} \equiv \mathcal{F}_{U} \ket{\Phi_{\tilde{m}}} = \sum_{m\tilde{m}\tilde{m}'\check{m}} \left( \mathcal{F}_{U}\right)_{\tilde{m}m} \ket{m} \bra{\tilde{m}}\,\frac{1}{\sqrt{2}}\ket{\tilde{m}'} \ket{\check{m}}\delta_{\tilde{m}'\check{m}} = \frac{1}{\sqrt{2}} \sum_{m\check{m}} \left( \mathcal{F}_{U}\right)_{\check{m}m} \ket{m}\ket{\check{m}} \,,
\end{equation}
where we adopted the triplet state $\ket{\Phi_{\tilde{m}}}= \sum_{\tilde{m}'\check{m}}\ket{\tilde{m}'}\ket{\check{m}}/\sqrt{2}$, defined in the Hilbert space of two MPO indices $\tilde{m}$ and $\check{m}$. After these manipulations, $\mathcal{F'}_{U}$ is the required state living in the space $\mathcal{H}_{e/o}(d,m) \otimes \mathcal{H}_{o/e}(\tilde{d},\check{m})$, where the two subspaces correspond to ingoing and outgoing vertical bonds, such that one is even and the other odd depending on the coordinates $(x_w,x_2)$.

Differently from $\mathcal{S}$, $\mathcal{H}$ includes only states with an odd fermionic parity. From Eqs. \eqref{trfeven},\eqref{trfodd},\eqref{trfeven2} and \eqref{trfodd2}, we derive that all the states in $\mathcal{H}$ must fulfill:
\begin{align}
 &{\Theta}_g^{\tilde{d}\dag}\otimes \widetilde{\Theta}_g^{\tilde{d}'} \ket{f,\check{m}}_e  = D_{\check{m}\check{m}'}(g^{-1}) \ket{f,\check{m}'}_e  \,, \label{H1}\\
 &\widetilde{\Theta}_g^{\tilde{d}}\otimes {\Theta}_g^{\tilde{d}'\dag} \ket{f,\check{m}}_o =D_{\check{m}\check{m}'}(g^{-1}) \ket{f,\check{m}'}_o  \label{H3} \,,\\
 &{\Theta}_g^{d\dag}\otimes \widetilde{\Theta}_g^{d'} \ket{f,m}_o = D_{mm'}^\intercal(g)\ket{f,m'}_o \label{H4}\\
 &\widetilde{\Theta}_g^{d}\otimes {\Theta}_g^{d'\dag} \ket{f,m}_e = D_{mm'}^\intercal(g)\ket{f,m'}_e \label{H2}\,,
\end{align}
where $\ket{f,m}$ indicate a state with a fermionic component $\ket{f}$ and an MPO index $m$ and we must distinguish the behavior of even and odd bonds, characterizing $\mathcal{H}_{e/o}$ (see Appendix \ref{app:transfer} for more detail on the derivation of the previous equations).

Let us investigate more carefully the structure of $\mathcal{H}_e(\tilde{d},\check{m})$ and $\mathcal{H}_o(\tilde{d},\check{m})$ corresponding to outgoing bonds. We can rewrite Eqs. \eqref{H1} and \eqref{H3} in terms of the generators of the SU(2) transformations acting on the fermionic degrees of freedom and on the Pauli operators acting on the MPO index-state. We obtain:
\begin{align}
 &\left[ -R^a(\tilde{d})+L^a(\tilde{d}')\right] \ket{f,\check{m}}_e = -\frac{1}{2}\sigma^{a\intercal}({\rm MPO}) \ket{f,\check{m}}_e\,, \\
 &\left[ L^a(\tilde{d})-R^a(\tilde{d}')\right] \ket{f,\check{m}}_o = -\frac{1}{2}\sigma^{a\intercal}({\rm MPO}) \ket{f,\check{m}}_o\,.
\end{align}
In both cases, these constraints are compatible only with states with an odd fermionic number such that, in terms of representations of the group, the states in $\mathcal{H}_e$ must be included in $\left[ \left(\frac{1}{2} \otimes 0 \right) \oplus \left(0 \otimes \frac{1}{2} \right)\right] \otimes \frac{1}{2}$, where the representations in the parentheses describe the fermionic components.

Let us consider first $\mathcal{H}_e(\tilde{d},\check{m})$. In this case we have  $[R^z(\tilde{d})-L^z(\tilde{d}')]\ket{f,m}=\check{m}\ket{f,\check{m}}$. By solving also the equations for the $x$ and $y$ components we obtain that $\mathcal{H}_e(\tilde{d},\check{m})$ is spanned by:
\begin{equation} \label{He}
 \mathcal{H}_e(\tilde{d},\check{m}) = {\rm span} \left\lbrace \frac{ \ket{0_i,\downarrow,1}- \ket{0_i,\uparrow,2}}{\sqrt{2}},\frac{\ket{\uparrow,0_j,1}+ \ket{\downarrow,0_j,2}}{\sqrt{2}} \,; \; {\rm with} \; i=1,2\,, j=1,2\right\rbrace \,.
\end{equation}
Here the states are ordered in a basis which follows the $\tilde{d},\tilde{d}',\check{m}$ degrees of freedom. $\ket{\uparrow},\ket{\downarrow}$ label the fermionic eigenstates of $R^z=L^z$ in the $j=1/2$ representation, whereas $0_1$ and $0_2$ are the singlets obtained from the empty and fully occupied state. We label the MPO index by $\check{m} = 1,2$, corresponding to the $+1$ and $-1$ eigenstates of $\sigma^z$, respectively.
The dimension of $\mathcal{H}_e(\tilde{d},\check{m})$ is thus 4. For $\mathcal{H}_o(\tilde{d},\check{m})$, the role of $\tilde{d}$ and $\tilde{d}'$ are exchanged and we obtain:
\begin{equation} \label{Ho}
 \mathcal{H}_o(\tilde{d},\check{m}) = {\rm span} \left\lbrace \frac{-\ket{\downarrow,0_i,1}+ \ket{\uparrow,0_i,2}}{\sqrt{2}},\frac{\ket{0_j,\uparrow,1}+ \ket{0_j,\downarrow,2}}{\sqrt{2}} \,;\; {\rm with} \; i=1,2\,, j=1,2\right\rbrace.
\end{equation}
We observe that Eqs. \eqref{H4} and \eqref{H2} provide a totally equivalent description in terms of the degrees of freedom $d,d'$ and $m$. Thus, the Hilbert spaces for ingoing and outgoing modes coincide. $\mathcal{H}_{e/o}$ can be equivalently described in terms of $d,d'$ and $m$, leading to the same result of Eqs. \eqref{He} and \eqref{Ho}. Furthermore, we observe that the basis of the spaces $\mathcal{H}_e$ and $\mathcal{H}_o$ are related by the mapping $\ket{f,\check{m}}_o=\epsilon_{\check{m}\check{m}'}\ket{f,\check{m}'}_e$ as expected from Eq. \eqref{UeUe}.

\subsection{The Wilson loops and the physical properties of the phases in the pure gauge theory}

Both the analytical and the numerical results for the pure lattice gauge theory are consistent with the existence of only two phases: a gapped and a gapless phase. To understand their physical properties we start discussing the behavior of the expectation value of the Wilson loop.
The Wilson loop operator is obtained by a closed Wilson line of the kind \eqref{Wline} where the initial and final link of the line share the same vertex and the indices $i$ and $f$ are contracted with each other:
\begin{equation}
 \mathcal{W}(\mathcal{C}) = {\rm tr} \left[ \mathcal{L(\mathcal{C})}\right]
\end{equation}
where $\mathcal{C}$ is a closed loop on the lattice.

\begin{figure}
 \includegraphics[width=0.49\textwidth]{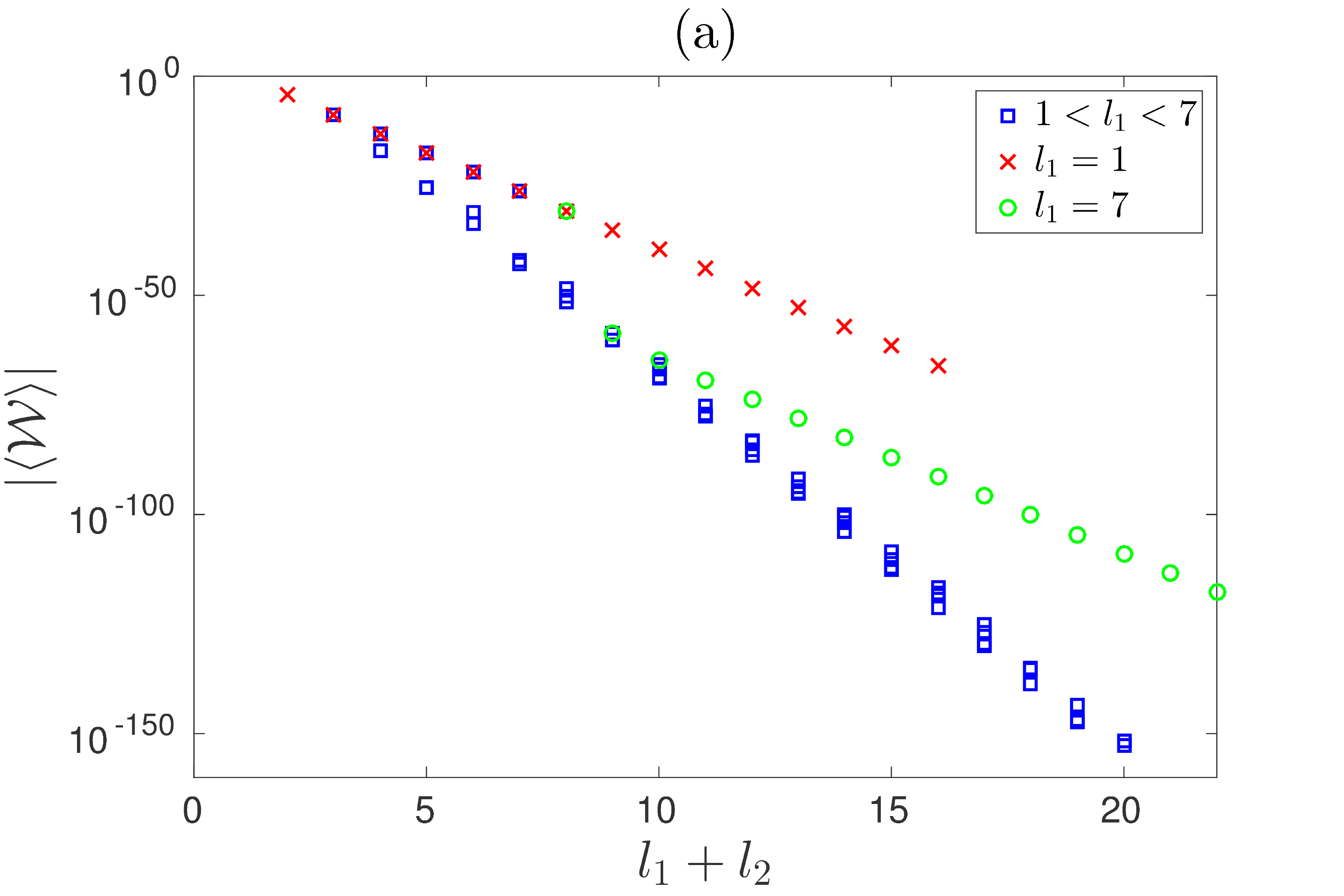}
 \includegraphics[width=0.49\textwidth]{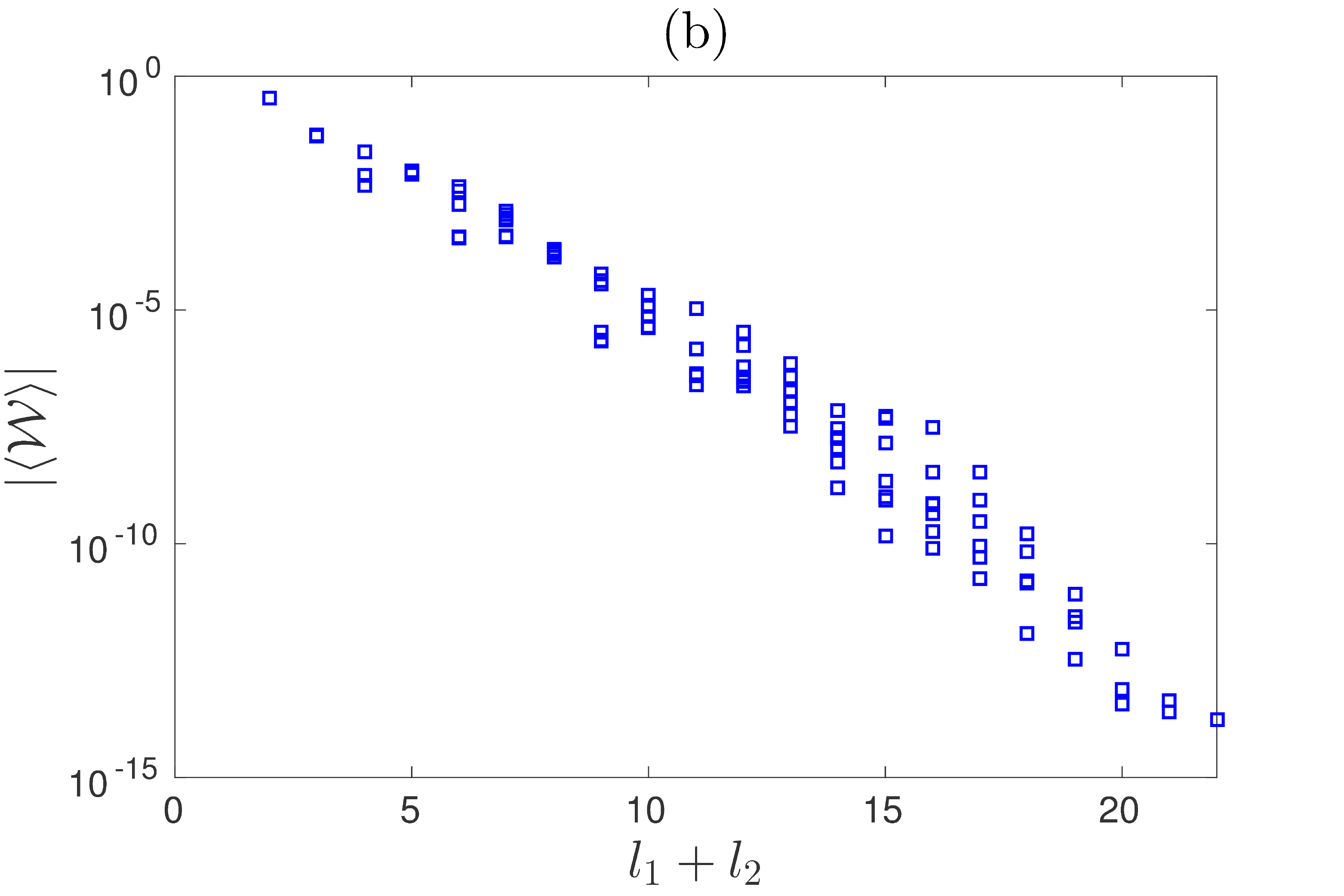}

 \caption{Magnitude of the Wilson loop as a function of its perimeter length $l_1 + l_2$ in (a) the gapped phase at $x = 0$, $z = 0.1$ and (b) the gapped phase at $x = z = 2.5$ for cylinder circumference $L_1 = 8$ and length $L_2 = 81+l_2$. The Wilson loop shows a perimeter law in the gapped phase for $1 < l_1 < 7$. The branches for $l_1 = 1, 7$ are probably due to the periodic boundary condition. In the gapless phase the Wilson loop also decays according to a (noisy) perimeter law. \label{fig:Wilson}}
\end{figure}

Fig. \ref{fig:Wilson} shows the numerical estimation of the expectation value of Wilson loops of different sizes for particular points in the phase diagram taken in the gapped and gapless phase. The loops are embedded in a cylinder of width $L_1=8$ and we can evaluate the extent of the finite size of the system by considering, for example, violations of the rotational symmetry. In an infinite lattice, Wilson loops associated with rectangles of dimensions $l_1 \times l_2$ and $l_2 \times l_1$ have, thanks to our construction of the fiducial states, exactly the same expectation value. The introduction of periodic boundary conditions, though, breaks the rotational symmetry of the system. Despite that, our numerical results show that, in the gapped phase, the anisotropy in the expectation value for rectangles of size $1 \times l$ and $l \times 1$ with $l<8$, defined as $\left| 1-\left\langle \mathcal{W}(1 \times l) \right\rangle /  \left\langle \mathcal{W}(l \times 1) \right\rangle \right|$ is always well below $10^{-3}$ for points reasonably far from the phase transition. This verifies that, as expected, the finite size has only minor effects in the gapped phase. For the gapless phase the situation is extremely different and for $l=6,7$ the expectation value of the rotated rectangles may vary of a factor 3. Therefore the numerical results for the gapless phase are heavily influenced by finite size effects.

Beside this direct effect, the influence of the periodic boundary conditions are obviously very strong when the Wilson loops have a width $l_1=6,7$ and are embedded in a system with $L_1=8$. In this case the distance between the vertical edges is respectively 2 and 1 plaquettes, thus strong lattice effects arise which limit the possibility of evaluating the thermodynamic behavior of the Wilson loops from the numerical data. It is indeed particularly evident that for thin loops $(l_1=1,2)$ or wide loops $(l_1=6,7)$ the scaling behavior is characterized by a different decay with respect to the intermediate and more reliable widths (see Fig. \ref{fig:Wilson}). This is due to the lattice effects which limit the possibility of evaluating the behavior in the continuum.

Despite these strong limitations, our data seem to suggest a decay of the Wilson loop with the perimeter of the loops (see Fig. \ref{fig:Wilson}). Such behavior would be expected in a Higgs phase, in the presence of a mass gap, and in a Coulomb phase in a gapless system \cite{Fradkin1979}. Therefore we can identify the phase at $|x+z|<1$ as a gapped Higgs phase, and the gapless phase as a Coulomb phase.

Given the unavoidable finite size effects of the numerical analysis, however, it is useful to gain an intuitive description of the origin of the perimeter law decay of the Wilson loop. To this purpose we consider first a perturbative analysis of the limiting case $x=0$ (with $z \neq 1$), which represents the gapped Higgs phase, and then we extend our observations to $x>0$, and show that we expect a perimeter law decay independently of the value of the variational parameters.

\subsubsection{Perturbative analysis of the limit $t=x=0$}

To investigate the behavior of a Wilson loop along a path $\mathcal{C}$, we begin by observing that all the operators $U_{mn}$ acting on the links belonging to $\mathcal{C}$ have the effect of changing the representation index of the physical gauge field degrees of freedom $\ket{jm'n'}$ from $j=0$ to $j=1/2$ and vice versa. This is true independently of the indices $m$ and $n$ of the operator $U$ and of the state $\ket{jm'n'}$ and it suggests to consider a simplified description of our states in which we consider only the representation of the gauge fields.

\begin{figure}
 \includegraphics[width=2.25cm]{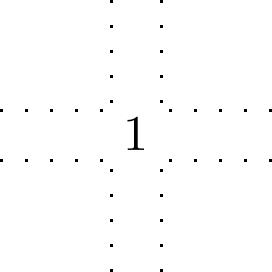} \hspace{0.4cm}
 \includegraphics[width=2.25cm]{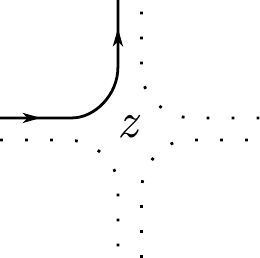} \hspace{0.4cm}
 \includegraphics[width=2.25cm]{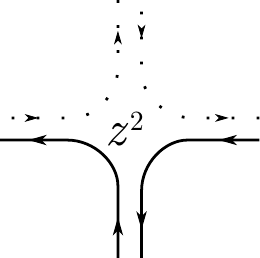}  \hspace{0.4cm}
 \includegraphics[width=2.25cm]{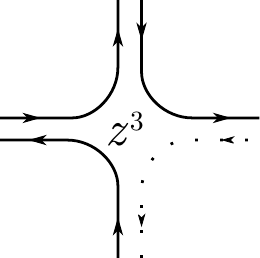} \hspace{0.4cm}
 \includegraphics[width=2.25cm]{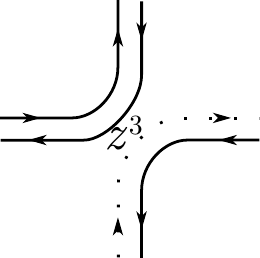} \hspace{0.4cm}
 \includegraphics[width=2.25cm]{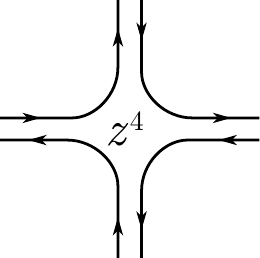}
 \caption{Graphical representation of some representative terms entering in Eq. \eqref{purefid}. The full and dotted lines represent respectively full and empty virtual fermionic modes. The arrows represent their $G_z$ charge consistently with the Gauss law for the third generator of SU(2). The first and last vertices correspond to trivial bosonic states along all the edges (vacuum configurations), whereas the intermediate configurations generate non trivial patterns of electric fluxes. Only the dependence on $z$ is specified, without numerical coefficients.}
 \label{fig:zterms}
\end{figure}

The representation index $j(\mathbf{l})$ on the link $\mathbf{l}$ is associated to the fermionic parity of the virtual modes created on the link $\mathbf{l}$ by one of the two local fiducial states delimiting the link. When only a single virtual mode is present, $j=1/2$. When the virtual modes are both empty or full, $j=0$.
Therefore it is convenient to represent the fiducial state $\ket{A^G(\mathbf{x})}$ just in terms of the occupation of its virtual modes, before the introduction of the operators $U$ which rotates the group states (see Eqs....). For $x=0$ the local fiducial state has a form of the kind:
\begin{equation} \label{purefid}
\ket{A} \propto \left( 1 + \frac{z}{\sqrt{2}} \sum a^\dag b^\dag + \frac{z^2}{2} \sum a^\dag a^\dag b^\dag b^\dag + \frac{z^3}{2\sqrt{2}} \sum a^\dag a^\dag a^\dag b^\dag b^\dag b^\dag + z^4 a^\dag a^\dag a^\dag a^\dag  b^\dag b^\dag b^\dag b^\dag\right) \ket{\Omega},
\end{equation}
up to a normalization. In this expression $a^\dag$ and $b^\dag$ represent creation operators of virtual fermions taken from the set of negative and positive $G_z$ charges respectively, consistently with Eq. \eqref{Ageneric}. The first sum is over the 8 quadratic terms corresponds to pairs of virtual fermions $a^\dag b^\dag$ with opposite $G_z$ charge, as, for example, $l^\dag_\Up d^\dag_\Dn$. Each of these terms is defined by a pair of modes lying on two edges of a corner. The parameter $z$, indeed, is associated to virtual electric flux lines describing corners across the vertex associated to the local fiducial state. Therefore all the terms in this first sum can be schematically represented as the second picture in Fig. \ref{fig:zterms}, whereas the vacuum state, associated to the amplitude $1$ in Eq. \eqref{purefid}, corresponds just to an empty set of links (represented by dotted lines in the figure).

In the same way, the second sum in the right hand side of Eq. \eqref{purefid} is over the 20 possible quartic terms, the third sum is over the 16 terms involving six virtual modes and the last term describes the configuration in which all the virtual modes are occupied in the fiducial state, which can be obtained in four different ways. An example for all these terms is represented in Fig. \ref{fig:zterms}.

In the case $z \ll 1$ we can approximate the previous expression by considering the first two term only, $\ket{A} \approx \left( 1 + z/\sqrt{2} \sum a^\dag b^\dag \right) \ket{\Omega}$. This approach helps us to estimate the scaling with $z$ of some particular example of Wilson loops.

Let us consider first the expectation value of a $1 \times 1$ Wilson loop (corresponding to a single plaquette operator in the Kogut-Susskind Hamiltonian): when $z \ll 1$ the dominating component in the fiducial state is the one without virtual fermions. Its most relevant excitation has an amplitude proportional to $z^4$ and corresponds to  a state where the four links of a plaquette $P$ of the square lattice are occupied by single virtual fermions, such that the representation $j$ of the physical gauge field states along these links is $1/2$. Therefore, by considering a single plaquette, we can schematically represent the dominant components of the physical state as
\begin{equation}
 \ket{\Psi(P)} \propto \ket{\Omega}_b + \frac{z^4}{4} \ket{\mathcal{W}}_b +  \frac{z^4}{4} \ket{\mathcal{W}^{\dag}}_b + \ldots
\end{equation} 
where $\ket{\Omega}_b$ is the product state with trivial links states in the four bonds, whereas $\ket{\mathcal{W}}_b$ and $\ket{\mathcal{W}^{\dag}}_b$ are the two states with representation $j=1/2$ on the four plaquette links which are obtained by applying a Wilson loop (or its adjoint) to the plaquette: $\ket{\mathcal{W}}_b=\mathcal{W}({\rm P})\ket{\Omega}$ with $\mathcal{W}({\rm P})=\mr{Tr}\left(U_1^s U_2^t {U_3^s}^\dag {U_4^t}^\dag\right)$
composed by the four oriented $U$ operators surrounding a single plaquette as depicted by the blue operator in Fig. \ref{fig:MPO}a.

\begin{figure}
 \includegraphics[width=0.6\columnwidth]{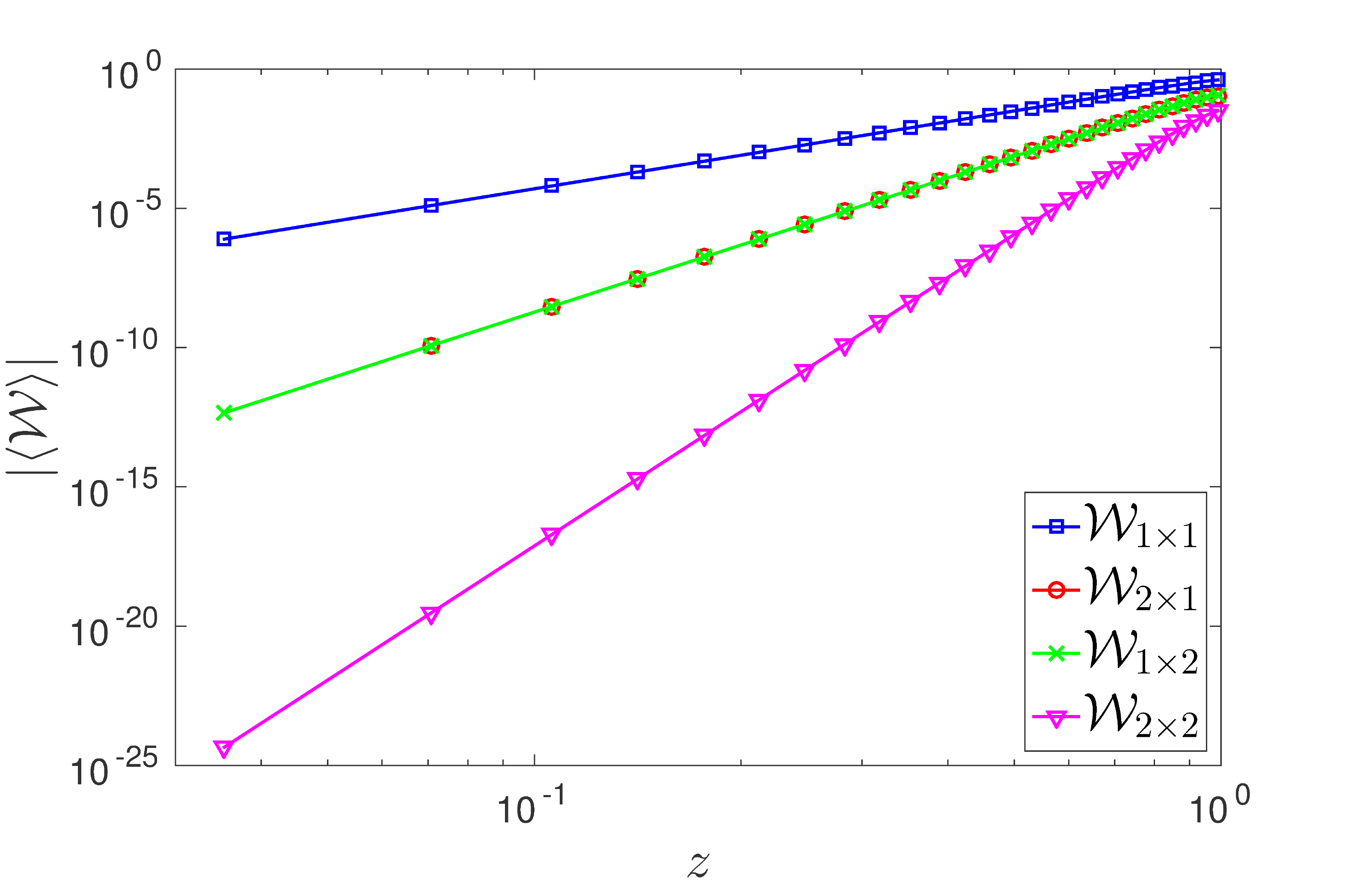}
 \caption{The expectation value of the Wilson loop for loops of area $A_\mathcal{W}=1,2,4$ for $L_1 = 8$, $L_2 = 82,83$ and $x=t=0$ as a function of $z$. It decays according to the area law $z^{4A_W}$ for small $z$ as predicted by analytical considerations.}\label{Wilsonsmallz}
\end{figure}

Therefore we can approximate the expectation value of the Wilson loop on the plaquette $P$ as:
\begin{equation}
 \left\langle \mathcal{W}(P) \right\rangle = \bra{\Psi(P)} \mathcal{W}(P) \ket{\Psi(P)} \approx \frac{z^4}{4}\; {}_b\bra{\mathcal{W}} \mathcal{W}(P) \ket{\Omega}_b +  \frac{z^4}{4}\; {}_b\bra{\Omega} \mathcal{W}(P) \ket{\mathcal{W}^\dag}_b +\ldots \approx  \frac{z^4}{2}
\end{equation}
which matches our numerical data with high accuracy for a broad range of $z<1$ (see Fig. \ref{Wilsonsmallz}).

A similar estimate can be extended to the case of Wilson loops for rectangles of the kind $1\times l$: in this case we must consider also contributions of order $z^2$ in Eq. \eqref{purefid}. The most relevant components with a single occupancy $(j=1/2)$ of the virtual states along such rectangles are obtained through a double occupancy of all the inner edges. Such configuration is obtained by considering components of the third kind depicted in Fig. \ref{fig:zterms} along the edges of the loops, which is consistent with taking the product of $l$ operators exciting all the single plaquettes composing the loop. Consistently with this description, our numerical analysis shows that, with good precision, $\left\langle \mathcal{W}(1 \times l) \right\rangle = \left\langle \mathcal{W}(l \times 1) \right\rangle \approx 4(3z^4/8)^l/3$ (the error is smaller than $10^{-6}$ up to $l=15$ and $z=0.14$).

\begin{figure}
 \includegraphics[width=5cm]{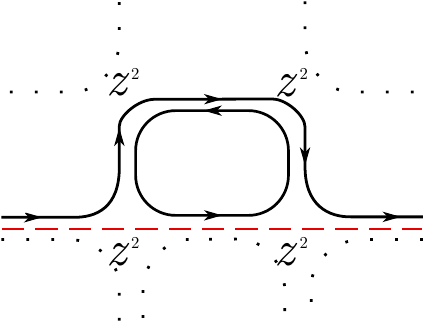}
 \caption{This scheme depicts a possible configuration of a set of virtual links in the fiducial state which gives rise to a straight line (the red dotted line) occupied by single virtual fermions, thus associated to the representation $j=1/2$ at the level of the physical states on the links. Like in Fig. \ref{fig:zterms} continuous black lines label occupied virtual modes on the links and the arrows specify their $G_z$ charge.} \label{fig:straight}
\end{figure}

 The Wilson loop for a square of size $2\times 2$ is consistent with this behavior as well, obtained effectively by multiplying single plaquette operators to obtain a larger loop. In particular $\left\langle \mathcal{W}(2 \times 2) \right\rangle \propto z^{16}$ (see Fig. \ref{Wilsonsmallz}). The mechanism of obtaining larger loops from the product of single plaquette contributions breaks down, though, when considering larger loops. For larger loops it is possible to determine other fiducial state configurations which rely on terms proportional to $z^2$ in Eq. \eqref{purefid} and allow to obtain closed paths on the lattice in which each virtual link is populated by a single fermion. The main element for this construction is showed in Fig \ref{fig:straight} and relies on the possibility of obtaining physical states on the links with $j=0$ through a double occupancy of the underlying virtual fermionic modes. This kind of configurations allow to obtain closed loops of flux lines with $j=1/2$ with an amplitude scaling roughly as $z^{4P_W}$, where $P_W$ is the perimeter of the loop. For small loops, or loops of the kind $1\times l$, such configurations are less relevant than the ones with an area law scaling obtained by the single plaquettes. For large loops, though, such perimeter behavior dominates. This allows us to confirm, for $x=0$, that the expectation value of the Wilson loop follows a perimeter decay in the thermodynamic limit, consistently with the data in Fig. \ref{fig:Wilson}.

We observe that, due to our perturbative approach, we considered the case $z<1$. However, thanks to the virtual symmetry \eqref{Hyptran}, we know that the parameters $(z,x=0)$ and $(1/z,x=0)$ describe the same physical state. It is indeed possible to build also an expansion in $1/z$ of the fiducial state for large $z$. The results are totally equivalent to our previous analysis as expected.

\subsubsection{Wilson loops for $x \neq 0$}

The introduction of the parameter $x$ does not change drastically the previous qualitative picture. We observe though that setting $t=z=0$ and $x\neq 0$, the corresponding physical state for the gauge fields is just a product state of links in the $j=0$ representation because all the virtual links are either empty or occupied by singlets. Therefore it is necessary to evaluate the simultaneous presence of both $z$ and $x$.

\begin{figure}
 \includegraphics[width=2.25cm]{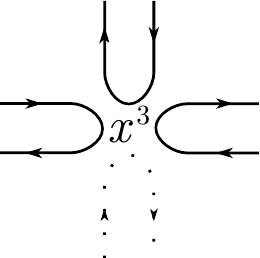} \hspace{0.4cm}
 \includegraphics[width=2.25cm]{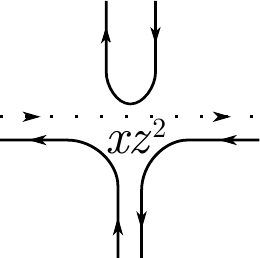} \hspace{0.4cm}
 \includegraphics[width=2.25cm]{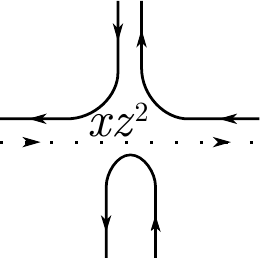}  \hspace{0.4cm}
 \includegraphics[width=2.25cm]{corner2.pdf} \hspace{0.4cm}
 \includegraphics[width=2.25cm]{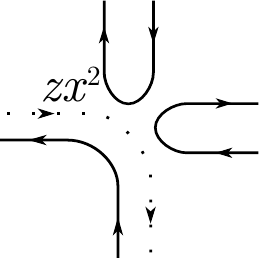}
 \caption{Graphical representation of some paradigmatic examples of the third order terms appearing in Eq. \eqref{fiducialmixed}. The $x^3$ component gives rise to a trivial gauge field configuration. The $xz^2$ terms generate a straight line of electric flux crossing the vertex and the terms $z^3$ and $zx^2$ define instead a flux corner. As in the previous figures, full and dotted lines represent respectively full and empty virtual fermionic modes and the amplitudes are not normalized.}
 \label{fig:xzterms}
\end{figure}

When considering both $x\neq 0$ and $z\neq 0$, several additional configurations of the virtual fermionic modes in the local fiducial state appear which determine a perimeter decay of the Wilson loop. These characterize, in particular, the gapless phase, and can be examined by considering the limit $z\approx x \gg 1$. To analyze the Wilson loop behavior in this limit, we can neglect all the components in the (non-normalized) local fiducial state but the ones at the third at fourth order in the parameters $x$ and $z$:
\begin{equation} \label{fiducialmixed}
 \ket{A} \approx \left[ \left(x^4 + z^4 \right)  l_\Up^\dag l_\Dn^\dag u_\Up^\dag u_\Dn^\dag r_\Up^\dag r_\Dn^\dag d_\Up^\dag d_\Dn^\dag+ x^3 \sum  a^\dag a^\dag a^\dag b^\dag b^\dag b^\dag + \frac{z^2x}{2} \sum  a^\dag a^\dag a^\dag b^\dag b^\dag b^\dag + \frac{z^3  + x^2z}{\sqrt{2}}\sum  a^\dag a^\dag a^\dag b^\dag b^\dag b^\dag \right] \ket{\Omega}
\end{equation}
where the fourth order term correspond to the totally full configuration of the virtual modes (thus to trivial $j=0$ physical link states), whereas the third order terms correspond to four classes of configurations of the kinds showed in Fig. \ref{fig:xzterms}

\begin{figure}
 \includegraphics[width=0.5\textwidth]{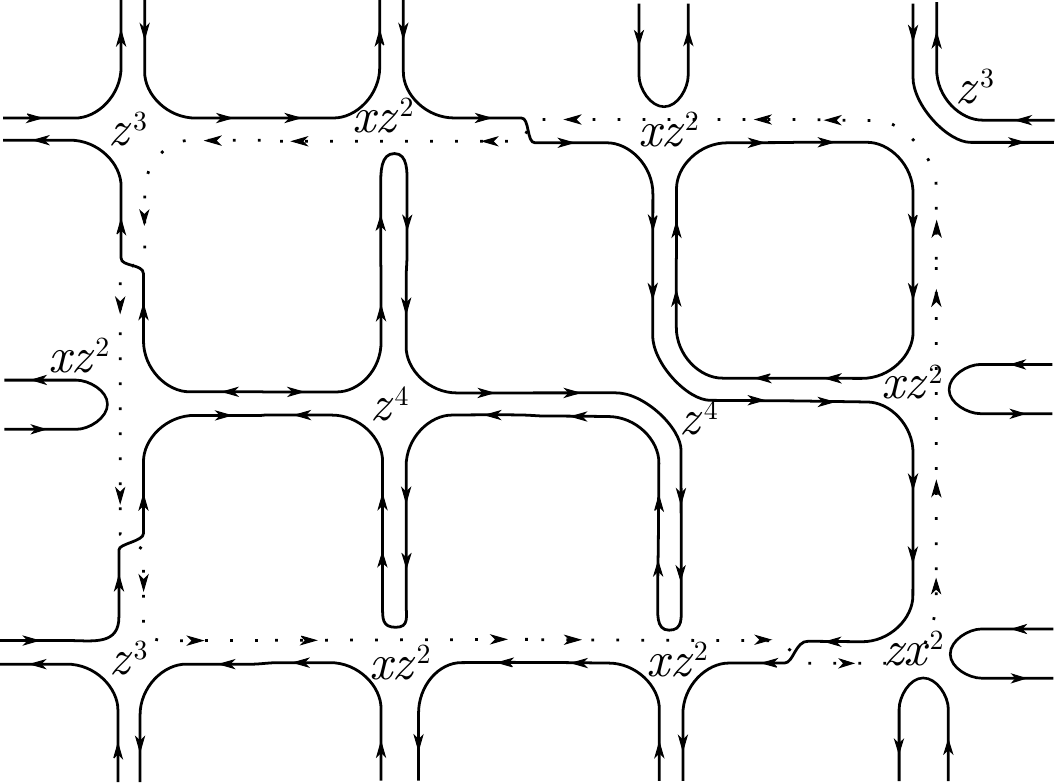}
 \caption{A configuration representing a non-trivial $2\times 3$ electric flux loop is depicted. This kind of configuration implies a perimeter law decay of the Wilson loop in the gapless phases for $x,z\gg 1$. All the vertices are described by a third or a fourth order component of the local fiducial state \eqref{fiducialmixed}. The non-trivial plaquette is characterized by a perimeter whose edges are occupied by single virtual fermions (a single full line along all the edges)}
 \label{fig:loop}
\end{figure}

In particular the terms with amplitude $z^2x/2$, for $x,z \gg 1$ are responsible for configurations in which a non-trivial electric flux crosses the lattice vertex in a straight line. Similar configurations drive the system into a phase with a perimeter-law behavior of the Wilson loop. We can consider indeed the most relevant perturbations appearing on top of the physical vacuum component, which is determined by the first two terms in \eqref{fiducialmixed}. By considering a single plaquette, the most relevant perturbation is created by four corner terms of the kind $z^3$ or $zx^2$ (see the last terms in Fig. \ref{fig:xzterms}), such that its amplitude scales roughly as $\left[ (z^3+zx^2)/(\sqrt{2}\mathcal{N})\right]^4$ where $\mathcal{N}$ is the norm of the local fiducial state \eqref{fiducialmixed}. When considering allowed, gauge-invariant, configuration on larger (rectangular) loops, however,  the terms $z^2x$ contribute too, and are responsible for components with an amplitude which is proportional to $\left(z^2x\right)^{P_W-4}\left(z^3+zx^2\right)^4/\mathcal{N}^{P_W}$ (see Fig. \ref{fig:loop} for the example of a $2\times3$ loop). These components describe, indeed, states were an electric flux lies on the perimeter $P$ of a rectangular loop and the corresponding amplitudes determine a perimeter law behavior which dominates over (all) the area law contributions for large enough rectangles.

\subsection{Static charges and flux lines}

We conclude the analysis of the thermodynamical phases in the pure gauge theory by evaluating numerically the behavior of the electric flux lines in the presence of static charges in the system.
A pair of static charges can be added by modifying two lattice sites. This modification must be done accordingly to the symmetries of the system, in particular the gauge invariance and the isotropy of these static charges. Therefore we exploit the form of the fiducial states in the presence of matter and we consider, in particular, the two fiducial states given in Eqs.(\ref{FST1},\ref{FST2}). In the limit $t\to \infty$, each of these states populates a single physical fermion, either $\Up$ or $\Dn$, in a given site of the lattice. To introduce a dipole for the $G_z$ charge, we add a pair of these fermions, located in an even and odd position. Applying the gauging procedure with the bosonic operators $U$ and $\bar{U}$ we get:
\begin{align}
 \ket{{\rm St.C.}}_e&=\psi_{\Up}^{\dagger}\left(l_\Dn^{\dagger}+\eta_{p}^{-1}d_\Dn^{\dagger}+iU^s_{\Up m}r_m^{\dagger}+\eta_{p}U^t_{\Up n}u_n^{\dagger}\right)\left|\Omega\right\rangle \label{St1}\,,\\
 \ket{{\rm St.C.}}_o&=\psi_{\Up}^{\dagger}\left(l_\Dn^{\dagger}+\eta_{p}^{-1}d_\Dn^{\dagger}+i\overline{U}^s_{\Up m}r_m^{\dagger}+\eta_{p}\overline{U}^t_{\Up n}u_n^{\dagger}\right)\left|\Omega\right\rangle \label{St2}\,.
\end{align}
Since the physical fermionic operators factor out in the construction of these fiducial state, the physical system obtained by this modification of two lattice site correspond to the presence of two local static charges, carrying a representation $j=1/2$. In the pure gauge model, the gauge transformations act on the gauge fields only and the introduction of this fermionic matter yields to a systematic violation of the chargeless Gauss law corresponding to the presence of the required static charges.

\begin{figure}
 \includegraphics[width=0.5\textwidth]{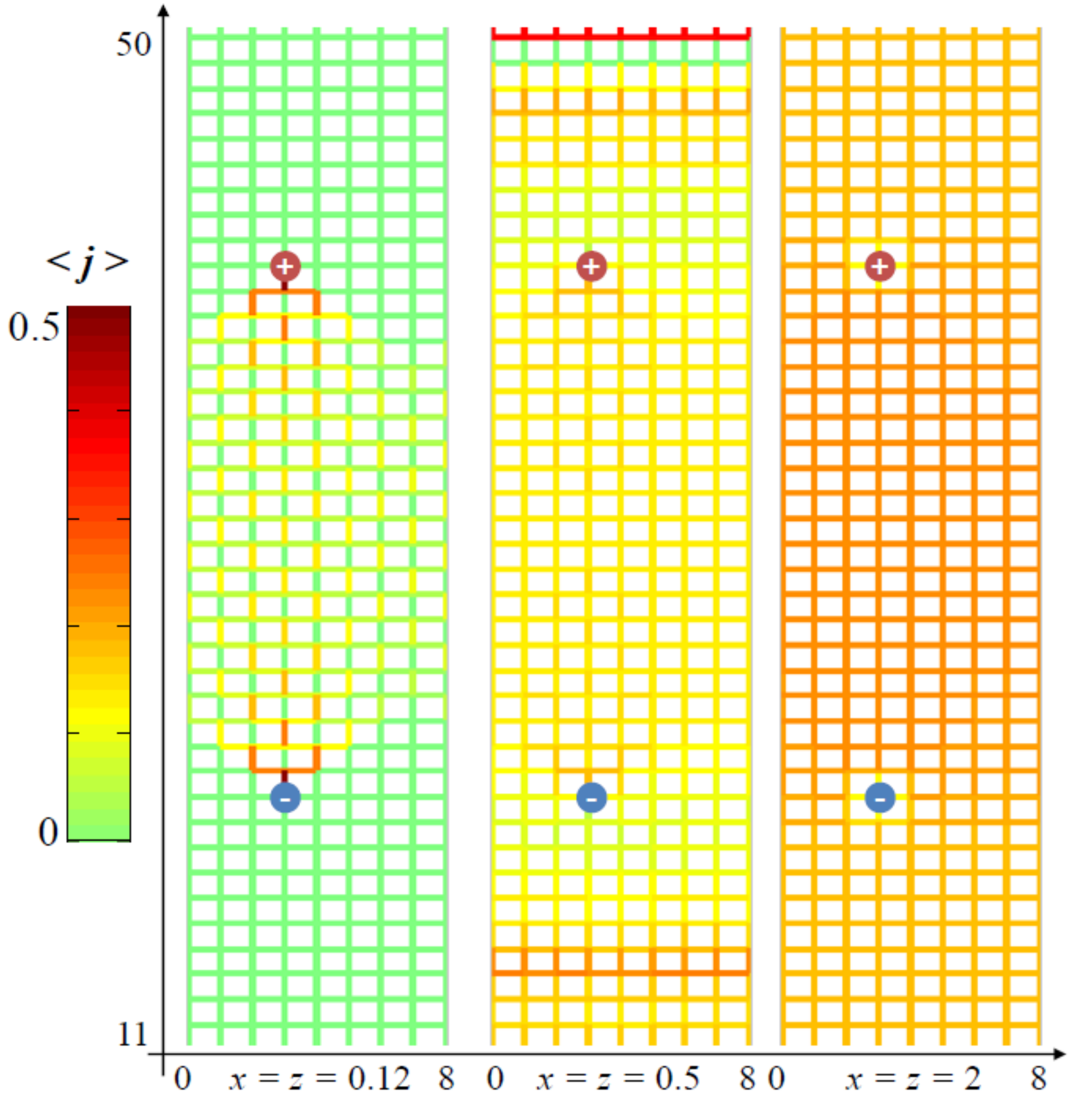}
 \caption{The expectation value of the electric flux representation $j$ is depicted for the whole system with dimension $8 \times 62$. The lowest panel shows an example taken from the Higgs phase $(x=z=0.1)$, the intermediate corresponds to the phase transition $(x=z=0.5)$ and the upper is taken in the Coulomb phase $(x=z=2)$. } \label{fig:fluxpure}
\end{figure}

In the pure gauge model, these static charges generate electric fields surrounding them and, to represent them, we can evaluate the average value of the representation $j$ in all the links of the system as a function of $x$ and $z$. Typical results for the Higgs and Coulomb phase, and for the critical point $x=z=1/2$ are shown in Fig. \ref{fig:fluxpure}, where we depict the expectation value of $j$ on all the links of a system with dimension $8 \times 62$ such that the two static charges are displaced in the vertical direction by a distance $21$.

In the Higgs phase, the electric fluxes propagate mainly in the region between the two static charges. Such charges may be considered as sources and the flux lines originate from one charge and bend towards the other. In the particular case represented in Fig. \ref{fig:fluxpure} the distance between the charges is larger than the width $L_1=8$ of the system, such that the electric lines propagate in the horizontal direction around the whole surface of the cylinder, before being reabsorbed in the other static charge. The vertical links in the intermediate region are characterized by a strong even-odd effect.
In the region outside the two static charges, instead, the expectation of $j$ drops to a small and uniform value after a distance of two sites in the vertical direction from the static charges.

In the Coulomb phase, instead, the electric field is well spread across the full system. The field is more intense in between the two charges, and in the region outside it assumes an almost uniform expectation value which depends on the chosen parameters $x$ and $z$ and has a minimum around $x=z\approx0.71$.

The critical points display a less uniform behavior, with a likely strong finite size effects which cause a visible modulation of the electric field in the $x_2$ direction.

\section{The matter - gauge theory} \label{sec:matter}

Upon introducing dynamical matter in the system by considering $t>0$, the conservation of the matter particle number implies that $x=z=0$. We observed that, in the non-interacting case with a global U(2) symmetry, this corresponds to the creation of two BCS states composed by maximally localized Cooper pairs. The introduction of the gauge fields creates a non-trivial coupling between these states, but does not modify their localized nature: the matter - gauge state corresponding to $t>0$ can indeed be decomposed into a superposition of components characterized by all the possible configurations of mesons localized on single links $[\mathbf{x},\mathbf{x}+\hat{\mathbf{e}}_j]$ of the lattice and generated by physical operators of the form $\psi_m^\dag (\mathbf{x}) U_{mn} \psi^\dag_n (\mathbf{x}+\hat{\mathbf{e}}_j)$, with $U$ acting on the intermediate link. Therefore the physical state $\ket{\psi}$ constitutes a meson condensate described by a wavefunction of the form:
\begin{equation}
 \ket{\psi} = \prod_{m,n,\left\langle \mathbf{x},\mathbf{x}'\right\rangle } \left[ 1+t^2 f(m,n,\mathbf{x},\mathbf{x}')  \psi^\dag_m(\mathbf{x})U_{mn}(\mathbf{x},\mathbf{x}')\psi^\dag_n(\mathbf{x}')\right] \ket{\Omega}\,,
\end{equation}
where $\mathbf{x}$ and $\mathbf{x'}$ are nearest neighbor sites and $f$ is a suitable pairing function encoding the rotation properties of the state. Such state generalizes the BCS states discussed in the non-interacting case to the presence of the bosonic modes and, for $t\to \infty$, describes a non trivial bosonic state in a background of filled fermionic sites.

We observe that a single fermionic site can be characterized by occupation numbers $0,1$ or $2$. The empty state $0$ and the singlet state $2$ correspond to a representation index $j=0$, whereas the singly occupied states behave non-trivially under gauge transformations, following the representation $j=1/2$ of the gauge group.
Due to the form of the matrix $T$ in Eq. \eqref{T0matrix} and the related fiducial state, the density of physical fermions increases monotonically with $t$. The most significant observable to the purpose of the description of the system, though, is the expectation value of the fermionic parity of the matter sites, which corresponds to their representation $j$ and to the density of singly occupied sites.

\begin{figure}
 \includegraphics[width=7cm]{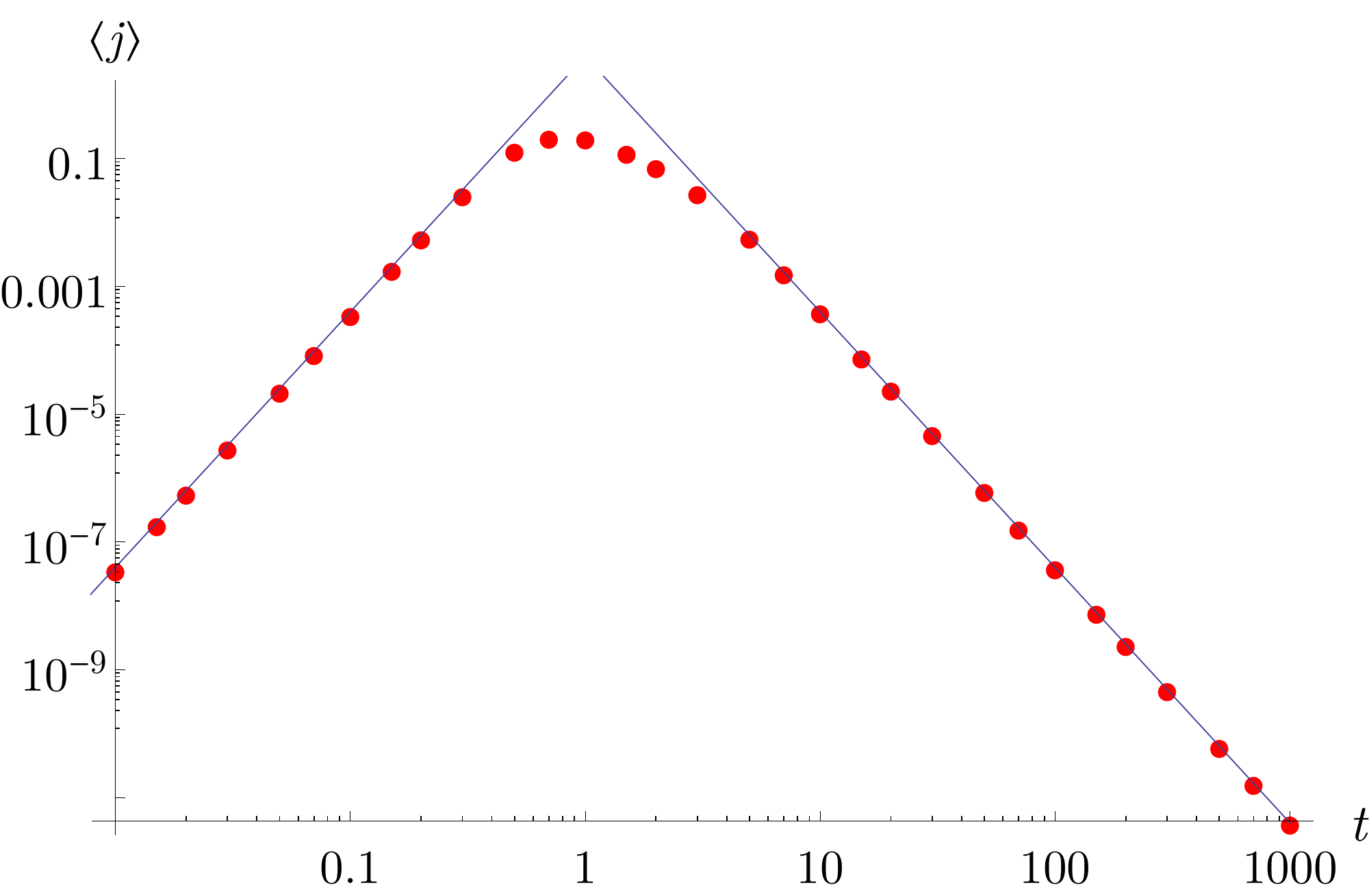}
 \caption{Logarithmic plot of the expectation value of the representation $\left\langle j \right\rangle$ evaluated in a lattice site in the bulk of a system with $L_1=8$ as a function of $t$. The red dots are the numerical data. The blue lines correspond to the asymptotes $4t^4$ and $4/t^4$ for small and large $t$. }
 \label{fig:parity}
\end{figure}

Let us consider the expectation value of $j$ in the bulk of the system: the limit $t=0$ corresponds to the vacuum; after the introduction of a small $t$, a finite density of singly occupied states appears and, for $t \ll 1$, $\left\langle j\right\rangle  \approx 4t^4$. The expectation value $\left\langle j \right\rangle$ reaches its maximum around $t \approx 0.7$ and then, for larger $t$, the doubly occupied sites become more and more relevant, thus decreasing $\left\langle j\right\rangle$. For $t\gg1$, the average representation index behaves asymptotically as $\left\langle j \right\rangle \approx 4/t^4$ (see Fig. \ref{fig:parity}). The limit of small $t$ is asymptotically the same as the non-interacting case, whose average density can be evaluated from the correlation function \eqref{corrfun}. For large $t$, instead, the coupling introduced by the gauge field becomes essential and brings to a systematic shift of the full theory from the global gauge invariant case.

This behavior has an interesting interpretation when we consider the relation \eqref{tma} derived in the non-interacting regime which relates $t$ with the mass $m$ of the dynamical fermion with $t \propto 1/\sqrt{ma}$. For $t \to 0$, the mass of the matter particles diverges, thus explaining the vanishing number of physical fermions. In particular, as long as the mass term dominates over the kinetic energy, the density of singly occupied states is small and proportional to $1/m^2$. With increasing $t$, $m$ decreases, the kinetic energy and mass terms become comparable and $\left\langle j\right\rangle$ reaches its maximum. For larger $t$, the kinetic term dominates, therefore it is more and more energetically favorable to increase the density of particles until one saturates the state at density $2$, with the density of singly occupied states decreasing as $m^2$. As noticed before, however, the results about the fermionic density of the systems with local and global gauge symmetry significantly differ for large $t$, therefore the interpretation of these results in terms of the mass $m$ of the non-interacting case is just an intuitive, speculative picture which may not hold in this regime.

\begin{figure}
 \includegraphics[width=0.49\textwidth]{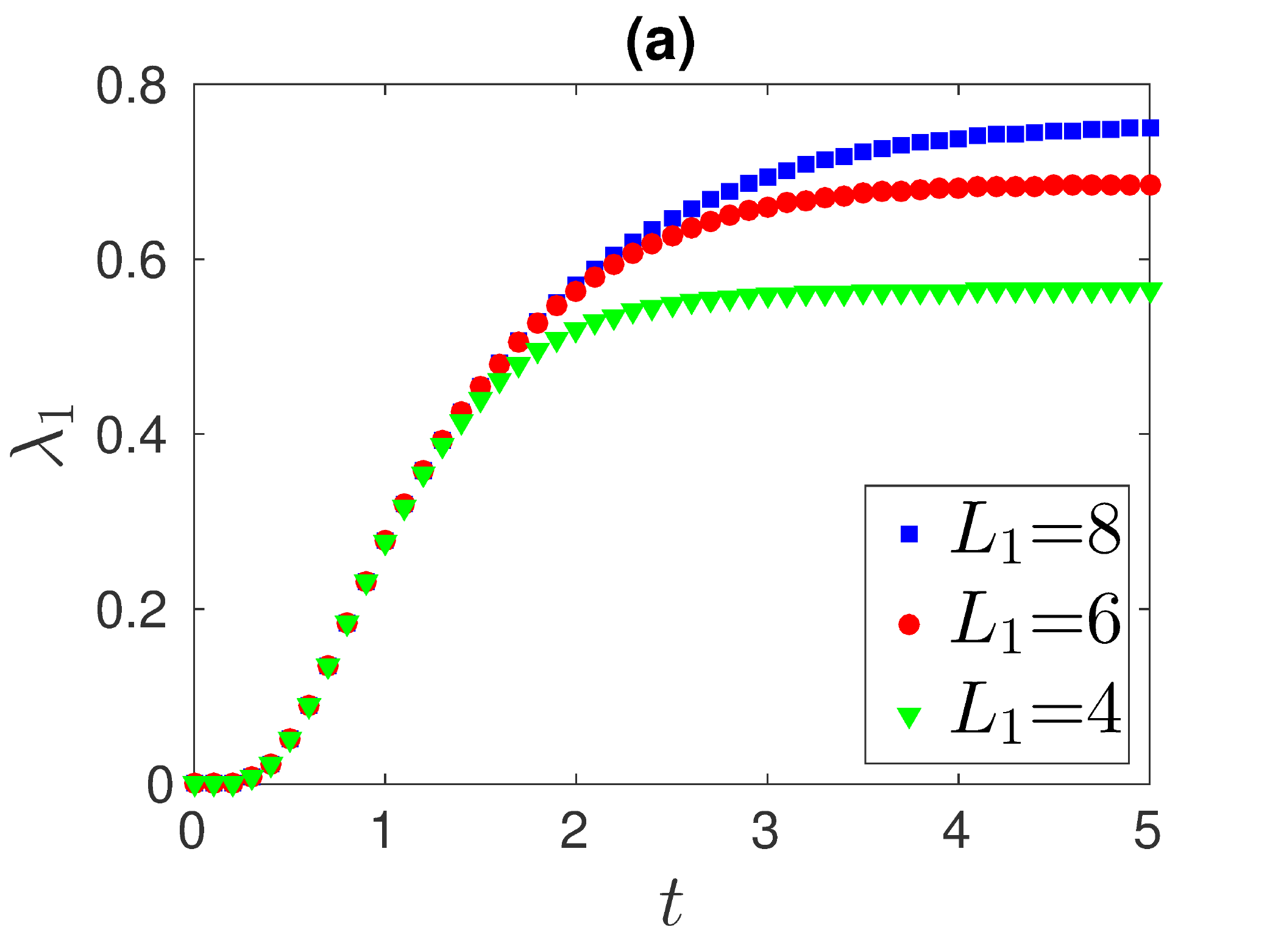}
  \includegraphics[width=0.49\textwidth]{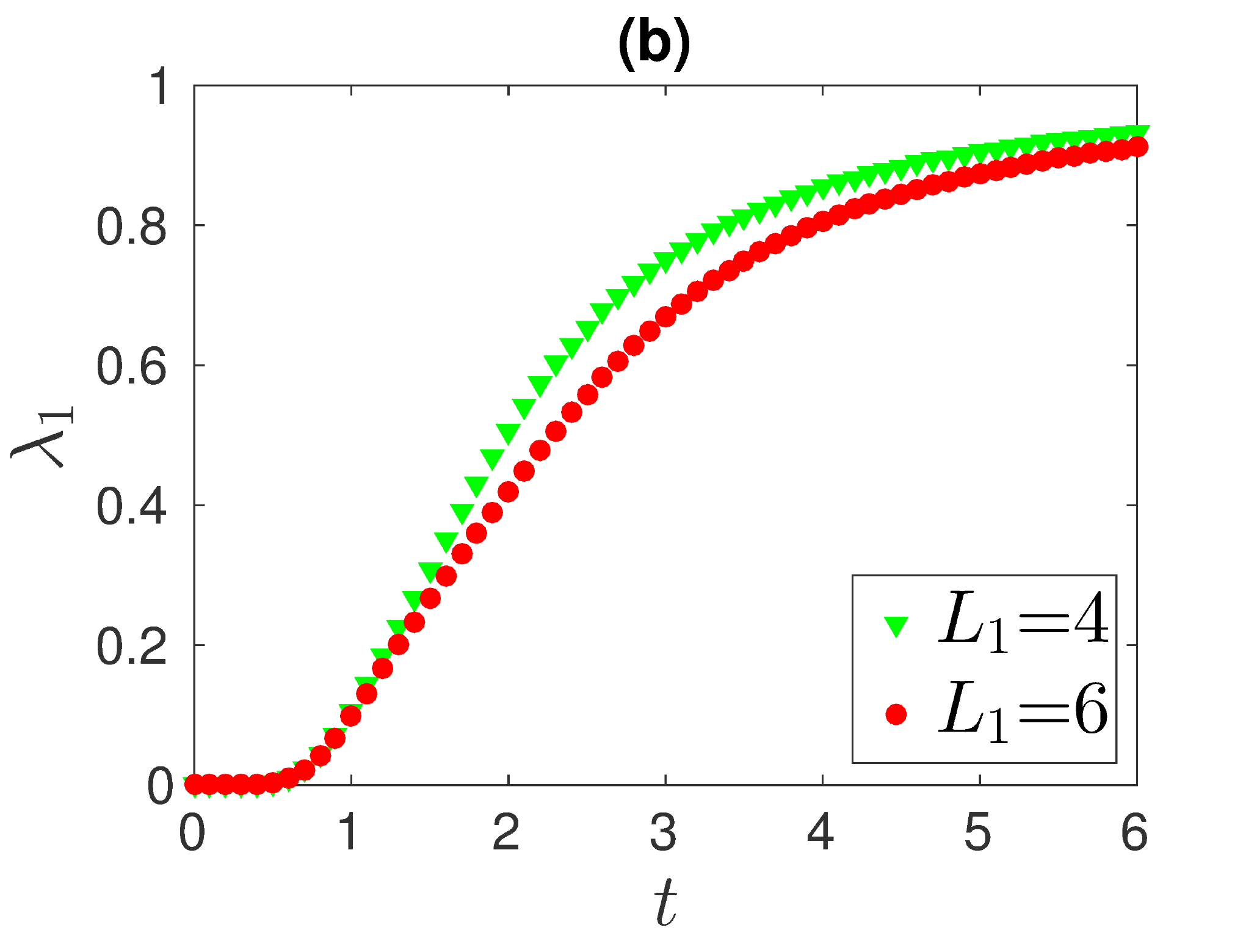}
 \caption{(a) Magnitude of the second highest eigenvalue $\lambda_1$ of the transfer matrix $\mathcal{T}$ for $L_1 = 4,6,8$ as a function of $t$ for $x = z = 0$, where we set $\lambda_0 = 1$. The gap $1 - \lambda_1$ decreases with increasing $L_1$, but appears to saturate to a non-zero value for finite $|t|$. The lower eigenvalues $\lambda_{2,3,\ldots}$ (not shown) are well separated from $\lambda_1$.
(b) Same plot for transfer matrix $\Upsilon_{U}^e\Upsilon_{U}^o$ in the presence of the Wilson MPO string for $L_1 = 4,6$. All of its eigenvalues are two-fold degenerate. The main difference to (a) is that the gap increases with the cylinder circumference for all considered $t > 0$.
 \label{fig:gap_t}}
\end{figure}

The average value $\left\langle j \right\rangle$  presents no discontinuity, either for the matter sites, or for the gauge field links. This suggests that, independently of the value of $t$, the system is always in the same thermodynamic phase. We verified that, indeed, the gap $\varDelta$ of the transfer matrix $\mathcal{T}$ remains consistently different from zero for all the values of $t$ and all the system sizes we could probe ($L_1 = 4,6,8$), see Fig. \ref{fig:gap_t}. We observe, however, that $\varDelta$ decreases with $t$ and with the system size, and our numerical analysis cannot establish whether, in the limit $t \to \infty$ and $L_1 \to \infty$, the system becomes gapless. A similar behavior is found for the gap of the transfer matrix with an MPO string, the main difference being that the gap of $\Upsilon^e \Upsilon^o$ increases with the cylinder circumference $L_1$ for any finite $t$.

To characterize this gapped phase, we estimated the expectation values of the Wilson loop and of the gauge invariant mesonic strings. For the latter we considered the particular case of operators of the form $\mathcal{M}(\mathcal{P})=\psi^\dag_f(\mathbf{x}_f) \mathcal{L}(\mathcal{P})_{fi}\psi_i^\dag (\mathbf{x}_i)$ with $\psi^\dag_f(\mathbf{x}_f)$ and $\psi_i^\dag(\mathbf{x}_i)$ belonging to different sublattices.
$\mathcal{P}$ labels the path of the flux line defining the mesonic string.
For $i$ and $f$ located on the same sublattice, instead, we define the tunneling operator (which we also call $\mc{M}(\mc{P})$, but with $\mc{P}$ of even length), where $\psi^\dag_i(\mathbf{x}_i)$ is to be replaced by $\psi_i(\mathbf{x}_i)$ in order to respect the global $U(1)$ symmetry.

\begin{figure}
\includegraphics[width=0.33\textwidth]{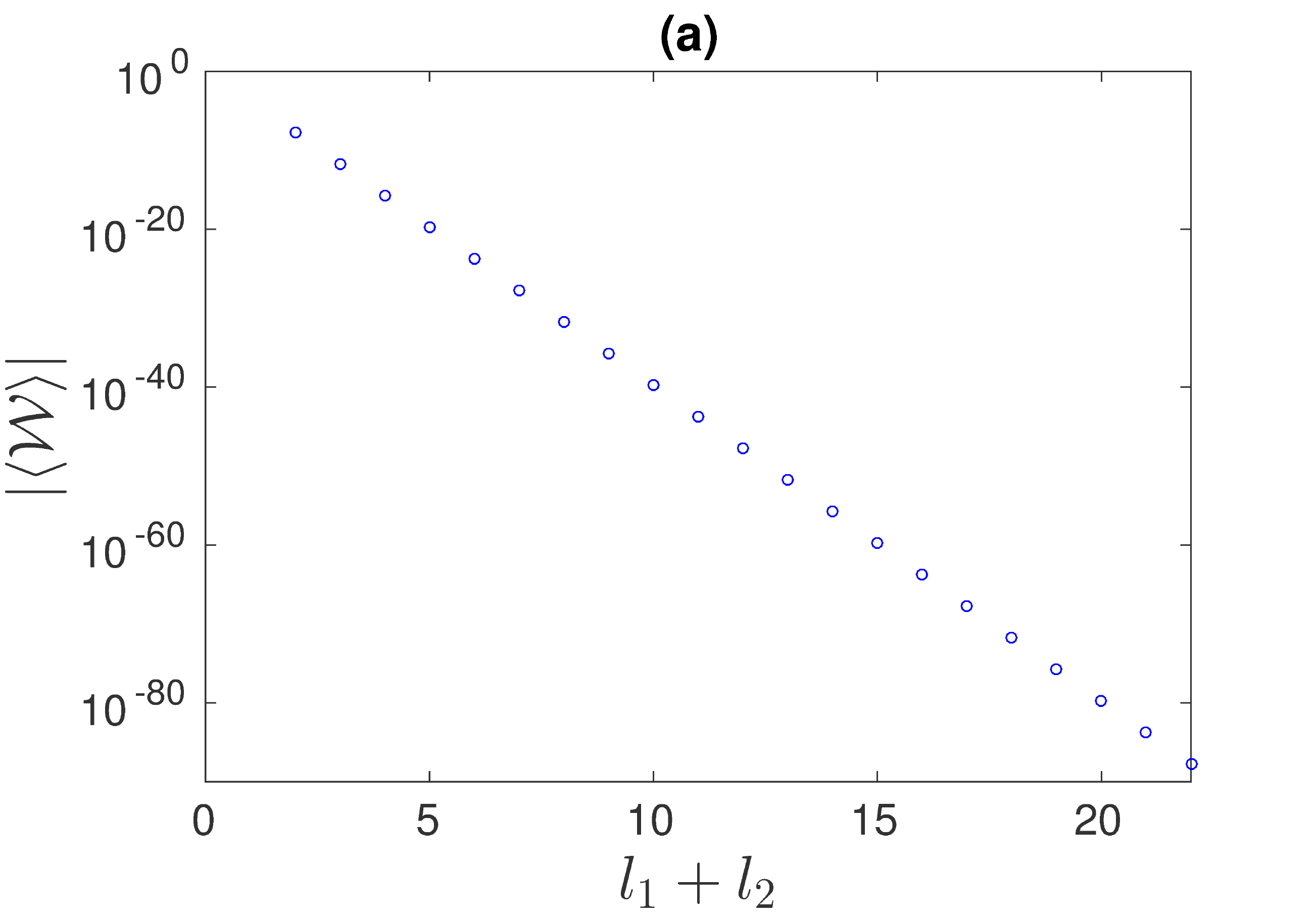}
\includegraphics[width=0.33\textwidth]{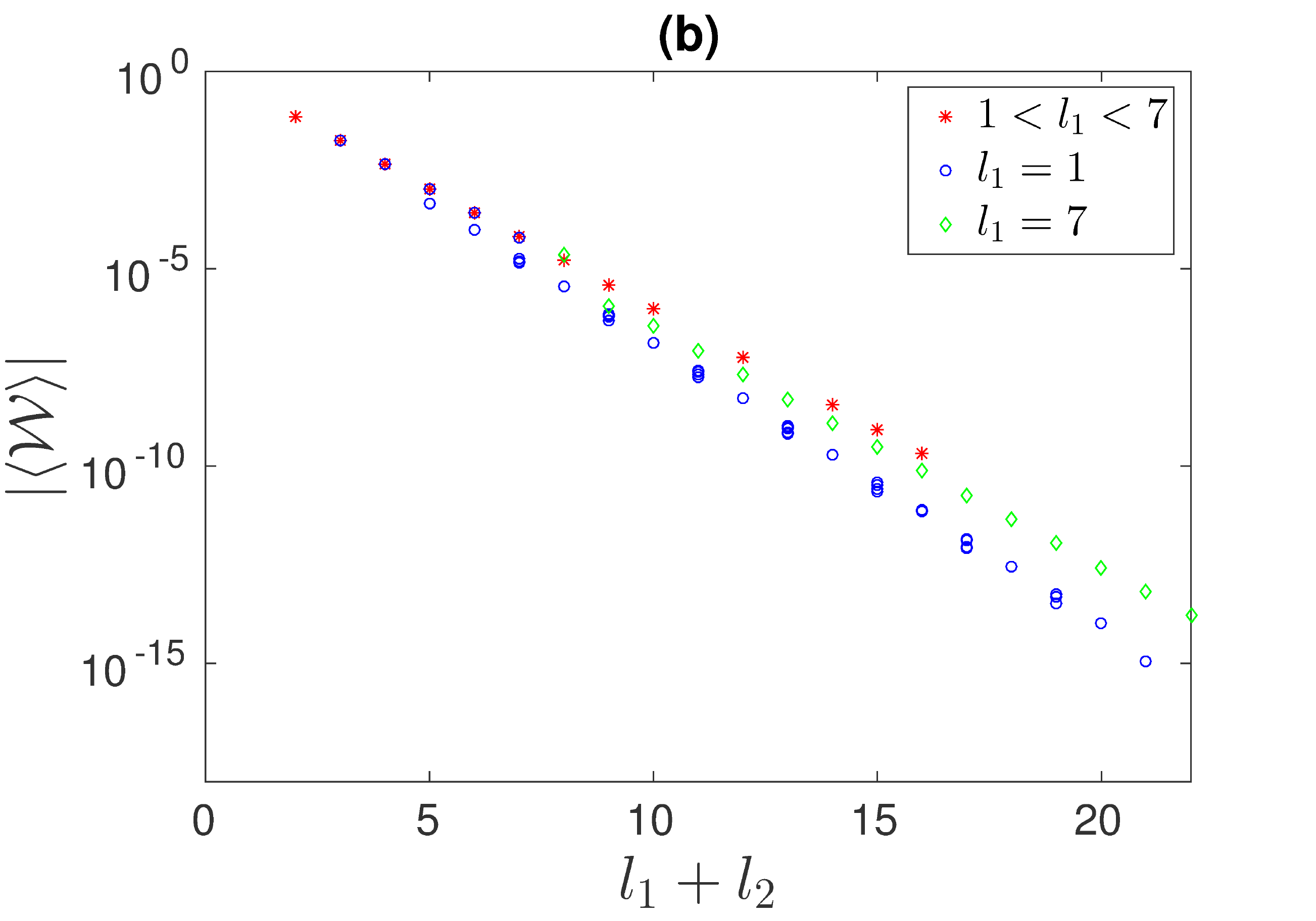}
\includegraphics[width=0.33\textwidth]{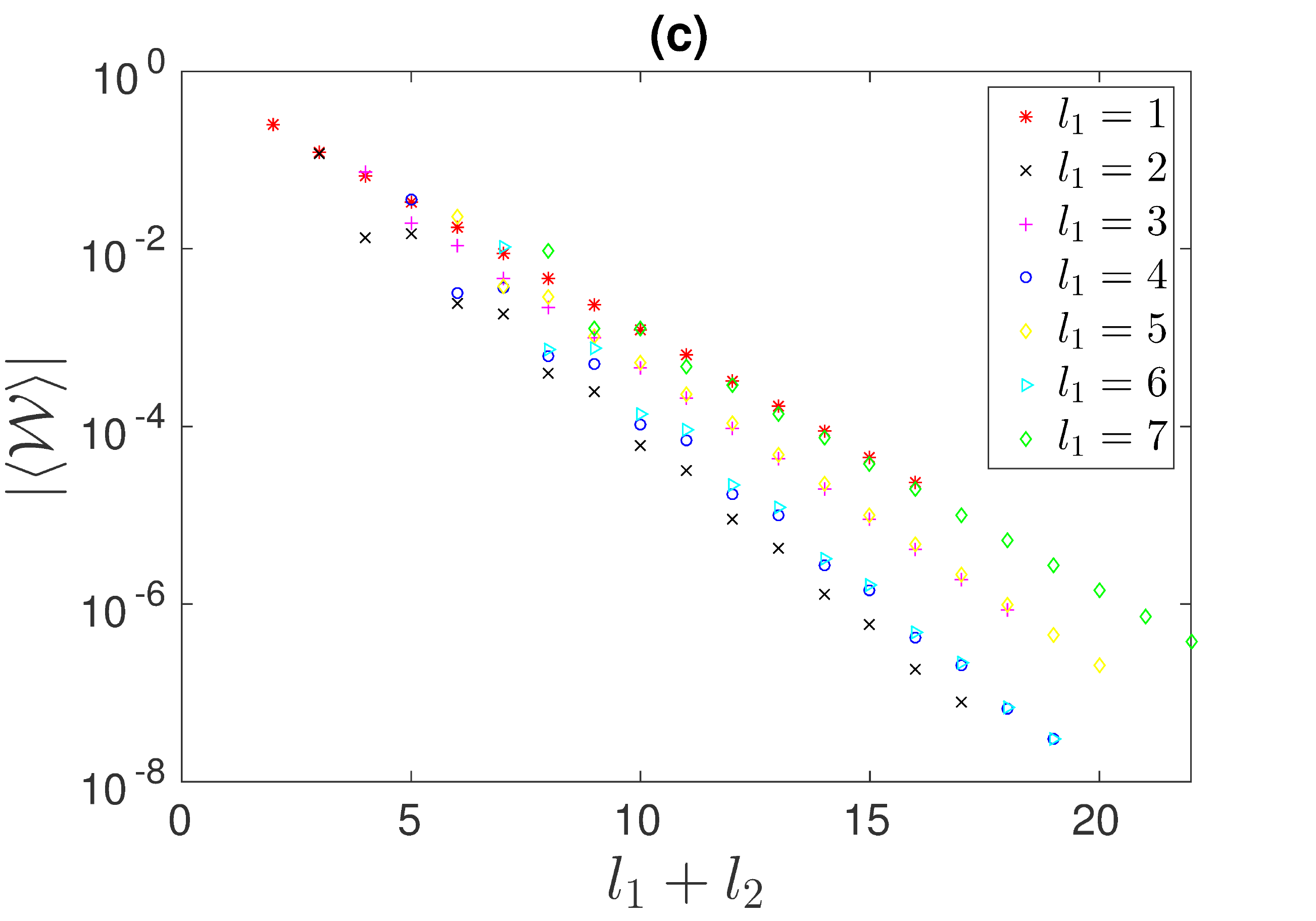}
 \caption{Magnitude of the expectation value of the Wilson loop operator as a function of its perimeter $l_1 + l_2$ for $t = 0.1$ (a), $t = 1$ (b) and $t = 10$ (c). The results were obtained on a cylinder of size $L_1 \times L_2 = 8 \times (81 + l_2)$. For $t = 0.1$ the decay follows a perimeter law, whereas from $t = 1$ on, several branches appear corresponding to different widths $l_1$ of the Wilson loop. Data points of $l_1 = 4 \pm n$ ($n = 1,2,3$) lie on top of each other, due to the periodic boundary conditions, suggesting that the deviation from a pure perimeter law for large $t$ is a result of the finite cylinder circumference.}
 \label{fig:Wilsont}
\end{figure}

\begin{figure}
 \includegraphics[width=0.48\textwidth]{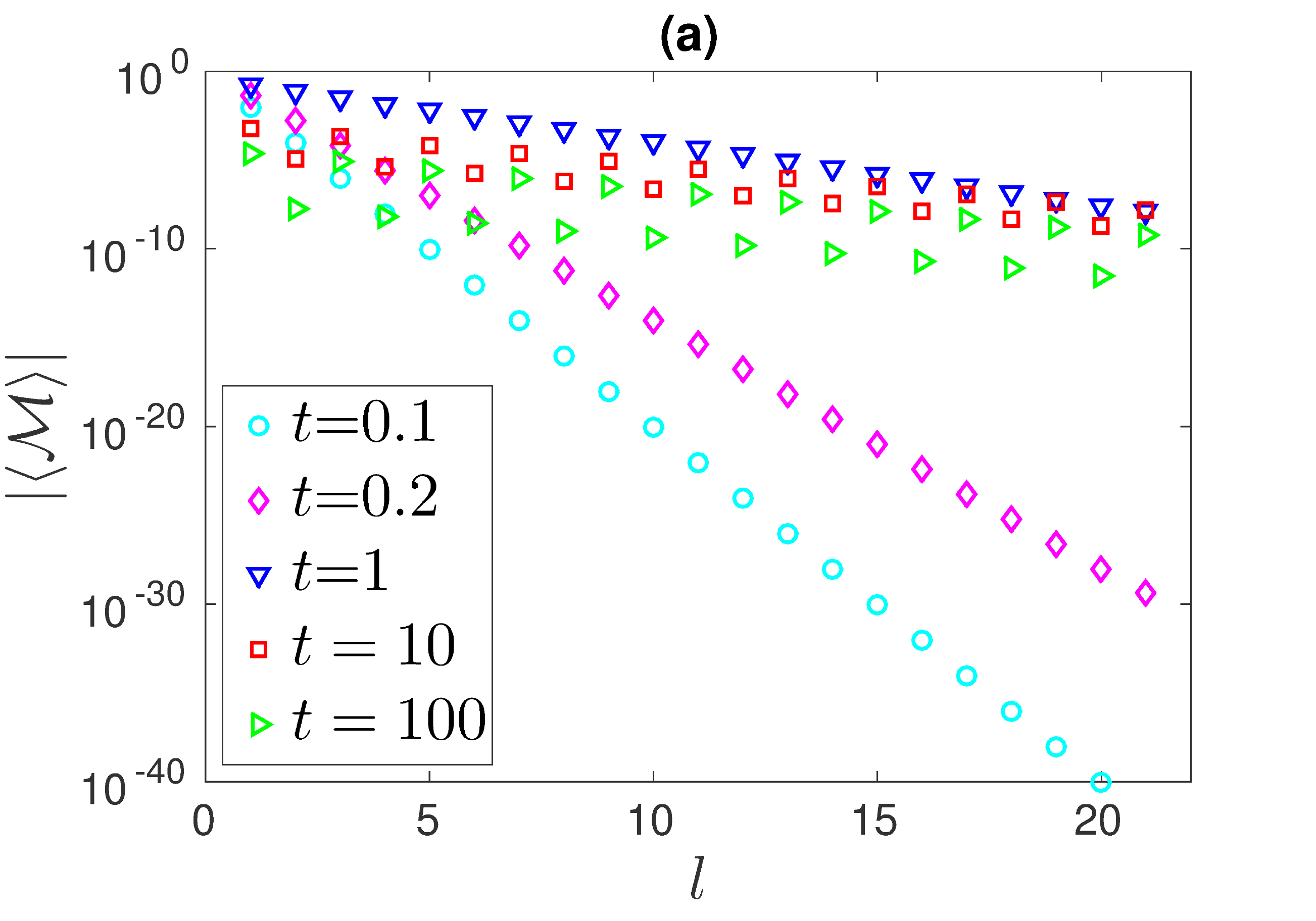}
 \includegraphics[width=0.50\textwidth]{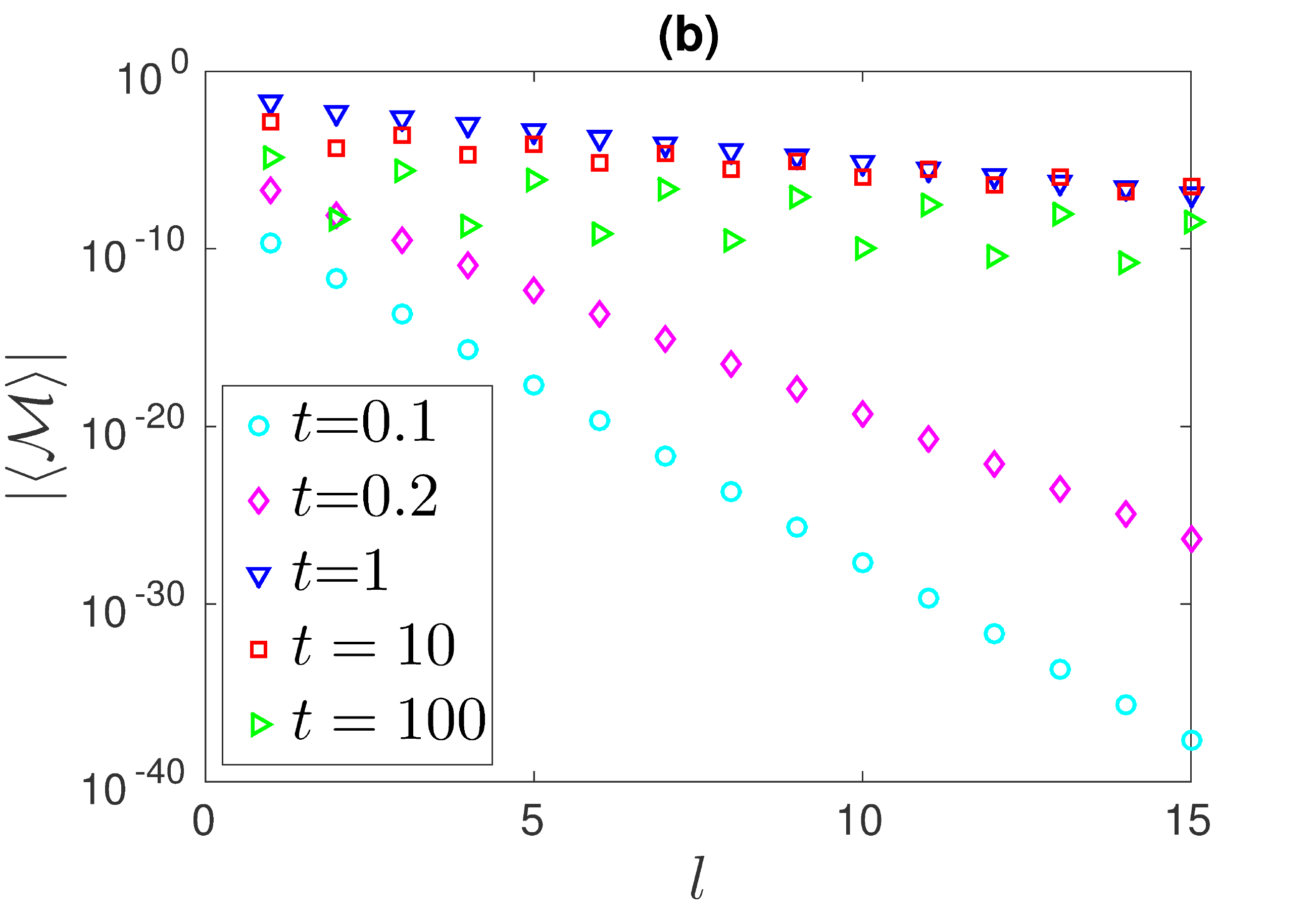}

 \caption{(a) Magnitude of the meson and tunneling operator for odd and even lengths $l$ of the string operator, respectively, for the shown values of $t$.  The results were obtained on a cylinder of size $L_1 \times L_2 = 8 \times (81 + l)$. We observe an exponential decay whose decay length increases with $t$ until $t = 1$ is reached, whereupon it starts to decrease. The meson and tunneling operator have the same decay length. The overall prefactor of their exponential scaling is the same until $t = 1$ and develops a mismatch for larger $t$, corresponding to the even-odd fluctuations in the figure. (b) Magnitude of the horseshoe shaped meson / tunneling operator $\mathcal{M}(\mathcal{P}=\mathcal{C}(l)/2)$ showing the same behavior.} \label{fig:mesons_horseshoe}
\end{figure}

The expectation value of the Wilson loop displays a perimeter law decay, with finite size effects which increase with $t$ (in agreement with the gap $\varDelta$ decreasing with $t$), cf. Fig. \ref{fig:Wilsont}.  Besides, the mesonic string for mesons oriented in the vertical direction always shows a well defined exponential decay with the distance, consistently with the gapped transfer matrix, although the decay length increases with increasing $t$ until $t = 1$, whereupon it starts to decrease again (and acquires a different overall prefactor as compared to the tunneling operator), see Fig. \ref{fig:mesons_horseshoe}.

The two sets of data can be combined to obtain the Fredenhagen-Marcu horseshoe order parameter \cite{Fredenhagen1986,Fredenhagen1988}). In analogy with the study \cite{Zohar2015b} of U(1) symmetric PEPS states, we define such order parameter for rectangular loops of width $l_1=4$ as a function of their height $l_2$:
\begin{equation} \label{FM}
 \rho\left(l_2\right) = \frac{\left|\left\langle \mathcal{M}(\mathcal{P}=\mathcal{C}(l_2)/2) \right\rangle \right| }{\sqrt{\left\langle \mathcal{W}(\mathcal{C}( l_2))\right\rangle }}
\end{equation}
where the Wilson loop $\mathcal{W}$ is associated to a rectangle $\mathcal{C}$ of size $4 \times l_2$ of height $l_2$ (with $l_2$ odd). The fermionic particle and antiparticle associated to $\mathcal{M}$ are created in the middle of the two horizontal edges of the rectangle $\mathcal{C}$, and the path $\mathcal{P}$ covers half of it.

\begin{figure}
\includegraphics[width=0.55\textwidth]{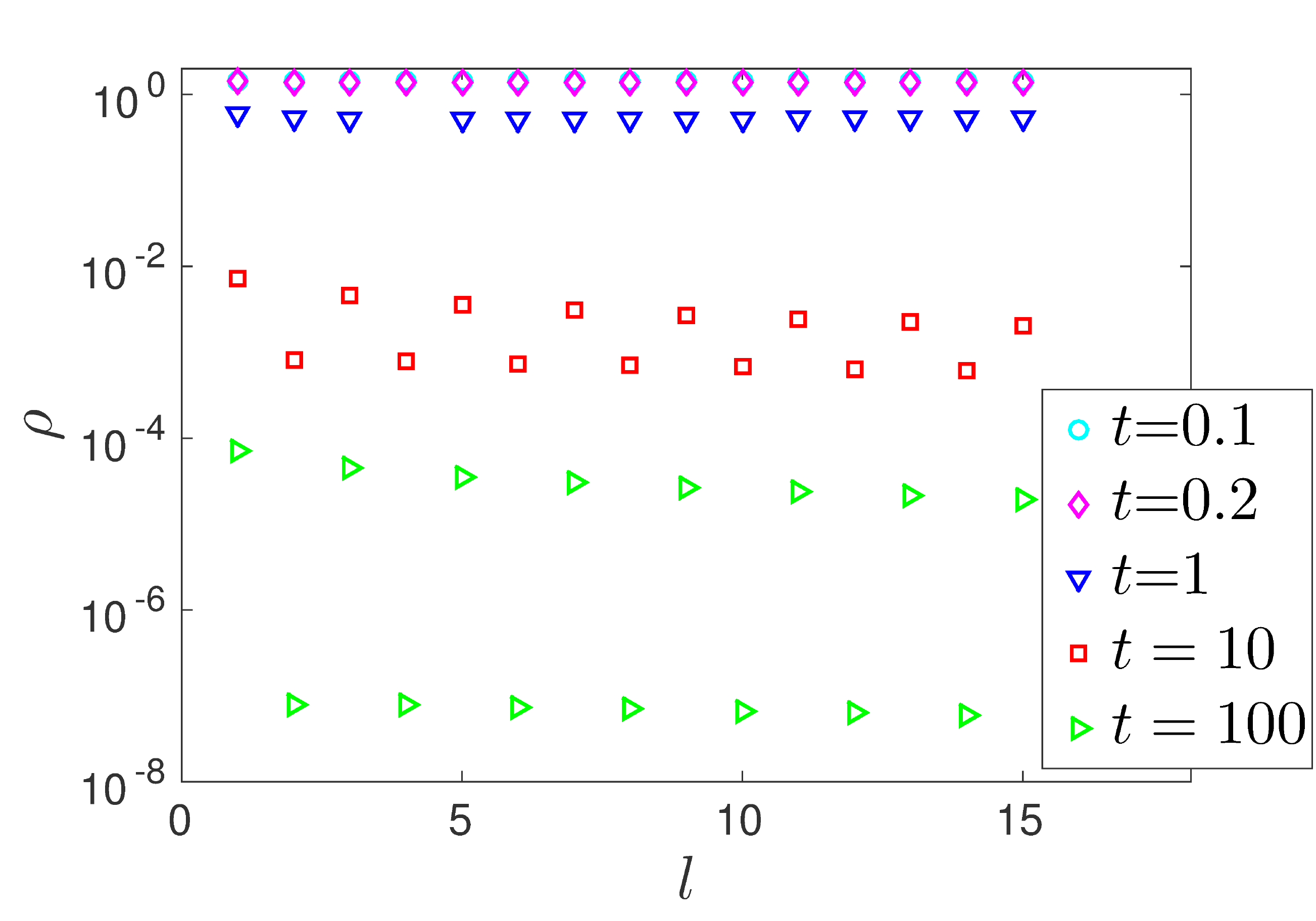}
 \caption{Fredenhagen-Marcu order parameter $\rho$ as a function of the length $l_2$ of the Wilson loop $\mathcal{W}(\mathcal{C}( l_2))$ and meson string $\mathcal{M}(\mathcal{P}=\mathcal{C}(l_2)/2)$ calculated from the data displayed in Figs. \ref{fig:Wilsont} and \ref{fig:mesons_horseshoe} according to Eq. \eqref{FM}. Apart from a strong even/odd effect, the parameter $\rho$ seems to converge to positive limits for all the values of $t$.} \label{fig:FM}
\end{figure}

The order parameter $\rho$ may be associated to the screening of dynamical charges \cite{Gregor2011,Fredenhagen1986,Greensite03}, and, in our case, it converges to a finite constant for increasing $l_2$ for every value of $t$, as show in Fig. \ref{fig:FM}. This signals that the same mechanism determines the decay of the mesonic string and the Wilson loop, therefore the phase should be consistent with the presence of screening of the dynamical charges.

\begin{figure}
 \includegraphics[width=0.85\textwidth]{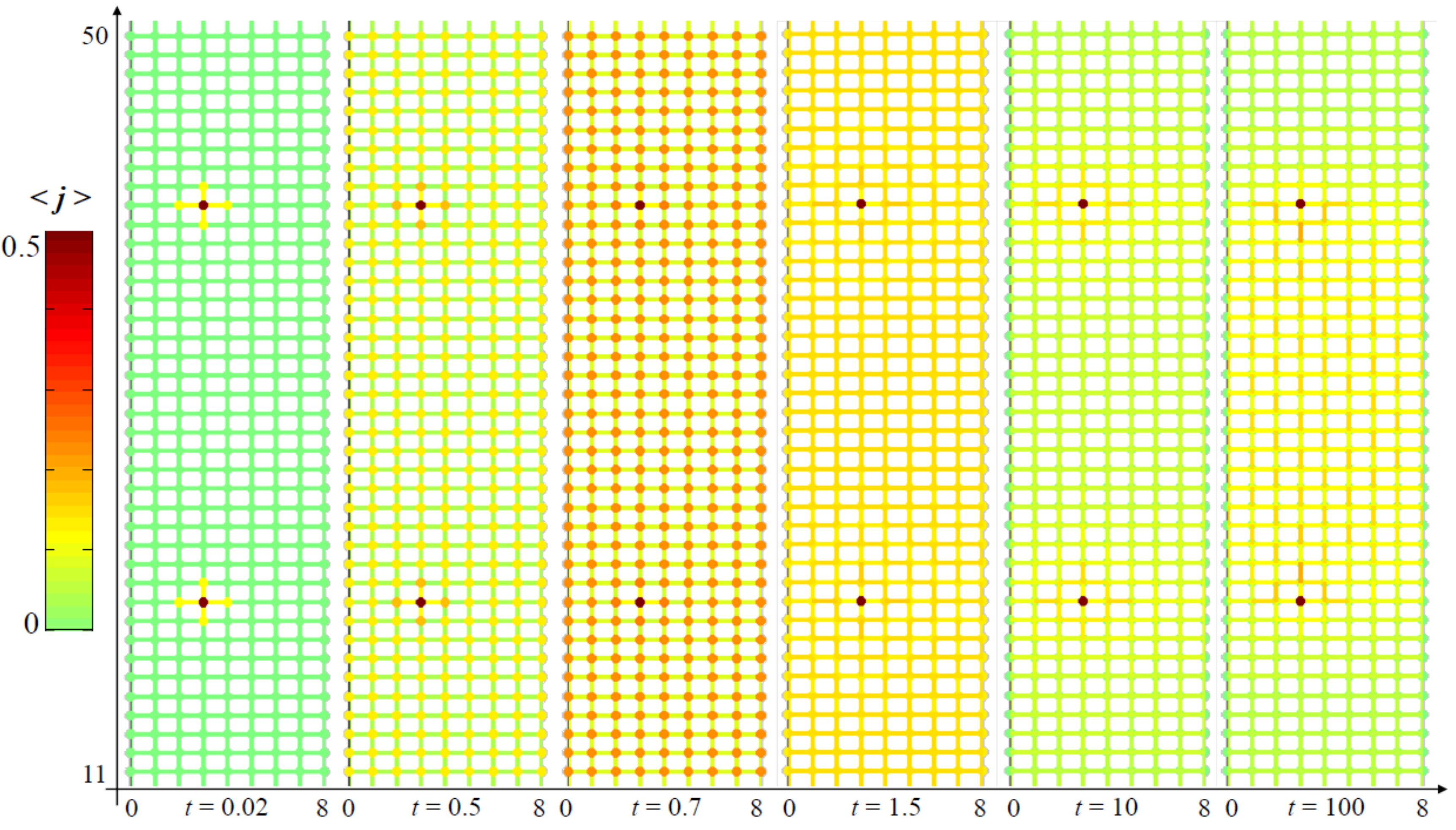}
 \caption{The expectation value of the representation of both the matter sites (circles) and the gauge fields (links) is displayed in the presence of two static charges for several value of $t>0$ for a cylinder of size $8 \times 62$. For small values of $t$ both the static charges and the fields are screened; for large values of $t$ the electric field propagates over longer distances, especially in the region included between the charges.}
 \label{Fig:TwostaticDynamic}
\end{figure}

This screening behavior appears to be consistent also for static charges in the system, which can be added by exploiting a pair of fiducial states of the kind (\ref{St1},\ref{St2}). Similarly to the study of the pure gauge case, we introduced such a pair in a column of the system, but, for $t>0$, the physics we obtain is considerably different. Our results are exemplified in Fig. \ref{Fig:TwostaticDynamic}. Each static charge gets effectively screened by a cloud of dynamical charges in the adjacent sites. For $t \ll 1$, one can observe that the four sites around the static charge assume an average fermionic density $1/4$, thus completely screening the static fermion: the expectation value of the electric field representation cannot be distinguished from its background value outside the four links surrounding each background charge (This results from the fact that we do not allow the presence of a dynamical charge in the site of the static one, due to the construction (\ref{St1},\ref{St2})). By increasing the value of $t$, however, the average occupation number of the physical fermions in the bulk increases and it is more and more difficult to distinguish the screening charges from the background; the electric field perturbation, however, still appears to be localized around the static charges until the parameter $t$ reaches a threshold $t \approx 2$ . For $t>2$ the electric field starts to percolate in a larger radius. The increase of this radius is related to the appearance of doubly occupied fermionic sites, which become increasingly more convenient from the energetic point of view and compete with the possibility of screening the static charges. When considering a configuration of static charges separated vertically by 21 links, the electric flux perturbations generated by these charges begin to merge for $t \gtrsim 15$. For large values of $t$, indeed, one observes that the intermediate region presents an average value of the representation for the electric fluxes sensibly different from the outer region (see Fig. \ref{Fig:TwostaticDynamic}).
In configurations in which two static charges are neighboring, instead, their effect is negligible up to $t \approx 0.7$. For larger values of $t$, however, a perturbation in the electric flux can be observed decaying in a radius of approximately 8 sites.

To summarize, the changing screening behavior from small to large $t$ may be a hint for a crossover between different regimes: for small $t$ the dynamical charges appear to be confined and they are introduced in the system only in the presence of static charges, causing their efficient screening. For large $t$, instead, dynamical charges are progressively introduced in the system until saturate at their maximal density for $t \to \infty$ (the maximal density corresponds to $2$ fermions per lattice site, thus $\left\langle j\right\rangle =0$); this reduces the screening of static charges, thus allowing a relevant spreading of the electric fluxes. We observe, however, that in this second regime with average fermionic density larger than $1$, the lattice effects become more and more relevant and the physics described close to saturation may have no physical counterpart in the standard field theory description \cite{Azcoiti2003}.

\section{Conclusions and Outlook}

In this work we discussed Projected Entangled Pair States (PEPS) in $2+1$ dimensions, with both global and local SU(2) symmetries, as a case study and a demonstration of applying PEPS techniques for the study of states of lattice gauge theories with a non-Abelian gauge group.

We began with a study of fermionic Gaussian PEPS with a global SU(2) symmetry; we parameterized such states for a bond dimension 2 and studied them using standard Gaussian techniques. We were able to obtain a continuum limit of this model, which may be seen as a non-relativistic limit of the free Dirac theory. Then we gauged the states to obtain PEPS which describe both gauge fields and fermionic matter, with a local SU(2) symmetry; finally, we studied their phase diagram using both analytical and numerical tools, exploiting special features of the PEPS.

The pure gauge theory exhibits two well known phases - the Higgs (gapped) and Coulomb (gapless) phases; the states involving both gauge fields and matter, instead, do not show any precise signature of phase transitions, but rather imply the possibility of a crossover, between two different physical behaviors, possibly having to do with screening of static charges and/or confinement of dynamical ones.

We emphasize that the SU(2) model discussed in this work is merely an example to illustrate the capabilities of the methodology we propose, which enables to exploit locally symmetric PEPS for the study of lattice gauge theories. We showed that the analysis of the symmetries of the PEPS may provide precise informations about the phase diagram of the model, as in the case of the pure gauge theory in which we could analytically determine the phase boundaries of the system. Furthermore we could verify such analytical results through the study of the PEPS transfer matrix and its spectrum. These are powerful tools at the basis of the study of many-body systems with PEPS, and we believe they can provide a new insight also for the study of lattice gauge theories.

In our paradigmatic case of a non-Abelian and local gauge symmetry, we showed that the corresponding symmetries of the fiducial states in the construction of the PEPS allows us to obtain transfer matrix operators with very strong symmetry requirements, able to simplify the analytical and numerical investigations of the system. Furthermore, we introduced an additional transfer matrix operator which includes an MPO for the description of Wilson line. We believe that this must be considered a central object in the study of the phase diagrams, due to its relationship with the gauge-invariant string order parameters customarily used for the study of lattice gauge theories.

We also remark that, in this work, the PEPS were contracted exactly; such exact contraction is numerically demanding but, on the practical level, it is also possible to employ advanced numerical methods as in \cite{Cirac2011,Corboz2014,Lubash2014} to increase the range of possibilities achieved with such methods. This may allow, in particular, to extend both the physical dimension of the systems under analysis and the bond dimension of the PEPS. These improvements in the numerical analysis may bring to achieve two different kind of results: on one hand, it would extend the variety of models and physical phenomena which can be systematically studied (for example to more demanding but realistic symmetries, such as SU(3), or to more exotic phases such as color superfluidity); on the other, this would open the path to variational studies of existing, known lattice gauge models.

Finally we mention that another possible use of these tensor network techniques is the study of quantum simulations of lattice gauge theories, based on cold atomic gases or other simulating systems, in which the physical realization of a gauge invariant system may rely on truncations similar to the one which characterizes our PEPS construction. More in general several quantum simulation schemes have been recently proposed \cite{Zohar2015a,Wiese2013} and even realized \cite{Martinez2016} and it is foreseeable that tensor network techniques may be used to validate and interpret the experimental results.

\section*{Acknowledgments}
EZ would like to thank the Humboldt foundation for its support. TBW gratefully acknowledges financial support by the Topological Protection and Non-Equilibrium States in Strongly Correlated Electron Systems (TOPNES) research programme.

\appendix

\setcounter{sttmnt}{0}
\renewcommand{\thesttmnt}{A\arabic{sttmnt}}

\section{SU(2) transformation properties} \label{app1}

In this Appendix, we discuss in further detail the transformation properties under SU(2) transformations. First, we consider the transformations of the fermionic fundamental SU(2) spinors $\alpha^{\dagger}_m$. The transformations may be either right ($\Theta_g$ as in eq. (\ref{Rtrans}), generated by the set of generators $R^a\left(\alpha\right)$ defined in eq. (\ref{Rgen})), or left ($\tilde{\Theta}_g$ as in (\ref{Ltrans}), generated by $L^a\left(\alpha\right)$ defined in (\ref{Lgen})).

Using the $j=1/2$ Wigner matrices, the results of such transformations, used throughout this work, is given in the following table for several common objects.

\begin{center}
\begin{tabular}{|c|c|c|c|c|}
\hline
$X$ & $\Theta_{g}X\Theta_{g}^{\dagger}$ & $\Theta_{g}^{\dagger}X\Theta_{g}$ & $\widetilde{\Theta}_{g}X\widetilde{\Theta}_{g}^{\dagger}$ & $\widetilde{\Theta}_{g}^{\dagger}X\widetilde{\Theta}_{g}$\tabularnewline
\hline
\hline
$\alpha_{m}$ & $D_{mn}\left(g^{-1}\right)\alpha_{n}$ & $D_{mn}\left(g\right)\alpha_{n}$ & $\alpha_{n}D_{nm}\left(g^{-1}\right)$ & $\alpha_{n}D_{nm}\left(g\right)$\tabularnewline
\hline
$\alpha_{m}^{\dagger}$ & $\alpha_{n}^{\dagger}D_{nm}\left(g\right)$ & $\alpha_{n}^{\dagger}D_{nm}\left(g^{-1}\right)$ & $D_{mn}\left(g\right)\alpha_{n}^{\dagger}$ & $D_{mn}\left(g^{-1}\right)\alpha_{n}^{\dagger}$\tabularnewline
\hline
$\widetilde{\alpha}_{m}\equiv\epsilon_{mn}\alpha_{n}$ & $\widetilde{\alpha}_{n}D_{nm}\left(g\right)$ & $\widetilde{\alpha}_{n}D_{nm}\left(g^{-1}\right)$ & $D_{mn}\left(g\right)\widetilde{\alpha}_{n}$ & $D_{mn}\left(g^{-1}\right)\widetilde{\alpha}_{n}$\tabularnewline
\hline
$\widetilde{\alpha}_{m}^{\dagger}\equiv\epsilon_{mn}\alpha_{n}^{\dagger}$ & $D_{mn}\left(g^{-1}\right)\widetilde{\alpha}_{n}^{\dagger}$ & $D_{mn}\left(g\right)\widetilde{\alpha}_{n}^{\dagger}$ & $\widetilde{\alpha}_{n}^{\dagger}D_{nm}\left(g^{-1}\right)$ & $\widetilde{\alpha}_{n}^{\dagger}D_{nm}\left(g\right)$\tabularnewline
\hline
$\alpha_{m}^{\dagger}\alpha_{m}$ & $\alpha_{m}^{\dagger}\alpha_{m}$ & $\alpha_{m}^{\dagger}\alpha_{m}$ & $\alpha_{m}^{\dagger}\alpha_{m}$ & $\alpha_{m}^{\dagger}\alpha_{m}$\tabularnewline
\hline
$\alpha_{m}^{\dagger}\widetilde{\alpha}_{m}^{\dagger}$ & $\alpha_{m}^{\dagger}\widetilde{\alpha}_{m}^{\dagger}$ & $\alpha_{m}^{\dagger}\widetilde{\alpha}_{m}^{\dagger}$ & $\alpha_{m}^{\dagger}\widetilde{\alpha}_{m}^{\dagger}$ & $\alpha_{m}^{\dagger}\widetilde{\alpha}_{m}^{\dagger}$\tabularnewline
\hline
\end{tabular}
\end{center}

Note that multiplication of a creation/annihilation operator by $\epsilon_{mn}$, as in $\widetilde{\alpha}_{m}^{\dagger}\equiv\epsilon_{mn}\alpha_{n}^{\dagger}$, results in an object which undergoes the same transformation as the respective annihilation/operators operator, or changes the orientation (right $\leftrightarrow$ left) while inverting the transformation ($g \leftrightarrow g^{-1}$). This can be seen from the table, and it is a useful feature for particle-hole transformations. This is the result of the relation
\begin{equation}
\epsilon M^\intercal\epsilon^\intercal=\mathrm{adj}\left(M\right)=\det\left(M\right)M^{-1}\label{eq:adjM}
\end{equation}
which holds for $2 \times 2$ matrices $M$. In particular, for  $M=D\in$ SU(2), since $\det\left(D\right)=1$,
\begin{equation}
\epsilon D^\intercal\epsilon^\intercal=\mathrm{adj}\left(D\right)=D^{-1}
\end{equation}
On the other hand, for Pauli matrices whose determinant is $-1$,
\begin{equation}
\epsilon \sigma^\intercal\epsilon^\intercal=-\sigma^{-1}=-\sigma
\end{equation}

Similarly, one may consider the transformation properties of the gauge field operators, $U_{mn}$:

\begin{center}
\begin{tabular}{|c|c|c|c|c|}
\hline
$X$ & $\Theta_{g}X\Theta_{g}^{\dagger}$ & $\Theta_{g}^{\dagger}X\Theta_{g}$ & $\widetilde{\Theta}_{g}X\widetilde{\Theta}_{g}^{\dagger}$ & $\widetilde{\Theta}_{g}^{\dagger}X\widetilde{\Theta}_{g}$\tabularnewline
\hline
\hline
$U_{mn}$ & $U_{mn'}D_{n'n}\left(g\right)$ & $U_{mn'}D_{n'n}\left(g^{-1}\right)$ & $D_{mm'}\left(g\right)U_{m'n}$ & $D_{mm'}\left(g^{-1}\right)U_{m'n}$\tabularnewline
\hline
$\overline{U}_{mn} \equiv \epsilon_{mm'}U_{m'n'}\epsilon_{nn'}$ & $\overline{U}_{mn'}D_{nn'}\left(g^{-1}\right)$ & $\overline{U}_{mn'}D_{nn'}\left(g\right)$ & $D_{m'm}\left(g^{-1}\right)\overline{U}_{m'n}$ & $D_{m'm}\left(g\right)\overline{U}_{m'n}$\tabularnewline
\hline
\end{tabular}
\end{center}

\section{Covariance matrix approach for the parametrization of the fPEPS} \label{app2}

Here we shall describe another method for the derivation of the parametrization of the global SU(2) invariant fermionic Gaussian PEPS, through the matrix $T$, which will lead to the result (\ref{TSU2}). This way is more mathematical than the one introduced in the main text, and it shows also the other direction, i.e. that a state $\left|\psi\left(T\right)\right\rangle$ built in the PEPS construction discussed above is invariant under the SU(2) global transformation (\ref{SU2invglob}) if and only if it has the parametrization (\ref{TSU2}).

This approach is very similar to the one used for the U(1) invariant
states discussed in \cite{Zohar2015b}. It is based on the fact that, as the generators of the transformation
satisfy the (right) $SU\left(2\right)$ algebra, they are simply rotations,
and as such they may be decomposed by three rotations using the Euler
angles, e.g.
\begin{equation}
\hat \Theta_{g}=\Theta\left(\alpha,\beta,\gamma\right)=e^{i\alpha G_{z}}e^{i\beta G_{x}}e^{i\gamma G_{z}}
\end{equation}

Thus, it is sufficient to demand that our fiducial state, or its creating operator $A$, is invariant under $G_{z}$ and $G_{x}$
transformations.

Let us discuss $G_z$ invariance first. This operator, up to a factor of $1/2$, contains only number operators of all the participating fermionic modes (both physical and virtual), with either positive or negative signs. We label the fermionic modes as "negative" and "positive"
with respect to $G_{z}$ - i.e., according to the sign of their number
operator within the generator: the negative modes, $\left\{ a_{i}^{\dagger}\right\} $,
are $\left\{ \psi_{1}^{\dagger},l_{1}^{\dagger},d_{1}^{\dagger},r_{2}^{\dagger},u_{2}^{\dagger}\right\} $,
while the positive ones, $\left\{ b_{i}^{\dagger}\right\} $, are
$\left\{ \psi_{2}^{\dagger},l_{2}^{\dagger},d_{2}^{\dagger},r_{1}^{\dagger},u_{1}^{\dagger}\right\} $,
where,  within each set, the operators which
transform with the right transformations appear before these which transform
with the left transformations.

In terms of the fermionic operators $\alpha_i,\alpha_i^{\dagger}$, one defines the covariance sub-matrices
\begin{equation}
Q_{ij}=\frac{i}{2}\left\langle \left[\alpha_{i},\alpha_{j}\right]\right\rangle \,; \quad R_{ij}=\frac{i}{2}\left\langle \left[\alpha_{i},\alpha_{j}^{\dagger}\right]\right\rangle
\end{equation}
out of which the fermionic covariance matrix is built:
\begin{equation}
\varGamma=\left(\begin{array}{cc}
Q & R\\
\overline{R} & \overline{Q}
\end{array}\right)
\label{fercov}
\end{equation}
the sub-blocks $Q,R$ may be further decomposed into blocks, corresponding
to correlations between positive and negative modes,
\begin{equation}
Q=\left(\begin{array}{cc}
Q_{aa} & Q_{ab}\\
Q_{ba} & Q_{bb}
\end{array}\right)\,; \quad R=\left(\begin{array}{cc}
R_{aa} & R_{ab}\\
R_{ba} & R_{bb}
\end{array}\right).
\end{equation}

Note that the $G_z$ invariance is an Abelian symmetry, identical (up to the $1/2$ factor and the appearance of two physical modes) to the U(1) symmetry discussed in \cite{Zohar2015b}. Thus, it is straightforward to use the parametrization presented there as the starting point:
the
state is invariant under $G_{z}$ transformations if and only if
\begin{equation}
Q_{aa}=Q_{bb}=R_{ab}=R_{ba}=0.
\end{equation}

Next we wish to demand invariance under $G_{x}$ transformations as well. For
that, what we have to do is a change of basis: the modes are given
in the $z$ basis, but if we rotate them to the $x$ basis, $G_{x}$
will be diagonal, and then one can demand a similar block structure
for the rotated covariance matrix. Fortunately, $\sigma_{x}^\intercal=\sigma_{x}$,
and therefore the right and left generators are similar and we do not
have to worry about the orientation. For that reason, for all the modes, the transformation
is
\begin{equation}
\begin{aligned}\alpha_{1}\longrightarrow\frac{1}{\sqrt{2}}\left(\alpha_{1}+\alpha_{2}\right)\\
\alpha_{2}\longrightarrow\frac{1}{\sqrt{2}}\left(\alpha_{1}-\alpha_{2}\right)
\end{aligned}
\end{equation}
which, if we align the modes such that negative modes are first and
positive second, corresponds to rotating the set of creation (or
annihilation) operators with the matrix
\begin{equation}
U_{x}=\frac{1}{\sqrt{2}}\left(\begin{array}{cccccccccc}
1 & 0 & 0 & 0 & 0 & 1 & 0 & 0 & 0 & 0\\
0 & 1 & 0 & 0 & 0 & 0 & 1 & 0 & 0 & 0\\
0 & 0 & 1 & 0 & 0 & 0 & 0 & 1 & 0 & 0\\
0 & 0 & 0 & -1 & 0 & 0 & 0 & 0 & 1 & 0\\
0 & 0 & 0 & 0 & -1 & 0 & 0 & 0 & 0 & 1\\
1 & 0 & 0 & 0 & 0 & -1 & 0 & 0 & 0 & 0\\
0 & 1 & 0 & 0 & 0 & 0 & -1 & 0 & 0 & 0\\
0 & 0 & 1 & 0 & 0 & 0 & 0 & -1 & 0 & 0\\
0 & 0 & 0 & 1 & 0 & 0 & 0 & 0 & 1 & 0\\
0 & 0 & 0 & 0 & 1 & 0 & 0 & 0 & 0 & 1
\end{array}\right)
\end{equation}
then, one obtains the blocks of the covariance matrix in the $x$
basis,
\begin{equation}
Q^{x}=U_{x}QU_{x}^\intercal\,; \quad R^{x}=U_{x}RU_{x}^{\dagger}
\end{equation}
and all we have to do is to demand that
\begin{equation}
Q_{aa}^{x}=Q_{bb}^{x}=R_{ab}^{x}=R_{ba}^{x}=0.
\end{equation}

In addition to this, one has to demand that $\varGamma\varGamma^{\dagger}=\frac{1}{4}\mathbf{1}$ for a pure state. In this way a complete parameterization of the covariance matrix of the SU(2) invariant fiducial states is achieved.

How shall the exponential operator constructing the fiducial state
be built then? First, as in the U(1) case, it is clear
that only $a$ and $b$ modes may be coupled, and thus
\begin{equation}
A=\exp\left(T_{ij}a_{i}^{\dagger}b_{j}^{\dagger}\right)
\end{equation}
will create a $G_{z}$ invariant state. However, $T_{ij}$ must obey a more
specific structure in order to comply with the full non-Abelian invariance.
For that, we write $A$ in the form $A=\exp\left(\tilde{T}_{ij}\alpha_{i}^{\dagger}\alpha_{j}^{\dagger}\right)$
with $\tilde{T}=\frac{1}{2}\left(\begin{array}{cc}
0 & T\\
-T^\intercal & 0
\end{array}\right)$. Then, one can rotate the modes to the $x$ basis again, and demand
that the new matrix obtained, $\hat{T}=U_{x}^\intercal\tilde{T}U_{x}$,  has
a block structure similar to $\hat{T}^{ij}_{mn}=\mu^{ij}\epsilon_{mn}$ (i.e.,
only oppositely signed modes are coupled with each other). This is
achieved if and only if eq. (\ref{TSU2}) is satisfied.

\section{Removing redundant phases in the parametrization} \label{app:phases}

In this Appendix, we show how to remove the phases $\eta_{r,u}$ in the PEPS parametrization (\ref{Tmatrix0}).
Denote:
\begin{equation}
t\equiv t_{0}e^{i\phi_{t}}\,,\quad z\equiv z_{0}e^{i\phi_{z}}\,, \quad x\equiv x_{0}e^{i\phi_{x}}\,,
\end{equation}
with $t_{0},z_{0},x_0\geq 0$,  and
\begin{equation}
\eta_{u}\equiv e^{i\phi_{u}}
\,,\quad
\eta_{r}\equiv e^{i\phi_{r}}.
\end{equation}
Define the matrix
\begin{equation}
S=\left(\begin{array}{ccccc}
1 & 0 & 0 & 0 & 0\\
0 & e^{\frac{i}{2}\left(\phi_{r}-\phi_{u}\right)} & 0 & 0 & 0\\
0 & 0 & e^{\frac{i}{2}\left(\phi_{r}+\phi_{u}\right)} & 0 & 0\\
0 & 0 & 0 & e^{-\frac{i}{2}\left(\phi_{r}-\phi_{u}\right)} & 0\\
0 & 0 & 0 & 0 & e^{-\frac{i}{2}\left(\phi_{r}+\phi_{u}\right)}
\end{array}\right)
\label{phtrans}
\end{equation}
which may represent a unitary transformation on the virtual fermions,
\begin{equation}
a_{i}^{\dagger}\longrightarrow S_{ij}a_{j}^{\dagger}
\,,\quad
b_{i}^{\dagger}\longrightarrow S_{ij}b_{j}^{\dagger}\,.
\end{equation}
The bond projectors will be invariant under this transformation,
and, on the other hand, $T$ will transform to
\begin{equation}
S^\intercal TS=\left(\begin{array}{ccccc}
0 & e^{\frac{i}{2}\left(\phi_{r}-\phi_{u}+2\phi_{t}\right)}t_{0} & \eta_{p}^{-1}e^{\frac{i}{2}\left(\phi_{r}-\phi_{u}+2\phi_{t}\right)}t_{0} & -\eta_{p}^{-2}e^{\frac{i}{2}\left(\phi_{r}-\phi_{u}+2\phi_{t}\right)}t_{0} & -\eta_{p}^{-3}e^{\frac{i}{2}\left(\phi_{r}-\phi_{u}+2\phi_{t}\right)}t_{0}\\
e^{\frac{i}{2}\left(\phi_{r}-\phi_{u}+2\phi_{t}\right)}t_{0} & e^{i\left(\phi_{r}-\phi_{u}+\phi_{x}\right)}x_{0} & e^{i\left(\phi_{r}+\phi_{z}\right)}z_{0}/\sqrt{2} & 0 & e^{i\left(\phi_{r}+\phi_{z}\right)}z_{0}/\sqrt{2}\\
\eta_{p}^{-1}e^{\frac{i}{2}\left(\phi_{r}-\phi_{u}+2\phi_{t}\right)}t_{0} & e^{i\left(\phi_{r}+\phi_{z}\right)}z_{0}/\sqrt{2} & e^{i\left(\phi_{r}-\phi_{u}+\phi_{x}\right)}x_{0} & -e^{i\left(\phi_{r}+\phi_{z}\right)}z_{0}/\sqrt{2} & 0\\
\eta_{p}^{-2}e^{\frac{i}{2}\left(\phi_{r}-\phi_{u}+2\phi_{t}\right)}t_{0} & 0 & e^{i\left(\phi_{r}+\phi_{z}\right)}z_{0}/\sqrt{2} & -e^{i\left(\phi_{r}-\phi_{u}+\phi_{x}\right)}x_{0} & -e^{i\left(\phi_{r}+\phi_{z}\right)}z_{0}/\sqrt{2}\\
\eta_{p}^{-3}e^{\frac{i}{2}\left(\phi_{r}-\phi_{u}+2\phi_{t}\right)}t_{0} & -e^{i\left(\phi_{r}+\phi_{z}\right)}z_{0}/\sqrt{2} & 0 & -e^{i\left(\phi_{r}+\phi_{z}\right)}z_{0}/\sqrt{2} & -e^{i\left(\phi_{r}-\phi_{u}+\phi_{x}\right)}x_{0}
\end{array}\right)\,.
\label{Tphases}
\end{equation}
Since the projectors are invariant, we have that for every such $S$,
\begin{equation}
\left|\psi\left(S^\intercal T S\right)\right\rangle = \left|\psi\left(T\right)\right\rangle\,.
\end{equation}
We may use this ``virtual symmetry'' to remove the phases $\eta_{u},\eta_{r}$. Let us choose:
\begin{equation}
\phi_{r}=-\phi_{z}\,,
\label{phase1}
\end{equation}
\begin{equation}
\phi_{u}=-\phi_{z}+2\phi_{t}\,.
\label{phase2}
\end{equation}
We obtain:
\begin{equation}
S^\intercal TS=\left(\begin{array}{ccccc}
0 & t_{0} & \eta_{p}^{-1}t_{0} & -\eta_{p}^{-2}t_{0} & -\eta_{p}^{-3}t_{0}\\
t_{0} & e^{i\left(\phi_{x}-2\phi_{t}\right)}x_{0} & z_{0}/\sqrt{2} & 0 & z_{0}/\sqrt{2}\\
\eta_{p}^{-1}t_{0} & z_{0}/\sqrt{2} & e^{i\left(\phi_{x}-2\phi_{t}\right)}x_{0} & -z_{0}/\sqrt{2} & 0\\
\eta_{p}^{-2}t_{0} & 0 & z_{0}/\sqrt{2} & -e^{i\left(\phi_{x}-2\phi_{t}\right)}x_{0} & -z_{0}/\sqrt{2}\\
\eta_{p}^{-3}t_{0} & -z_{0}/\sqrt{2} & 0 & -z_{0}/\sqrt{2} & -e^{i\left(\phi_{x}-2\phi_{t}\right)}x_{0}
\end{array}\right)
\end{equation}
i.e., we can redefine $z$ and $t$ as real, non-negative parameters,
as well as the phase of $x$, and obtain that the most general parameterization
is as given by Eq. (\ref{Tgen}).

In the pure gauge case, where $t=0$, $\phi_t$ is not a relevant quantity any more. Then, we may
revisit equation (\ref{Tphases}) and replace equation (\ref{phase2}) by
\begin{equation}
\phi_u=\phi_r+\phi_x=\phi_x-\phi_z
\end{equation}
still taking equation (\ref{phase1}) into account. Finally we obtain that the final parametrization for the pure gauge states is given by (\ref{Tgen}), with
$t=0;\, x,z \in\mathbb{R};\,z,x \geq 0$; in particular, it implies that the signs of $x,z$ are not important and the physical state is thus invariant under
$x \rightarrow -x$ or $z \rightarrow -x$.

\section{Gaussian mapping} \label{app3}

Here we will briefly sketch the Gaussian mapping procedure which leads to the result of Sec. \ref{BCSsection}. For further details, the reader should refer to \cite{Zohar2015b}, which is following the procedure of \cite{Kraus2010}.

For each fermionic mode of the fiducial state, described by the creation and annihilation operators $\alpha_i^{\dagger},\alpha_i$, let us define the Majorana operators
\begin{equation}
c_{2i-1}=\alpha_i+\alpha_i^{\dagger}\,;\quad c_{2i}= i\left( \alpha_i - \alpha_i^{\dagger}\right)
\end{equation}
out of which one may calculate the covariance matrix of the fiducial state,
\begin{equation}
M_{ij}=\frac{i}{2}\left\langle\left[c_i,c_j\right]\right\rangle = \left(
    \begin{array}{cc}
      A & B\\
      -B^\intercal & D \\
    \end{array}
  \right)
\end{equation}
with the blocks $A,B,D$ corresponding to the correlations of physical modes with physical modes, physical with virtual and virtual with virtual modes respectively (this can be obtained by transforming the $\varGamma$ of eq. (\ref{fercov}) into the basis of Majorana modes).

After demanding the global U(1) invariance, we are left with a single parameter $t$, and the PEPS decomposes into a product of two identical parts, and thus it suffices to consider the $M$ matrix of such a single part. $M$ is then very simple to calculate.

On the links, we have Gaussian states as well, $\left|H\right\rangle$ and $\left|V\right\rangle$, which have their own covariance matrices. These two are equal, and expressed by
\begin{equation}
\Gamma_0 = \left(
  \begin{array}{cc}
    0 & -\sigma_x \\
    \sigma_x & 0 \\
  \end{array}
\right)
\end{equation}
where the Majorana modes are ordered as follows (in pairs): $l,r$ in the horizontal case and $u,d$ in the vertical one.

Since the state has translation invariance, we expect everything to be block-diagonal in momentum space. After performing a Fourier transform, $M$ trivially keeps its form, but for the link states we obtain the covariance matrix $G_{\text{in}}\left(\mathbf{k}\right)$ for every $\mathbf{k}$ in the Brillouin zone,
\begin{equation}
G_{\mathrm{in}}\left(\mathbf{k}\right)=\left(\begin{array}{cc}
0 & -\sigma_{x}e^{ik_{1}}\\
\sigma_{x}e^{-ik_{1}} & 0
\end{array}\right)\oplus\left(\begin{array}{cc}
0 & -\sigma_{x}e^{-ik_{2}}\\
\sigma_{x}e^{ik_{2}} & 0
\end{array}\right)
\end{equation}

The blocks of the momentum space covariance matrix of the whole PEPS $\left|\psi\right\rangle$ may be obtained then from $M,G_{\mathrm{in}}$ using a Gaussian mapping \cite{Bravyi05,Kraus2010},
\begin{equation}
G_{\mathrm{out}}\left(\mathbf{k}\right)=A+B\left(D-G_{\mathrm{in}}\left(\mathbf{k}\right)\right)^{-1}B^\intercal
\end{equation}
and the result is
\begin{equation}
G_{out}\left(\mathbf{k}\right)=\left(\begin{array}{cc}
iP\left(\mathbf{k}\right) & R\left(\mathbf{k}\right)+iI\left(\mathbf{k}\right)\\
-R\left(\mathbf{k}\right)+iI\left(\mathbf{k}\right) & -iP\left(\mathbf{k}\right)
\end{array}\right)
\end{equation}
(all the functions are real).

 We define
\begin{equation}
\Delta\left(\mathbf{k}\right)=P\left(\mathbf{k}\right)-iI\left(\mathbf{k}\right)
\end{equation}
which satisfy \cite{Kraus2010}
\begin{equation}
R^{2}\left(\mathbf{k}\right)+P^{2}\left(\mathbf{k}\right)+I^{2}\left(\mathbf{k}\right)=1
\end{equation}
\begin{equation}
d\left(\mathbf{k}\right)\equiv\det\left(D-G_{\mathrm{in}}\left(\mathbf{k}\right)\right)
\end{equation}
\begin{equation}
R_{0}\left(\mathbf{k}\right)=d\left(\mathbf{k}\right)R\left(\mathbf{k}\right)
\end{equation}
\begin{equation}
P_{0}\left(\mathbf{k}\right)=d\left(\mathbf{k}\right)P\left(\mathbf{k}\right)
\end{equation}
\begin{equation}
I_{0}\left(\mathbf{k}\right)=d\left(\mathbf{k}\right)I\left(\mathbf{k}\right)
\end{equation}

This corresponds to the p-wave BCS state
\begin{equation}
\left|\psi\right\rangle =\underset{\mathbf{k}}{\otimes}\left(\alpha\left(\mathbf{k}\right)+\beta\left(\mathbf{k}\right)\psi^{\dagger}\left(\mathbf{k}\right)\psi^{\dagger}\left(-\mathbf{k}\right)\right)\left|\Omega\left(\mathbf{k}\right)\right\rangle
\end{equation}
(for each copy of the PEPS, or each color). The unpaired momentum modes in the Brillouin zone will be in the vacuum state (see \cite{Zohar2015b} for further explanations).

Define the physical Majorana modes,
\begin{equation}
c_1\left(\mathbf{x}\right) = \psi\left(\mathbf{x}\right)+\psi^{\dagger}\left(\mathbf{x}\right); \quad
c_2\left(\mathbf{x}\right) = i\left( \psi\left(\mathbf{x}\right)-\psi^{\dagger}\left(\mathbf{x}\right) \right)
\end{equation}
and their Fourier transforms
\begin{equation}
d_a\left(\mathbf{k}\right) = \frac{1}{L}\underset{\mathbf{x}}{\sum}e^{-i \mathbf{k} \cdot \mathbf{x}} c_a\left(\mathbf{x}\right)
\end{equation}
where $L=\sqrt{L_1L_2}$ is the geometric mean of the system's dimensions. Then,
\begin{equation}
\left(G_{\text{out}}\left(\mathbf{k}\right)\right)_{ij} = \frac{1}{2}\left\langle\left[c_i\left(\mathbf{k}\right),c_j\left(\mathbf{k}\right)\right]\right\rangle
\end{equation}
Thus, the correlations of quadratic operators may easily be derived from
the covariance matrix elements. In momentum space, one has
\begin{equation}
\left\langle \psi_{\alpha}^{\dagger}\left(\mathbf{k}\right)\psi_{\beta}\left(\mathbf{q}\right)\right\rangle =\frac{1}{2}\delta_{\alpha\beta}\delta_{\mathbf{k},\mathbf{q}}\left(1-R\left(\mathbf{k}\right)\right)
\end{equation}
\begin{equation}
\left\langle \psi_{\alpha}\left(\mathbf{k}\right)\psi_{\beta}\left(\mathbf{q}\right)\right\rangle =-\frac{1}{2}\delta_{\alpha\beta}\delta_{\mathbf{k},\mathbf{-q}}\Delta\left(\mathbf{k}\right)=-\frac{1}{2}\delta_{\alpha\beta}\delta_{\mathbf{k},\mathbf{-q}}\left(P\left(\mathbf{k}\right)-iI\left(\mathbf{k}\right)\right)
\end{equation}

Define the Fourier transform by
\begin{equation}
\begin{aligned}\psi_{\alpha}^{\dagger}\left(\mathbf{x}\right) & = & \frac{1}{L}\underset{\mathbf{k}}{\sum}e^{i\mathbf{k\cdot x}}\end{aligned}
\psi_{\alpha}^{\dagger}\left(\mathbf{k}\right)
\end{equation}
and then
\begin{equation}
\left\langle \psi_{\alpha}^{\dagger}\left(\mathbf{x}\right)\psi_{\beta}\left(\mathbf{y}\right)\right\rangle =\frac{1}{2}\delta_{\alpha\beta}\left(\delta_{\mathbf{x},\mathbf{y}}-\hat{R}\left(\mathbf{x-y}\right)\right)
\end{equation}
\begin{equation}
\left\langle \psi_{\alpha}\left(\mathbf{x}\right)\psi_{\beta}\left(\mathbf{y}\right)\right\rangle =-\frac{1}{2}\delta_{\alpha\beta}\hat{\Delta}\left(\mathbf{\mathbf{x-y}}\right)
\end{equation}
with
\begin{equation}
\hat{f}\left(\mathbf{x}\right)=\frac{1}{L^{2}}\underset{\mathbf{k}}{\sum}e^{i\mathbf{k\cdot x}}f\left(\mathbf{k}\right)
\end{equation}

Rotation invariance implies
\begin{equation}
R\left(\Lambda\mathbf{k}\right)=R\left(\mathbf{k}\right)\,;\quad \Delta\left(\Lambda\mathbf{k}\right)=-i\Delta\left(\mathbf{k}\right)
\end{equation}
(for an explanation of this and further properties of the physical covariance matrix and its elements, refer to \cite{Zohar2015b}).

For the PEPS, as discussed in the main text, everything can be solved analytically:
\begin{equation}
\alpha\left(\mathbf{k}\right)=1
\end{equation}
\begin{equation}
\beta\left(\mathbf{k}\right)=2t^{2}\left(\sin\left(k_{1}\right)-i\sin\left(k_{2}\right)\right)
\end{equation}
from which we get
\begin{equation}
R\left(\mathbf{k}\right)=\frac{\left|\alpha\left(\mathbf{k}\right)\right|^{2}-\left|\beta\left(\mathbf{k}\right)\right|^{2}}{\left|\alpha\left(\mathbf{k}\right)\right|^{2}+\left|\beta\left(\mathbf{k}\right)\right|^{2}}=\frac{1-4t^{4}\left(\sin^{2}\left(k_{1}\right)+\sin^{2}\left(k_{2}\right)\right)}{1+4t^{4}\left(\sin^{2}\left(k_{1}\right)+\sin^{2}\left(k_{2}\right)\right)}
\end{equation}
\begin{equation}
\Delta\left(\mathbf{k}\right)=\frac{2\overline{\alpha}\left(\mathbf{k}\right)\beta\left(\mathbf{k}\right)}{\left|\alpha\left(\mathbf{k}\right)\right|^{2}+\left|\beta\left(\mathbf{k}\right)\right|^{2}}=\frac{4t^{2}\left(\sin\left(k_{1}\right)-i\sin\left(k_{2}\right)\right)}{1+4t^{4}\left(\sin^{2}\left(k_{1}\right)+\sin^{2}\left(k_{2}\right)\right)}
\end{equation}

The parent Hamiltonian, of whose our PEPS is the ground state, is obtained from the covariance matrix too, with a proper choice of the dispersion relation.
We define the energy spectrum as
\begin{equation}
E\left(\mathbf{k}\right)=\left|\alpha\left(\mathbf{k}\right)\right|^{2}+\left|\beta\left(\mathbf{k}\right)\right|^{2}=1+4t^{4}\left(\sin^{2}\left(k_{1}\right)+\sin^{2}\left(k_{2}\right)\right)
\end{equation}
and, following the procedure of \cite{Zohar2015b}, obtain the parent Hamiltonian (\ref{parham}).

\section{Transformation rules for the transfer matrix} \label{app:transfer}

In this Appendix we provide a more detailed calculation for the transformation rules of the local transfer matrices $\mathcal{E}$ and $\mathcal{F}_U$.
To this purpose let us recall an explicit form for the unitary operators responsible for left and right transformation of the gauge fields degrees of freedom:
\begin{align}
 &\Theta_g^{t/s} = \sum_{jmnl} D^j_{nl}(g) \ket{jmn}\bra{jml} \,,\\
 &\widetilde{\Theta}_g^{t/s} = \sum_{jmnl} D^j_{mn}(g) \ket{jml}\bra{jnl} \,.
\end{align}

One of the crucial elements in the calculation of the transformation relations for both $\mathcal{E}$ and $\mathcal{F}_U$ are the symmetries of the gauge field triplet states $\bra{\Phi}_{s/t}$ defined in Eq. \eqref{triplet}. In particular let us consider the effect of generic transformation $\widetilde{\Theta}_g$ and $\widetilde{\Theta}_g'$ applied to the normal and primed degrees of freedom respectively. We have:
\begin{multline}
 \bra{\Phi}_{t/s}\, {\Theta}_g^{t/s} \otimes {\Theta}_{g'}^{t'/s'} = \frac{1}{\sqrt{5}} \sum_{jmnlj'm'n'l'}  \left( \bra{jmn} \otimes \bra{j'm'n'}\right)  \, \delta_{jj'}\delta_{mm'}\delta_{nn'} \left(  D^j_{nl}(g) \ket{jmn}\bra{jml} \otimes D^{j'}_{n'l'}(g') \ket{j'm'n'}\bra{j'm'l'}\right) = \\
 =\frac{1}{\sqrt{5}} \sum_{jmnll'} \bra{jml}\otimes \bra{jml'}\; D^j_{nl}(g) D^{j}_{nl'}(g')\,;
\end{multline}
from this equation we observe that, when $D^{j}_{nl'}(g')=D^{j}_{l'n}(g^{-1})={\bar D}^{j}_{nl'}(g)$, then the state $\bra{\Phi}_{s/t}$ is invariant. In particular, since only $j=0$ and $j=1/2$ are considered here, we can consider the group element $z$ such that $D^{j=1/2}(z)=\epsilon$ and we obtain that $D^{j*}(g)=\epsilon^\intercal D^j(g) \epsilon$. Therefore we obtain that:
\begin{equation} \label{phiinv}
 \bra{\Phi}_{t/s}\, {\Theta}_g^{t/s} \otimes {\Theta}_{z^{-1}gz}^{t'/s'}=\bra{\Phi}_{s/t}\,,
\end{equation}
and a similar relation hold for left transformations. We are now ready to discuss the symmetries of $\mathcal{E}$. Let us consider, for example, Eq. \eqref{treeven}:
\begin{multline}
\Theta_g^{\tilde{d}\dag}\otimes \widetilde{\Theta}_g^{\tilde{d}'} \mathcal{E}_e =  \bra{\Phi_t}\bra{\Phi_s}P_\psi \varpi \Theta_g^{\tilde{d}\dag} \zeta A^G \otimes \varpi' \widetilde{\Theta}_g^{\tilde{d}'} \zeta'A^{G\prime}\ket{\Omega} = \\
= \bra{\Phi_t}\bra{\Phi_s}P_\psi \varpi  \zeta \widetilde{\Theta}_g^{u} A^G \otimes \varpi'  \zeta' {\Theta}_g^{u'\dag} A^{G\prime}\ket{\Omega} = 
\bra{\Phi_t}\bra{\Phi_s}P_\psi \varpi  \zeta \widetilde{\Theta}_g^{u} A^G \otimes \varpi'  \zeta' \widetilde{\Theta}_{z^{-1}gz}^{u'} A^{G\prime}\ket{\Omega} = \\
=\bra{\Phi_t}{\Theta}_g^{t}{\Theta}_{z^{-1}gz}^{t'}\bra{\Phi_s}P_\psi \varpi  \zeta  A^G \otimes \varpi'  \zeta'  A^{G\prime}\ket{\Omega} =\bra{\Phi_t}\bra{\Phi_s}P_\psi \varpi  \zeta  A^G \otimes \varpi'  \zeta'  A^{G\prime}\ket{\Omega}=\mathcal{E}\,,
\end{multline}
where we used first the transformation of the operator $\zeta$ in \eqref{transfproj2}, then the relation $\Theta^{u'\dag}_g=\widetilde{\Theta}_{z^{-1}gz}^{u'}$ and the transformations \eqref{Urrel} acting on $A^G$ and $A^{G\prime}$ for even sites, and finally the invariance relation of $\Phi^t$ in Eq. \eqref{phiinv}. We emphasize that the previous result is based on $\ket{\Omega}$ being the global vacuum such that $\Theta^{\tilde{d}}_g\ket{\Omega}=\widetilde{\Theta}^{u\dag}_g\ket{\Omega}=\Theta^{t\dag}_g\ket{\Omega}=\ket{\Omega}$. The other relations in Eqs. \eqref{treeven} and \eqref{treodd} can be obtained in an analogous way. Also the relations \eqref{treeven2} and \eqref{treodd2} are can be obtained in a similar way but they rely also on the invariance of the operator $P_\psi$: 
\begin{multline}
 P_\psi \, \Theta^p_g \otimes \widetilde{\Theta}^{p'\dag}_g = \frac{1}{2} \prod_n \left(\psi_n\psi_n^\dag \psi_n'\psi_n^{\prime\dag}\right) \prod_{mm'}\left(1+\psi_m\psi_m' \right)\delta_{mm'} \,  \Theta^p_g \otimes \widetilde{\Theta}^{p'\dag}_g = \\
 =\frac{1}{2} \prod_n \left(\psi_n\psi_n^\dag \psi_n'\psi_n^{\prime\dag}\right) \prod_{mm'}\left(1+\Theta^{p\dag}_g\psi_m\Theta^p_g \widetilde{\Theta}^{p'}_g\psi_m'\widetilde{\Theta}^{p'\dag}_g \right)\delta_{mm'} = \frac{1}{2} \prod_n \left(\psi_n\psi_n^\dag \psi_n'\psi_n^{\prime\dag}\right) \prod_{ll'm}\left(1+\psi_l\psi'_{l'} \right) D_{lm}(g) D_{ml'}(g^{-1})=P_\psi\,.
\end{multline}
Let us finally consider the transformations of $\mathcal{F}_U$ and, in particular, Eq. \eqref{trfeven}:
\begin{multline}
\Theta_g^{\tilde{d}\dag}\otimes \widetilde{\Theta}_g^{\tilde{d}'} \left( \mathcal{F}_U^e\right)_{\tilde{m}m}  =  \bra{\Phi_t}\bra{\Phi_s}P_\psi U^t_{m\tilde{m}} \varpi \Theta_g^{\tilde{d}\dag} \zeta A^G \otimes \varpi' \widetilde{\Theta}_g^{\tilde{d}'} \zeta'A^{G\prime}\ket{\Omega} = \\
= \bra{\Phi_t}\bra{\Phi_s}P_\psi U^t_{m\tilde{m}} \varpi  \zeta \widetilde{\Theta}_g^{u} A^G \otimes \varpi'  \zeta' {\Theta}_g^{u'\dag} A^{G\prime}\ket{\Omega} = 
\bra{\Phi_t}\bra{\Phi_s}P_\psi U^t_{m\tilde{m}} \varpi  \zeta \widetilde{\Theta}_g^{u} A^G \otimes \varpi'  \zeta' \widetilde{\Theta}_{z^{-1}gz}^{u'} A^{G\prime}\ket{\Omega} = \\
=\bra{\Phi_t}{\Theta}_{z^{-1}gz}^{t'}\bra{\Phi_s}P_\psi U^t_{m\tilde{m}}{\Theta}_g^{t} \varpi  \zeta  A^G \otimes \varpi'  \zeta'  A^{G\prime}\ket{\Omega} =\bra{\Phi_t}\bra{\Phi_s}P_\psi \left[ {\Theta}_g^{t\dag}U^t_{m\tilde{m}}{\Theta}_g^{t}\right]  \varpi  \zeta  A^G \otimes \varpi'  \zeta'  A^{G\prime}\ket{\Omega} =\\
=\bra{\Phi_t}\bra{\Phi_s}P_\psi \left[ U^t_{ml}D_{l\tilde{m}}(g^{-1})\right]  \varpi  \zeta  A^G \otimes \varpi'  \zeta'  A^{G\prime}\ket{\Omega}=\left( \mathcal{F}_{UD(g^{-1})}^e\right)_{\tilde{m}m}=D^\intercal_{\tilde{m}\tilde{m}'}(g^{-1})\left( \mathcal{F}_U^e\right)_{\tilde{m}'m}\,.
\end{multline}
Here we repeated the same steps as in the calculations of $\mathcal{E}$, but additional care must be taken due to the presence of the observable $U^t$.
The other transformation relations in Eqs. (\ref{trfeven}-\ref{trfodd2}) can be derived in an analogous way.

Finally, we address the transformation relations of the state $\mathcal{F}_U'$ defined in \eqref{Fprime}. Let us consider first the effect of Eq. \eqref{trfeven}:
\begin{equation}
 \Theta_g^{\tilde{d}\dag}\otimes \widetilde{\Theta}_g^{\tilde{d}'} \ket{\mathcal{F}_U^{e\prime}}=\frac{1}{\sqrt{2}} \sum_{m\check{m}}\Theta_g^{\tilde{d}\dag}\otimes \widetilde{\Theta}_g^{\tilde{d}'} \left( \mathcal{F}_U^e\right)_{\check{m}m}\ket{m}\ket{\check{m}}=\frac{1}{\sqrt{2}} \sum_{m\check{m}\check{m}'} D^\intercal_{\check{m}\check{m}'}(g^{-1}) \left( \mathcal{F}_U^e\right)_{\check{m}'m}\ket{m}\ket{\check{m}}=\frac{1}{\sqrt{2}} \sum_{m\check{m}\check{m}'}\left( \mathcal{F}_U^e\right)_{\check{m}'m}\ket{m}D_{\check{m}'\check{m}}(g^{-1})\ket{\check{m}}\,,
\end{equation}
which shows that the state $\ket{\check{m}}$ is mapped into $D_{\check{m}'\check{m}}(g^{-1})\ket{\check{m}}$ leading to Eq. \eqref{H1}. Analogously we can derive Eq. \eqref{H2} starting from Eq. \eqref{trfodd2}:
\begin{equation}
 \widetilde{\Theta}_g^{d}\otimes {\Theta}_g^{d'\dag} \ket{\mathcal{F}_U^{o\prime}} = \frac{1}{\sqrt{2}}\widetilde{\Theta}_g^{l\dag}\otimes {\Theta}_g^{l'}\sum_{mm'\check{m}}\left( \mathcal{F}_U^o\right)_{\check{m}m'} D^\intercal_{m'm}(g)\ket{m}\ket{\check{m}} \,,
\end{equation}
which shows that  $\ket{m}$ is mapped into $D^\intercal_{m'm}(g)\ket{m}$ up to the gauge transformation of the degrees of freedom on the left link, which, however, is not
influencial due to the property \eqref{treeven} of the neighboring transfer matrix block $\mathcal{E}$. This leads to Eq. \eqref{H2}. The relations \eqref{H3} and \eqref{H4} for odd bonds can be derived in an analogous way.

\newpage

\bibliography{ref}

\end{document}